\documentclass[%
 reprint,
 amsmath,amssymb,
]{revtex4-1}

\usepackage{graphicx}
\usepackage{dcolumn}
\usepackage{bm}
\usepackage{subcaption}
\usepackage{caption}

\newcommand{\etal}{{\itshape et al.}}

\begin{document}

\title{Prediction of shear-thickening of particle suspensions in viscoelastic fluids 
\\
by direct numerical simulation}%

\author{Yuki Matsuoka\(^{ab}\)}
\email{ymatsuoka@sumibe.co.jp}
\author{Yasuya Nakayama\(^{b}\)}
\email{nakayama@chem-eng.kyushu-u.ac.jp}
\author{Toshihisa Kajiwara\(^{b}\)}

\affiliation{%
\(^{a}\)
Corporate Engineering Center, Sumitomo Bakelite Co., Ltd., Shizuoka 426-0041, Japan}%
\affiliation{%
\(^{b}\)Department of Chemical Engineering,
Kyushu University,
Nishi-ku,
Fukuoka 819-0395,
Japan
}%

\date{\today}

\begin{abstract}
{To elucidate the key factor for the quantitative prediction of the shear-thickening in suspensions in viscoelastic fluids, direct numerical simulations of many-particle suspensions in a multi-mode Oldroyd-B fluid are performed using the smoothed profile method.} 
Suspension flow under simple shear flow is solved under periodic 
boundary conditions by using Lees--Edwards boundary conditions for particle dynamics and a time-dependent oblique coordinate system that evolves with mean shear flow for fluid dynamics.
Semi-dilute many-particle suspensions up to a particle volume fraction of 0.1 are investigated.
The presented numerical results regarding the bulk rheological properties of the shear-thickening behavior agree quantitatively with recent experimental results of semi-dilute suspensions {in a Boger fluid. 
The presented result clarifies that an accurate estimation of the first normal stress difference of the matrix in the shear-rate range where the shear-thickening starts to occur is crucial for the quantitative prediction of the suspension shear-thickening in a Boger fluid matrix at around the Weissenberg number $\rm{Wi}=1$ by an Oldroyd-B model. Additionally, the effect of suspension microstructures on the suspension viscosity is examined.}
The paper concludes with a discussion on how the flow pattern and the elastic stress development change with the volume fraction and Weissenberg number.
\end{abstract}

\maketitle

\section{\label{sec:level1} Introduction}

\noindent Suspension systems consisting of solid particles and a 
polymeric host fluid are widely used in industrial materials and 
products such as inks, paints, polymer composites. In the 
manufacturing processes, such suspensions are subject to various 
types of flow, hence understanding and controlling the 
rheological properties of them are crucial for efficient 
productivity. In a polymeric fluid, including polymer solutions 
and melts, viscoelasticity originates from the change in the 
conformation of polymer molecules caused by flow history. Since 
the polymeric host fluid exhibits viscoelasticity, the 
interaction between the particles and flow in suspensions in 
viscoelastic fluid flow is elusive. For instance, unique behavior 
not observed in Newtonian media has been reported, such as 
shear-thickening even in a dilute particle concentration under 
simple shear flow~\citep{Tanner2019,Shaqfeh2019} and the 
formation of a string of particles under shear 
flow~\citep{Michele1977,Scirocco2004}.

To examine the medium's elastic effects on the suspension 
rheology, suspensions in Boger fluids have been used 
experimentally. Boger fluids show the constant shear viscosity 
and finite normal stress difference~(NSD), which is preferable for 
separating the effects of the medium's elasticity from the 
non-linear effects in the shear viscosity. Experimentally 
measured shear-thickening in suspensions in Boger fluids has been 
reported where the suspension viscosity increases with shear-rate 
or shear stress, even at dilute particle concentrations where the 
inter-particle interactions are 
negligible~\citep{Zarraga2001,Scirocco2005,Tanner2013,Tanner2015}. 
The shear-thickening mechanism has been discussed 
theoretically~\citep{Koch2016,Einarsson2018} and 
numerically~\citep{Yang2016,Yang2018,Shaqfeh2019,Vazquez-Quesada2019,Matsuoka2020}. These theoretical and numerical studies reveal 
that this shear-thickening in dilute viscoelastic suspensions is 
mainly originated by the development of polymeric stress around 
the particles. While the qualitative shear-thickening mechanism 
has become progressively clearer, there are still some 
discrepancies between numerical calculations and measurements in 
the quantitative prediction of shear-thickening behaviours in 
viscoelastic suspensions.

To evaluate the complex responses of a viscoelastic suspension 
under different types of flow, direct numerical simulations~(DNS) 
are carried out, in which the fluid flow around finite-volume 
solids rather than point masses is solved, to accurately treat 
hydrodynamic interactions. A few computational studies have 
reported the dynamics of many-particle systems in viscoelastic 
suspensions~\citep{Hwang2004,Jaensson2015,Vazquez-Quesada2017,Vazquez-Quesada2019,Yang2018a}.
Experimentally measured and DNS obtained shear-thickening in viscoelastic suspensions were compared.
A scaling relation between the shear-thickening part and the 
suspension stress up to semi-dilute particle volume fraction 
$\phi_p\leq 0.1$ has been discussed based on the results of 
immersed-boundary many-particle DNS using a Giesekus fluid 
mimicking a Boger fluid from~\citet{Tanner2013}~\citep{Yang2018a}. 
However, the relative suspension viscosity predicted by using the 
scaling relation and the numerical result from a single-particle 
dilute suspension in an Oldroyd-B medium resulted in an 
underestimation of the experimental shear-thickening at 
$\phi_{p}\leq 0.1$~\citep{Yang2018a}.
To explain the discrepancy, 
a lack of constitutive modelling 
of the elongational response in the fluid was pointed out.
\citet{Vazquez-Quesada2019} performed a smoothed particle 
hydrodynamics~(SPH) simulation using an Oldroyd-B medium up to 
$\phi_{p}\leq 0.3$, and showed that the relative suspension 
viscosity from a many-particle simulation is larger than that from 
a single-particle simulation even at a dilute particle volume 
fraction, thus indicating that the interaction between particles 
is important even in dilute suspensions.
The corresponding numerical result for the suspension viscosity agrees quantitatively with experimental data for a dilute suspension ($\phi_{p}=0.05$) but was diverted for semi-dilute conditions ($\phi_{p}=0.1, 0.3$).
It is still unclear whether the Oldroyd-B model can 
quantitatively predict shear-thickening in semi-dilute 
suspensions in Boger fluids.

In this study, the smoothed profile method~(SPM), which is a DNS 
method originally developed for Newtonian suspension systems, is 
extended to study the bulk shear rheology of a suspension in a 
viscoelastic medium in a three-dimensinal~(3D) space. To impose simple shear flow on 
a suspension under periodic boundary conditions rather than 
wall-driven shear flow in a confined system, a time-dependent 
oblique coordinate system is used for the fluid; its formulation 
conforms to Lees--Edwards boundary conditions for particle 
dynamics and is preferred for examining the bulk stress as well 
as local stress in suspensions without wall effects.

To elucidate the key factor for the quantitative prediction of 
the shear-thickening in suspensions in Boger fluids, DNS of 
many-particle suspensions in a multi-mode Oldroyd-B fluid is 
performed using SPM. The suspension viscosity and the NSD are 
compared with published experimental results~\citep{Yang2018a} at 
dilute to semi-dilute conditions. {Additionally, the effect of 
suspension microstructures on the suspension viscosity is 
examined by comparing a many-particle system with a 
single-particle system which corresponds to a cubic array 
suspension in our DNS. Next, the contribution of each polymer 
relaxation mode to the suspension shear-thickening is evaluated. 
The suspension stress decomposition into the stresslet and the  
particle-induced fluid stress is conducted to discuss scaling 
relations for these contributions.
Finally, the change in the flow pattern and elastic stress development in many-particle suspensions is discussed.

The paper is organized as follows. In Sec.~\ref{sc:numerical}, 
our numerical method is explained. The governing equations for a 
suspension in a viscoelastic medium based on a smoothed profile 
of particles are described in Sec.~\ref{sc:governing_eq}. The 
calculation of stress for the rheological evaluation in SPM is 
described in Sec.~\ref{sc:stress}. {The boundary conditions are 
explained in Sec.~\ref{sc:bc}.} In Sec.~\ref{sc:apps}, the 
numerical results are presented. First, our DNS method is 
validated by the rheological evaluation for a single-particle 
system in a single-mode Oldroyd-B fluid in 
Sec.~\ref{sc:single_rheo}. Next, shear-thickening behviours in 
dilute and semi-dilute viscoelastic suspensions are studied by 
performing a many-particle calculation in a multi-mode Oldroyd-B 
fluid in Sec.~\ref{sc:many_rheo}. The results are summarized in 
Sec.~\ref{sc:summary}.

\section{\label{sc:numerical}Numerical Method}

\noindent In SPM, the fluid--solid interaction is treated by applying the smoothed profile function of a solid particle~\citep{Nakayama2005,Nakayama2008}.  
Since a regular mesh rather than a surface-conforming mesh can be 
used for continuum calculations in SPM, the calculation cost of 
fluid fields, which is dominant in total calculation costs, is 
nearly independent of the number of particles~\citep{Nakayama2008}, 
thus making the direct simulation of a many-particle system feasible. SPM has been applied to suspensions in Newtonian fluids to evaluate the shear viscosity~\citep{Iwashita2009,Kobayashi2011,Molina2016},  complex modulus~\citep{Iwashita2010}, and particle coagulation rate~\citep{Matsuoka2012} of Brownian suspensions up to $\phi_p \leq 0.56$.
The application of SPM was extended to complex host fluids, such as electrolyte solutions~\citep{Kim2006,Nakayama2008,luo2010} and to active swimmer suspensions~\citep{Molina2013}.

\subsection{Governing equations}
\label{sc:governing_eq}

\noindent Consider the suspension of $N$ neutrally buoyant and non-Brownian 
spherical particles with radius $a$, mass $M$, and moment of inertia {$\bm{I}_p=2Ma^2\bm{I}/5$} in a viscoelastic fluid, where $\bm{I}$ is the unit tensor. 
In SPM, the velocity field $\bm{u}(\bm{r}, t)$ at position $\bm{r}$ and time $t$ is governed as follows:
\begin{align}
\label{eq:navier-stokes}
\rho\left(\frac{\partial}{\partial t}+\bm{u}\cdot\nabla\right)\bm{u}&=\nabla\cdot(\bm{\sigma}_n+\bm{\sigma}_p)+\rho\phi\bm{f}_p, \\
\label{eq:cnt_eq}
\nabla\cdot\bm{u}&=0,
\end{align}
where $\rho$, 
$\bm{\sigma}_n=-p\bm{I}+2\eta_s\bm{D}$, $\bm{\sigma}_p$, $\bm{D}=(\nabla\bm{u}+\nabla\bm{u}^T)/2$, and $p$ 
are the fluid mass density, 
Newtonian solvent stress, polymer stress, strain-rate tensor, and pressure, respectively. {In this study, the polymer stress term is newly incorporated into the previous hydrodynamic equation for a Newtonian fluid in SPM.}
In SPM, the particle profile field is introduced as 
$\phi(\bm{r},t)\equiv\sum_{i=1}^N\phi_i$, where $\phi_i\in[0,1]$ 
is the $i$-th particle profile function having a continuous diffuse 
interface domain with thickness $\xi$; the inside and outside 
of the particles are indicated by $\phi=1$ and $\phi=0$, respectively. Details
on the specific definition and the properties of the profile
function were reported by~\citet{Nakayama2008}. 
The body force $\rho\phi\bm{f}_p$ in Eq.~(\ref{eq:navier-stokes}) enforces particle rigidity in the velocity field~\citep{Nakayama2008,Molina2016}.
In SPM, the continuum velocity field is defined in the entire domain, including the fluid and solids. The velocity field $\bm{u}$ is interpreted as
\begin{align}
\bm{u}(\bm{r},t)=(1-\phi)\bm{u}_f+\phi\bm{u}_p,
\end{align}
where $\bm{u}_f$ and $\bm{u}_p$ are the fluid and particle velocity fields, respectively.  
The specific implementation of $\bm{u}_f,\bm{u}_p$, and $\phi\bm{f}_p$ is explained in Appendix~\ref{sc:numerical_implement}.

For the time evolution of polymer stress $\bm{\sigma}_p$, any constitutive equations proposed to reproduce the rheological behavior of real viscoelastic fluids can be used. In this study, the single- or multi-mode Oldroyd-B model, which is a minimal viscoelastic model for Boger fluids, is applied: 
\begin{align}
\label{eq:conformation}
\left(\frac{\partial}{\partial t}+\bm{u}\cdot\nabla\right)\bm{C}^{(k)}&=(\nabla\bm{u})^T\cdot\bm{C}^{(k)}+\bm{C}^{(k)}\cdot(\nabla\bm{u})-\frac{\bm{C}^{(k)}-\bm{I}}{\lambda^{(k)}},
\\
\label{eq:polymer_stress}
\bm{\sigma}_p&=\sum_k\bm{\sigma}_p^{(k)}=\sum_k\frac{\eta_p^{(k)}}{\lambda^{(k)}}(\bm{C}^{(k)}-\bm{I}),
\end{align}
where $\bm{C}^{(k)}(\bm{r},t),\lambda^{(k)}$, and $\eta_p^{(k)}$ 
are the conformation tensor, relaxation time, and polymer 
viscosity of the $k$-th relaxation mode, respectively. The 
conformation tensor of each relaxation mode $\bm{C}^{(k)}$ obeys 
an independent but same form of the constitutive equation as expressed by Eq.~(\ref{eq:conformation}). The total polymer stress is obtained by summing up the polymer stress of each mode $\bm{\sigma}_p^{(k)}$ by using Eq.~(\ref{eq:polymer_stress}). In the single-mode Oldroyd-B model, the mode index $k$ is omitted for simplicity. 
Microscopically, an Oldroyd-B fluid corresponds to a dilute 
suspension of dumbbells with a linear elastic spring in a 
Newtonian solvent~\citep{Bird1987}.   The conformation tensor is 
related to the average stretch and orientation of the dumbbells. 
The first and second terms on the right-hand side~(RHS) of 
Eq.~(\ref{eq:conformation}) represent the affine deformation of 
$\bm{C}^{(k)}$, by which $\bm{C}^{(k)}$ is rotated and stretched, 
and the last term is the irreversible relaxation of 
$\bm{C}^{(k)}$. At steady state in simple shear flow, the shear 
viscosity and the first and second NSDs are 
$\eta_0=\eta_s+\sum_k\eta_p^{(k)}$, 
$N_1=2\sum_k\eta_p^{(k)}\lambda^{(k)}\dot{\gamma}^2$, and zero, 
respectively, where $\dot{\gamma}$ indicates the applied shear 
rate. The steady-shear property of the Oldroyd-B model mimics 
that of Boger fluids and is characterized by rate-independent 
viscosity and finite elasticity. Boger fluids are often used to 
experimentally evaluate the effect of fluid elasticity separately 
from that of viscosity~\citep{Boger1977,James2009}. 

The individual particles evolve by 
\begin{align}
\dot{\bm{R}}_i&=\bm{V}_i,\label{eq:newton}\\
M_i\dot{\bm{V}}_i&=\bm{F}^H_i+\bm{F}^C_i,\label{eq:newton2}\\
\bm{I}_{p,i}\cdot\dot{\bm{\Omega}}_i&=\bm{N}^H_i,\label{eq:eular}
\end{align}  
where $\bm{R}_i$, $\bm{V}_i$, and $\bm{\Omega}_i$ are the 
position, velocity, and angular velocity of the $i$-th particle, 
respectively,
$\bm{F}_i^H$ and $\bm{N}_i^H$ are the hydrodynamic force and 
torque from the fluid~\citep{Nakayama2008,Molina2016}, 
respectively, and $\bm{F}_i^C$ is the inter-particle potential 
force due to the excluded volume that prevents particles from 
overlapping. {The non-slip boundary condition for the velocity 
field is assigned at particle surfaces.} The specific 
implementation of $\bm{F}_i^H,\bm{N}_i^H$, and $\bm{F}_i^C$ is 
explained in Appendix~\ref{sc:numerical_implement}. 

The governing equations can be non-dimesionalized by length unit 
$a$, velocity unit $a\dot{\gamma}$, and stress unit 
$\eta_0\dot{\gamma}$. 
In the following, a tilde variable ($\tilde{\cdot}$) indicates a 
non-dimensional variable. For the fluid momentum equation,
\begin{align}
{\rm Re}\left(\frac{\partial}{\partial \tilde{t}}+\tilde{\bm{u}}\cdot\tilde{\nabla}\right)\tilde{\bm{u}}=\tilde{\nabla}\cdot(\tilde{\bm{\sigma}}_n+\tilde{\bm{\sigma}}_p)+{\rm Re}\phi\tilde{\bm{f}}_p,
\end{align} where $\tilde{\bm{\sigma}}_n=-\tilde{p}\bm{I}+2\beta\tilde{\bm{D}}$ and the Reynolds number is defined as ${\rm Re}=\rho a^2\dot{\gamma}/\eta_0$. In this study, $\rm{Re}$ is kept small to exclude inertial effects from the rheological evaluations.
For the single-mode Oldroyd-B constitutive equation,
\begin{align}
\left(\frac{\partial}{\partial \tilde{t}}+\tilde{\bm{u}}\cdot\tilde{\nabla}\right)\bm{C}=(\tilde{\nabla}\tilde{\bm{u}})^T\cdot\bm{C}+\bm{C}\cdot(\tilde{\nabla}\tilde{\bm{u}})-\frac{\bm{C}-\bm{I}}{{\rm Wi}},
\end{align} where $\tilde{\bm{\sigma}}_p=(1-\beta)(\bm{C}-\bm{I})/{\rm Wi}$.
A single-mode Oldroyd-B fluid is characterized by two 
non-dimensional parameters: $\beta$ and $\rm{Wi}$. The viscosity 
ratio $\beta=\eta_s/\eta_0=\eta_s/(\eta_s+\eta_p)$ reflects the 
relative contribution of the solvent viscosity to the total 
zero-shear viscosity. The Weissenberg number is defined as ${\rm 
Wi}=\dot{\gamma}\lambda$ and measures the relative shear rate to 
the relaxation rate $1/\lambda$.

\subsection{Stress calculation}
\label{sc:stress}
\noindent {The momentum equation for the suspension is formally expressed as,
\begin{align}
\label{eq:eq_disp}
    \frac{D}{Dt}(\rho\bm{u})=\nabla\cdot\bm{\Sigma}^{\rm sus},
\end{align}
where $D/Dt$ is the material derivative and $\bm{\Sigma}^{\rm 
sus}$ represents the dispersion stress tensor, including the 
pressure, stresslet and fluid~(viscous and polymer) stress.}
To analyze the effect of solid inclusion in the suspension 
rheology, the instantaneous volume-averaged stress of the 
suspension {$\bm{\Sigma}^{\rm sus}$} is decomposed according 
to~\citet{Yang2016} as follows:
\begin{align}
\bm{\sigma}^{\rm sus}&=\frac{1}{V}\int_{D_V}\bm{\Sigma}^{\rm sus}d\bm{r}\\
&=\bm{\sigma}^{F0}+\frac{N}{V}(\bm{\Sigma}+\bm{S}),\label{eq:stress_decomp}\\
\label{eq:Sigma}
\bm{\Sigma}&=\frac{1}{N}\int_{D_V}(\bm{\sigma}^F-\bm{\sigma}^{F0})\mathrm{d}\bm{r},\\
\label{eq:stresslet}
\bm{S}&=\frac{1}{N}\int_{S_p}(\bm{r}(\bm{n}\cdot\bm{\sigma}^F))^{\rm sym}\mathrm{d}S.
\end{align} 
Here $D_V$ is the entire domain, and $V$ is the volume of $D_V$, 
and $S_p$ is the surface of the particles; $\bm{\sigma}^F$ is the 
stress in the fluid region and $\bm{\sigma}^{F0}$ is the fluid 
stress without particles under simple shear flow;
$(\bm{A})^{\rm sym}$ denotes the symmetric part of tensor 
$\bm{A}$; $\bm{\Sigma}$ represents the stress induced by particle 
inclusion per particle in the fluid region; and $\bm{S}$ is the 
stresslet. 
In the SPM formalism, by comparing 
Eq.(\ref{eq:navier-stokes}) with Eq.~(\ref{eq:eq_disp}), the 
following relation is obtained:
\begin{align}
\label{eq:suspension_stress_relation}
    \nabla\cdot\bm{\Sigma}^{\rm sus}=\nabla\cdot(\bm{\sigma}_n+\bm{\sigma}_p)+\rho\phi\bm{f}_p.
\end{align} Therefore, $\bm{\sigma}^{\rm sus}$ is evaluated as ~\citep{Nakayama2008,Iwashita2009,Molina2016},
\begin{align} 
\bm{\sigma}^{\rm sus}=\frac{1}{V}\int_{D_V}[\bm{\sigma}_n+\bm{\sigma}_p-\bm{r}\rho\phi\bm{f}_p]d\bm{r},\label{eq:suspension_stress}
\end{align}
where an identity for a second-rank tensor, 
$\bm{\sigma}=[\nabla\cdot(\bm{r\sigma})]^T-\bm{r}\nabla\cdot\bm{\
sigma}$ is used for the derivation. In this study, the Reynolds 
stress term is not considered due to the small-Reynolds-number 
conditions.
By assuming ergodicity, the ensemble average of the stress $\langle\bm{\sigma}^{\rm sus}\rangle$ is equated to the average over time. 

Evaluation of Eqs.~(\ref{eq:Sigma}) and (\ref{eq:stresslet}) requires surface or volume integrals. 
To calculate these integrals numerically using the immersed boundary method,
the appropriate location of the particle--fluid interface should be carefully examined~\citep{Yang2018a}. 
In contrast, in SPM, due to the diffuse interface of the smoothed profile function, 
both \(\bm{\Sigma}\) and \(\bm{S}\) are evaluated by the volume integral as follows.
By comparing Eq.~(\ref{eq:suspension_stress}) and Eqs.~(\ref{eq:stress_decomp})-(\ref{eq:stresslet}), we have
\begin{align}
\label{eq:Sigma_SP}
\bm{\Sigma}&\approx\frac{1}{N}\int_{D_V}
[\left(1-{\lfloor\phi\rfloor}\right)(\bm{\sigma}_n+\bm{\sigma}_p)^F-(\bm{\sigma}_n+\bm{\sigma}_p)^{F0}]d\bm{r} ,\\
\label{eq:S_SP}
\bm{S}&\approx-\frac{1}{N}\int_{D_V}\bm{r}\rho\phi\bm{f}_pd\bm{r}.
\end{align} 
Equation~(\ref{eq:S_SP}) indicates the relation between the 
stresslet and SPM body force $\rho\phi\bm{f}_p$. Since the 
stresslet is originated from the stress within a particle, it is 
calculated with $\rho\phi\bm{f}_p$ that originates from the 
particle rigidity.
Note that, in the particle region, there is no viscous stress or 
polymer stress, {\itshape i.e.}, $\bm{\sigma}^F=0$ in principle. 
In Eq.(\ref{eq:Sigma_SP}), this property is explicitly accounted 
for with the prefactor {$(1-\lfloor\phi\rfloor)$, where 
$\lfloor\cdot\rfloor$ is the floor function.} In practice, this 
prefactor is also effective in explicitly suppressing the 
accumulated numerical error in the stress field in the particle 
region when calculating $\bm{\Sigma}$.
This method of calculating the stress components in SPM was examined 
in our previous paper~\citep{Matsuoka2020}, and the results 
agreed with those determined by a surface-conforming mesh 
method~\citep{Yang2018}.

\begin{figure}    
\centerline{\includegraphics[scale=1.2]{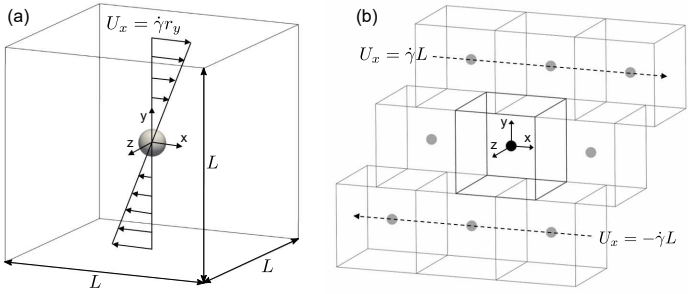}}
\caption{\label{fig:system} Schematic diagrams of the 
simulation setup: (a) single-particle system and (b) the sliding 
cell interpretation of Lees--Edwards boundary conditions. 
Here $U_x=\dot{\gamma}r_y$ represents the velocity of the mean shear 
flow; and $L$ is the box length of the cubic domain. In (b), the 
image cells along the vorticity direction are not shown for 
simplicity.}
\end{figure}

\subsection{Boundary conditions}
\label{sc:bc}\noindent To explain the boundary conditions of the 
sheared system, the single-particle system that is applied in 
Sec.~\ref{sc:single_rheo} is taken as an example. 
Fig.~\ref{fig:system} shows schematic diagrams of the simulation 
system. One particle is located in the center $(r_x=r_y=r_z=0)$ 
of a cubic domain of $[-L/2,\,L/2]^3$, where $L$ is the box 
length of the domain. Here $x,y$, and $z$ indicate the flow, 
velocity-gradient, and vorticity directions, respectively. Then, 
simple shear flow {$\bm{U}=\dot{\gamma}r_y\bm{e}_x$ is imposed by 
the time-dependent oblique coordinate system explained in 
Appendix~\ref{sc:tensor_express}, where $\bm{e}_i~(i=x,y,z)$ is 
the Cartesian basis set.} The corresponding velocity boundary 
conditions at the faces of the system are naturally established 
by the periodicity as follows: 
\begin{align}
\label{eq:pbc_x}
\bm{u}(L/2,r_y,r_z)&=\bm{u}(-L/2,r_y,r_z),\\
\label{eq:pbc_y}
\bm{u}(r_x,L/2,r_z)&=\bm{u}(r_x-\gamma L,-L/2,r_z)-\dot{\gamma}L\bm{e}_x,\\
\label{eq:pbc_z}
\bm{u}(r_x,r_y,L/2)&=\bm{u}(r_x,r_y,-L/2),
\end{align} 
where the simple periodic boundary conditions for the 
flow~(Eq.~(\ref{eq:pbc_x})) and vorticity~(Eq.~(\ref{eq:pbc_z})) 
directions and the shear periodic boundary condition for the 
velocity-gradient~(Eq.~(\ref{eq:pbc_y})) direction are 
established. The periodic boundary conditions for the 
conformation tensor are the same as 
Eqs.(\ref{eq:pbc_x})-(\ref{eq:pbc_z}) except that the last term 
in Eq.(\ref{eq:pbc_y}) is not included.
Lees--Edwards boundary conditions for particles can be 
interpreted as a sliding cell expression, as shown in 
Fig.~\ref{fig:system}(b). Initially, the image cells are aligned 
along all directions infinitely. Under simple shear flow, the 
upper and lower image cell layers stacked in the 
velocity-gradient direction slide in the flow direction with 
velocity $U_x=\pm\dot{\gamma}L$. The position and velocity of the 
particle going across the top and bottom faces of the main cell 
are modified as if the particle moved into the sliding image cell. 
These periodic boundary conditions in our method are preferred in 
evaluating bulk suspension rheology without the influence of the 
shear-driving walls. In our previous study, using this boundary 
condition, 3D steady shear simulations for a single-particle 
viscoelastic suspension system were conducted~\citep{Matsuoka2020}. 
Similar periodic boundary conditions were adopted for 
two-dimensional~(2D) steady shear flow 
simulations~\citep{Hwang2004,Jaensson2015} and 3D dynamic shear 
flow simulations~\citep{DAvino2013} of viscoelastic suspensions. 
In contrast to recent 3D steady shear flow simulations for 
many-particle systems which utilize walls to impose the shear 
flow~\citep{Yang2018a,Vazquez-Quesada2019}, this study presents 
for the first time wall-free 3D steady shear flow simulations for 
a many-particle viscoelastic suspension system. The details of 
the numerical solution procedure are described in 
Appendix~\ref{sc:numerical_implement}.

\section{Results and discussion}
\label{sc:apps}\noindent In this section, the developed DNS 
method is applied to the rheological evaluations of sheared 
viscoelastic suspensions. 
First, to show the validity of rheological evaluations by our 
developed DNS method, the suspension viscosity of the 
single-particle dilute system is evaluated and compared to 
previously reported numerical and theoretical results. {Further 
examinations of our DNS method are explained in Appendix 
\ref{sc:validation}.} Next, detailed rheological evaluation is 
conducted for a semi-dilute viscoelastic suspension, which 
contains many particles immersed in a multi-mode Oldroyd-B fluid, 
and the results are compared with previously reported 
experimental results.

\subsection{Suspension rheology of single-particle system}
\label{sc:single_rheo}

\begin{figure}    
\centerline{\includegraphics[scale=0.85]{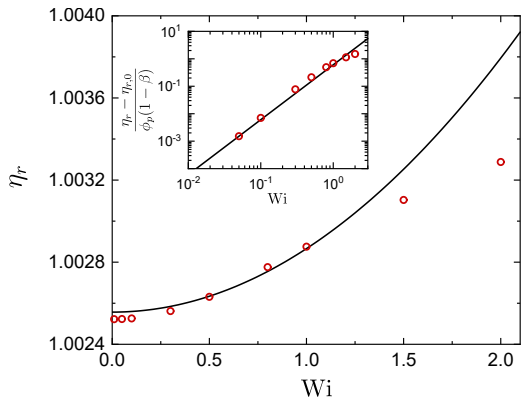}}
\caption{\label{fig:eta_r_einarrson} The $\rm{Wi}$ dependence of the 
relative viscosity of a dilute Oldroyd-B suspension at 
$\beta=0.5$. The inset shows the $\rm{Wi}$ dependence of the 
thickening part of $\eta_r$. 
Red open circles represent results from this work.
The black lines correspond to the 
theoretical prediction by~\citet{Einarsson2018} using 
Eq.~(\ref{eq:einarsson}).}
\end{figure}

\begin{figure}   
  \begin{center}
\includegraphics[scale=0.75]{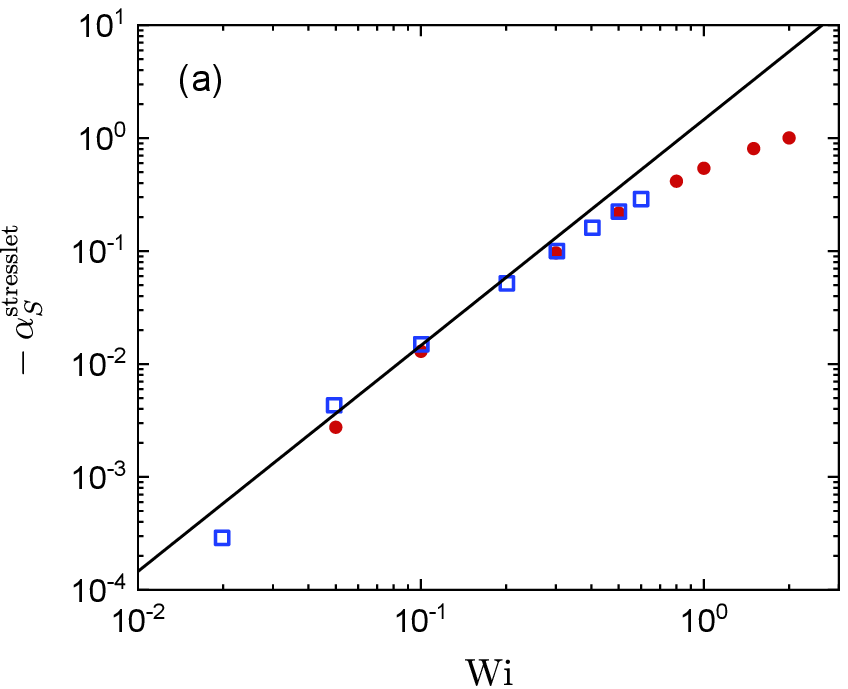}
\label{ein_a}
  \end{center}
  \begin{center}
\includegraphics[scale=0.75]{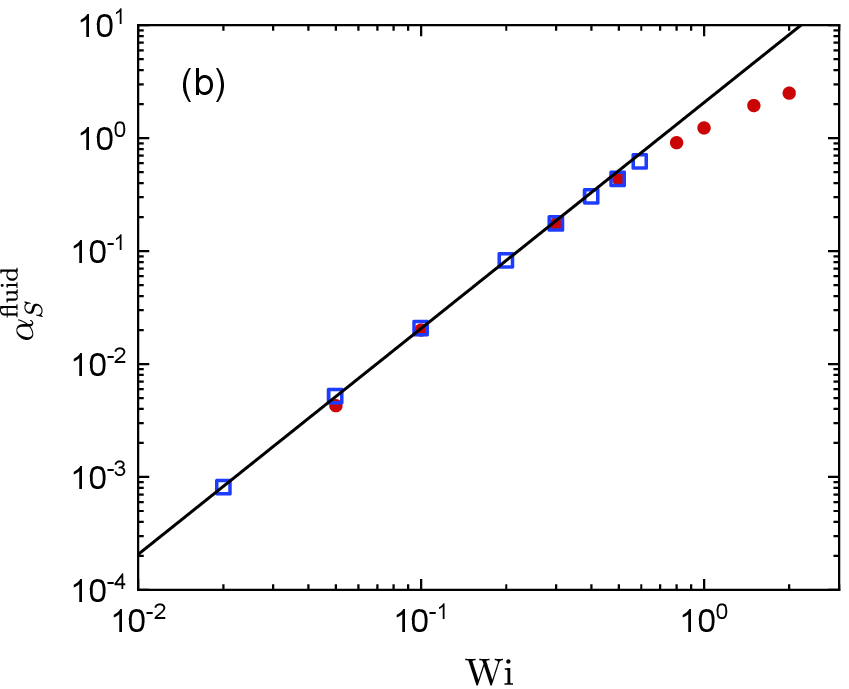}
\label{ein_b}
  \end{center}
\caption{\label{fig:compare_einarsson} The $\rm{Wi}$ dependence of (a) 
stresslet $\alpha_S^{\rm{stresslet}}$ and (b) particle-induced 
fluid stress $\alpha_S^{\rm{fluid}}$ contributions to the 
suspension viscosity at $\beta=0.5$. Red filled circles represent 
results from this work, and blue squares are the DNS results by 
\citet{Einarsson2018}. The black line is plotted according to the 
theory by \citet{Einarsson2018}.}
\end{figure}

\noindent A perturbation analysis of the suspension in a 
single-mode Oldroyd-B medium by~\citet{Einarsson2018} predicted 
the shear-thinning in the stresslet and the shear-thickening in 
the particle-induced fluid stress at $O(\phi_p\rm{Wi}^2)$:
\begin{align}\label{eq:einarsson}
    \eta_r=1+2.5\phi_p+\phi_p(1-\beta)(\alpha_S^{\rm{stresslet}}+\alpha_S^{\rm{fluid}}), 
\end{align} 
where 
$\alpha_S^{\rm{stresslet}}=-1.43\rm{Wi}^2-0.06(1-\beta)\rm{Wi}^2$ 
and $\alpha_S^{\rm{fluid}}=2.05\rm{Wi}^2+0.03(1-\beta)\rm{Wi}^2$ 
are the contributions from the stresslet and particle-induced 
fluid stress~(Sec.~\ref{sc:stress}), respectively.
DNS of a single particle in an Oldroyd-B medium by Yang and 
Shaqfeh~\citep{Yang2018} showed shear-thickening in the 
particle-induced fluid stress around a particle. 
To confirm that the method developed in this work can be applied 
for rheological evaluation, the viscosity and the bulk stress of 
a single-particle suspension in an Oldroyd-B medium is evaluated. 
The numerical setup is the same as that explained in 
Sec.\ref{sc:bc}~(Fig.~\ref{fig:system}(a)). The system size is 
$L=128\Delta$ and the particle radius and interfacial thickness 
are $a=8\Delta$ and $\xi=2\Delta$, respectively. This corresponds 
to $\phi_p=0.001023$. All calculations are conducted with a small 
Reynolds number $\rm{Re}\le0.051$, {\itshape i.e.}, the effect of 
inertia is negligible.
 
Fig.~\ref{fig:eta_r_einarrson} shows the $\rm{Wi}$ dependence of 
the steady-state relative shear viscosity, 
$\eta_r=\langle\sigma_{xy}^{\rm{sus}}\rangle/(\eta_0\dot{\gamma})
$, of the single-mode Oldroyd-B suspension at $\beta=0.5$. 
Shear-thickening is observed in the suspension viscosity for 
increasing $\rm{Wi}$. In the $\rm{Wi}\rightarrow 0$ limit, the 
relative viscosity ($\eta_{r,0}=1.002522$, which is obtained from 
fitting the numerical results at low $\rm{Wi}$ by using $\eta_r=\eta_{r,0}+b_{\rm{f}}\rm{Wi}^2$) approaches Einstein's theoretical value, $\eta_r=1+2.5\phi_p=1.002557$. The small discrepancy from the theoretical value in $\eta_{r,0}$ is mostly attributed to the stresslet contribution and is suggested to be due to the diffused interface of the particle surface in SPM. 
The developed method reveals the $\rm{Wi}^2$ dependence as predicted by Eq.~(\ref{eq:einarsson}) at roughly $\rm{Wi}<1$; the inset of Fig.~\ref{fig:eta_r_einarrson} shows this clearer, where the thickening part $\eta_r-\eta_{r,0}$ in the relative viscosity is shown.
However, at $\rm{Wi}\gtrsim 1$, shear-thickening is slower than $\rm{Wi}^2$ growth because the perturbation analysis is expected to be valid at $\rm{Wi}\ll 1$. 
For a more detailed comparison,  $\alpha_S^{\rm{stresslet}}$ and $\alpha_S^{\rm{fluid}}$ at $\beta=0.5$ are evaluated separately as
\begin{align}
    \alpha_S^{\rm{stresslet}}&=\frac{N\langle S_{xy}\rangle/V-\eta_0\dot{\gamma}(\eta_{r,0}-1)}{\eta_0\dot{\gamma}\phi_p(1-\beta)},\\
    \alpha_S^{\rm{fluid}}&=\frac{N\langle \Sigma_{xy}\rangle/V-\eta_p\dot{\gamma}}{\eta_0\dot{\gamma}\phi_p(1-\beta)},
\end{align}
as shown in Fig.~\ref{fig:compare_einarsson} with a previous DNS 
result obtained by using a surface-conforming 
mesh~\citep{Einarsson2018}; the results agree with the DNS by 
Einarsson \etal. By comparing with DNS results, the $O(\rm{Wi}^2)$ 
prediction (solid line) is found to be valid at $\rm{Wi} \lesssim 
0.3$ for $\alpha_S^{\rm{stresslet}}$ and $\rm{Wi} \lesssim 0.5$ 
for $\alpha_S^{\rm{fluid}}$. 
At higher $\rm{Wi}$ values, the $\rm{Wi}$ dependence is slower 
than $\rm{Wi}^2$ growth, which is observed both in 
$\left|\alpha_S^{\rm{stresslet}}\right|$ and in 
$\alpha_S^{\rm{fluid}}$. 
 
The agreement between the obtained results and those from 
perturbation theory and a previous DNS study verifies the 
capability of the developed SPM for the rheological evaluation of 
suspensions in viscoelastic media. By using the presented 
numerical method, the influence of $\beta$ on the rheology of a 
dilute suspension in an Oldroyd-B medium has been explored in 
detail~\citep{Matsuoka2020}.

\subsection{Suspension rheology of many-particle system}
\label{sc:many_rheo}\noindent For dilute and semi-dilute particle 
concentrations, the rheology of many-particle systems is studied 
in contrast to the single-particle system considered in 
Sec.~\ref{sc:single_rheo}. {The numerical condition in this study 
is decided in accordance with the experimental conditions 
previously reported by~\citet{Yang2018a}. They have performed 
detailed rheological measurements of a viscoelastic medium, 
including the elongation viscosity, in addition to the 
rheological measurements of a suspension system. Thus, their 
experimental results are likely to be the most complete dataset 
available for the quantitative rheological evaluation by DNS. 
Furthermore, as mentioned in their paper, wall effects for the 
rheological measurements are expected to be negligible in their 
experiments, which is suitable for our shear periodic boundary 
condition explained in Sec.~\ref{sc:bc}.}

\begin{table}
  \begin{center}
\begin{ruledtabular}
  \begin{tabular}{lcccccc}
\mbox{Mode $k$}&&\mbox{$\eta_p^{(k)}~\rm{(Pa\cdot s)}$}&&\mbox{$\lambda^{(k)}~\rm{(s)}$}&&\mbox{$\eta_p^{(k)}\lambda^{(k)}~\rm{(Pa\cdot s^2)}$}\\[3pt]
\hline
1&&0.67&&3.2&&2.144\\
2&&0.66&&0.26&&0.172\\
3&&0.25&&0.032&&$8.0\times10^{{-3}}$\\
4&&0.44&&0.002&&$8.8\times10^{-4}$\\
\mbox{Solvent}&&1.46&&--&&--
\end{tabular}
\end{ruledtabular}
\caption{\label{tab:yang} Parameters for a four-mode Oldroyd-B fluid. The values are from Table~I of \citet{Yang2018a}, which are estimated from the small-amplitude oscillatory shear  
measurement of a Boger fluid.} \end{center}
\end{table}

\subsubsection{Numerical conditions}
\label{sc:many_cond}\noindent The system and particle sizes are 
the same as in Sec.~\ref{sc:single_rheo}, {\itshape i.e.}, 
$L=128\Delta$, $a=8\Delta, and \xi=2\Delta$.
Considering dilute to semi-dilute particle concentrations, 
one has $\phi_p=0.001,0.025, 0.05$, and $0.1$ by setting the number of particles to {1}, 24, 49, and 98, respectively.
The initial positions of the particles are set to be 
randomly distributed and non-overlapping, with the inter-surface 
distance set to at least $2\Delta$.
For each $\phi_p$ except for $\phi_p=0.001$~(single-particle system), at least three different realizations are calculated. 
An experimental result reported by \citet{Yang2018a} is 
considered where the rheology of a suspension in a Boger fluid 
consisting of polybutene, polyisobutylene, and kerosene was evaluated.
For the rheological characterization of the Boger fluid, both steady-shear and small-amplitude oscillatory shear~(SAOS) 
measurements were reported~\citep{Yang2018a}.
In principle, the parameters in the Oldroyd-B model can be 
estimated from either the steady-shear or SAOS data; however, due to the limited range of the rate window, the zero-shear first 
NSD was available only from the SAOS data. 
Furthermore, in their experiment, the suspension viscosity 
begins to show shear-thickening at $\dot{\gamma}\approx 0.2{\rm 
s}^{-1}$, a shear rate that is below the rate window of 
steady-shear $N_1$ data.
Therefore, the parameters estimated from the SAOS data listed 
in Table~\ref{tab:yang} are used here to solve the corresponding four-mode 
Oldroyd-B fluid as a suspending medium.
Note that Yang and Shaqfeh also reported the DNS prediction with 
experimental data~\citep{Yang2018a}, where, in contrast to this 
work, the single-mode Oldroyd-B model with parameters estimated 
from the steady-shear property of the suspending Boger fluids 
resulted in an underestimation of the suspending viscosity.
The discrepancy between their simulation and experimental results 
is discussed later~(Sec.~\ref{sc:many_mode}).

After the steady state is reached, 
the viscometric functions of the many-particle suspension are time-averaged 
over at least $\dot{\gamma}\Delta t=10$ from { $\dot{\gamma}t \geq 10\max\{1,\dot{\gamma}\lambda^{(1)}\}$}.
Finally, the time-averaged values are ensemble-averaged over different realizations to obtain the viscometric functions of bulk suspensions.
The error bars in the following figures correspond to three times the standard deviation from the sample mean.
The Weissenberg number is defined based on the longest relaxation time $\lambda^{(1)}=3.2~$s as ${\rm Wi}=\dot{\gamma}\lambda^{(1)}$.
All calculations were conducted at a small Reynolds number $\rm{Re}\le0.018$ where effect of inertia is not significant.

\begin{figure}
\centerline{\includegraphics[scale=0.9]{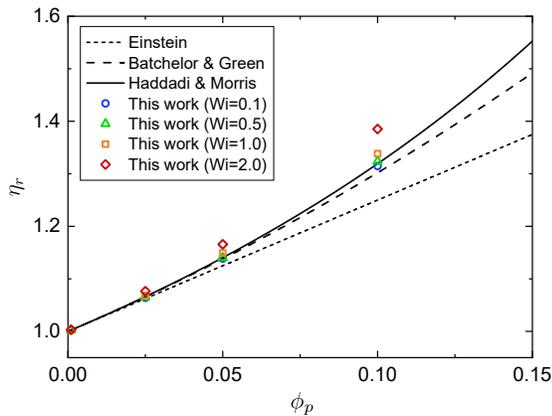}}
\caption{\label{fig:eta_r_phi} The $\phi_p$ dependence of the 
relative viscosity of suspensions at $\rm{Wi}=0.1$ (blue circles), 
$0.5$ (green triangles), $1.0$ (orange squares), and $2.0$ (red 
diamonds). The short-dashed and long-dashed lines correspond to 
the theoretical predictions for a Newtonian suspension by 
Einstein~\cite{einstein1911berichtigung} and 
Batchelor-Green~\cite{batchelor1972determination}, respectively. 
The empirical prediction from Haddadi and 
Morris~\cite{Haddadi2014} is shown as a solid line.}
\end{figure}

\begin{figure}   
  \begin{center}
\includegraphics[scale=0.8]{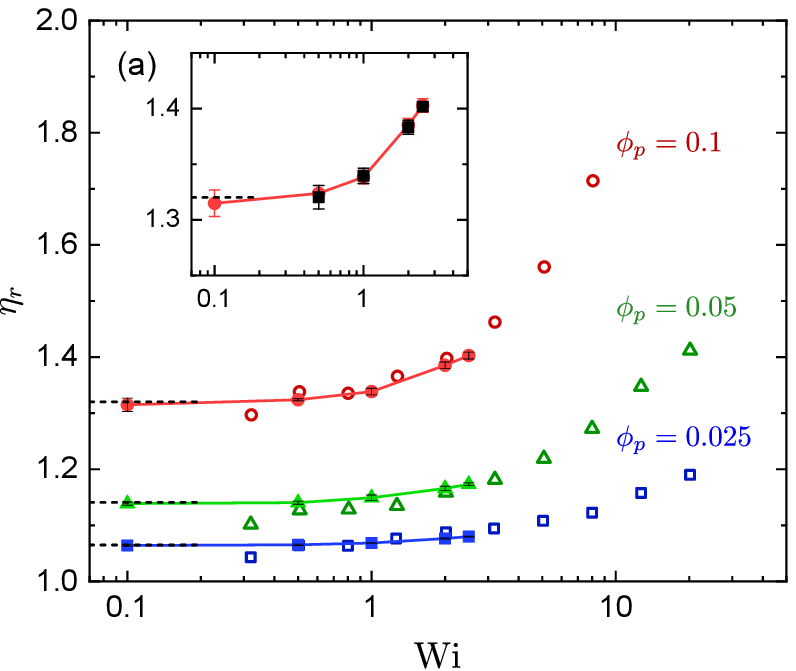}
  \end{center}
  \begin{center}
\includegraphics[scale=0.77]{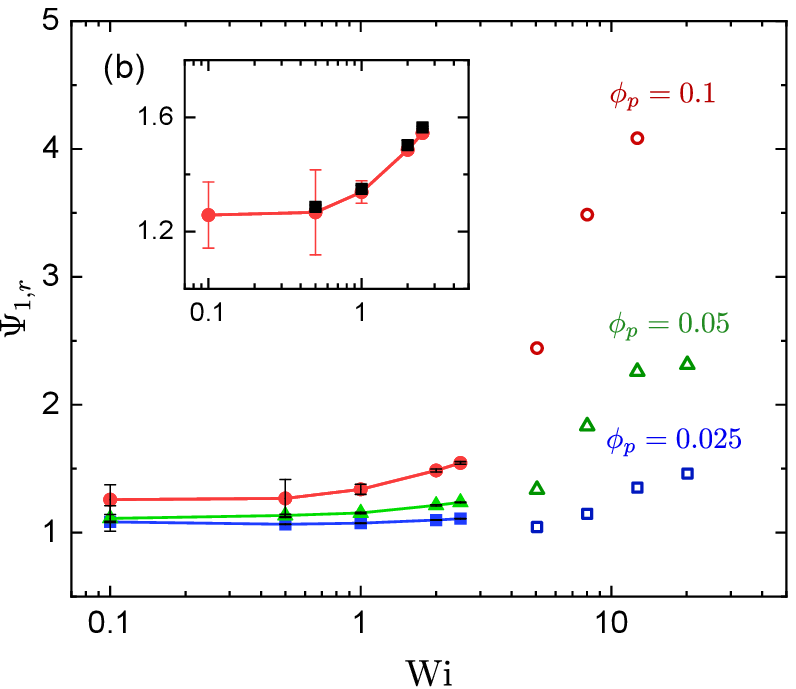}
  \end{center}
\caption{\label{fig:compare_Yang} Viscometric functions of 
suspensions as functions of $\rm{Wi}$ and $\phi_p$: (a)~relative 
viscosity and (b)~relative first NSD coefficient of suspensions. 
The closed symbols are simulated results from this work, and the 
open symbols are experimental results from~\citet{Yang2018a}. 
The blue squares, green triangles, and red circles correspond to 
the results for $\phi_p=0.025,\,0.05,$ and $0.1$, respectively. 
Experimental $\eta_{r}$ and $\Psi_{1,r}$ are calculated using 
$\eta(\phi_p, \dot{\gamma})$ and $\Psi_{1}(\phi_p, \dot{\gamma})$ 
reported by~\citet{Yang2018a}. The dashed lines in (a) are values 
predicted by the Eilers fit~\citep{Haddadi2014}. Solid lines are 
guides to the eye. The insets in (a) and (b) show the DNS 
results at $\phi_p=0.1$ by the multi-mode model~(red circles) and 
the effective single-mode model (black squares) explained in 
Sec.~\ref{sc:many_mode}.}
\end{figure}

\subsubsection{Suspension viscosity and first NSD coefficient}
\label{sc:suspensions_viscosity} Figure~\ref{fig:eta_r_phi} shows 
the steady-state suspension viscosity normalized by  $\eta_0$ for 
different $\rm{Wi}$ as functions of $\phi_p$; the theoretical 
trends for a Newtonian suspension in the creeping flow regime are 
also shown. Here, $\eta_r=1+2.5\phi_p+\alpha\phi_p^2$ where 
$\alpha=0$ for Einstein 
theory~\cite{einstein1911berichtigung}~(short-dashed line) and 
$\alpha=5.2$ for Batchelor--Green 
theory~\cite{batchelor1972determination}~(long-dashed line). In 
addition, the empirical Eilers fit for the numerical result of 
Newtonian suspensions by \citet{Haddadi2014}, 
$\eta_r=\left(1+\frac{1}{2}[\eta]\phi_p/(1-\phi_p/\phi_{p,m})\right)^2$, with $[\eta]=2.5$ and $\phi_{p,m}=0.63$, is also 
plotted~(solid line).
At $\rm{Wi}=0.1$, the suspension viscosity agrees well with the predictions by Batchelor--Green and Eilers fit for Newtonian suspensions. 
This is expected because the polymer stress is expected to fully relax at $\rm{Wi}\ll 
1$ to exhibit almost Newtonian behavior.
In contrast, as $\rm{Wi}$ increases, the suspension viscosity increases to be above the prediction for Newtonian suspensions.

In Fig.~\ref{fig:compare_Yang}, the viscosity (Fig.~\ref{fig:compare_Yang}(a)) and first NSD coefficient (Fig.~\ref{fig:compare_Yang}(b)) as functions of $\rm{Wi}$ are compared with the experimental result by \citet{Yang2018a} for different $\phi_p$.
The viscosity at the $\rm{Wi}\to 0$ limit calculated by Eilers fit in Fig.~\ref{fig:eta_r_phi} for each $\phi_{p}$ is also shown in Fig.~\ref{fig:compare_Yang}(a).
The numerical results of this work agree quantitatively with the experimental results up to a semi-dilute case 
of $\phi_{p}=0.1$.
The first NSD coefficient of the suspension, $\Psi_1=\langle\sigma_{xx}^{\rm{sus}}-\sigma_{yy}^{\rm{sus}}\rangle/\dot{\gamma}^2$, normalized by that of the medium is shown in Fig.~\ref{fig:compare_Yang}(b). As $\rm{Wi}$ increases, $\Psi_{1,r}$ also increases. Although the ranges of $\rm{Wi}$ of the experimental and numerical results do not overlap, the numerical results of this work smoothly connect with the experimental results. 

Note that, while the DNS results agree with the experimental 
$\eta_r$, the DNS using an Oldroyd-B model reported by~\citet{Yang2018a} underestimated it.
The main difference between this work and that of Yang--Shaqfeh is 
the estimation of the zero-shear $N_{1}$ of the suspending 
Boger fluid; $N_{1}$ from the SAOS measurement is approximately twice as
large as that from the steady-shear measurement; the difference occurs because the steady-shear measurement did not reach the terminal region and showed a decreased $N_{1}$.
These results suggest that predicting suspension shear-thickening 
at around $\rm{Wi}=1.0$ requires an accurate estimation of $N_1$ 
of the suspending medium in the shear-rate range where the 
shear-thickening starts to occur. For the Boger fluid used in 
~\citet{Yang2018a}, this range is supposed to be the terminal 
region, which cannot be reached by the steady-shear measurement.
The estimation of $N_1$ directly affects the level of polymer 
stress around the particles, because, as past studies on 
dilute systems have revealed~\citep{Yang2018,Matsuoka2020}, the 
elastic stress due to the stretched conformation nearby upstream 
of the particles contributes to the macroscopic shear stress.
In~\citet{Yang2018a}, their model's underestimation of the 
medium's elongational property is argued to be one reason why 
their DNS prediction underestimates the measured shear-thickening 
of suspensions. Although our four-mode Oldroyd-B model shows 
slightly higher elongational viscosity than that by the 
single-mode model used in~\citet{Yang2018a}, our multi-mode model 
still underestimates the measured elongational viscosity of the 
medium. This result suggests that suspension shear-thickening in 
Boger fluids at around $\rm{Wi}=1.0$ can be predicted with the 
Oldroyd-B model without additional modelling of the elongational 
response.

\begin{figure}   
  \begin{center}
\includegraphics[scale=0.8]{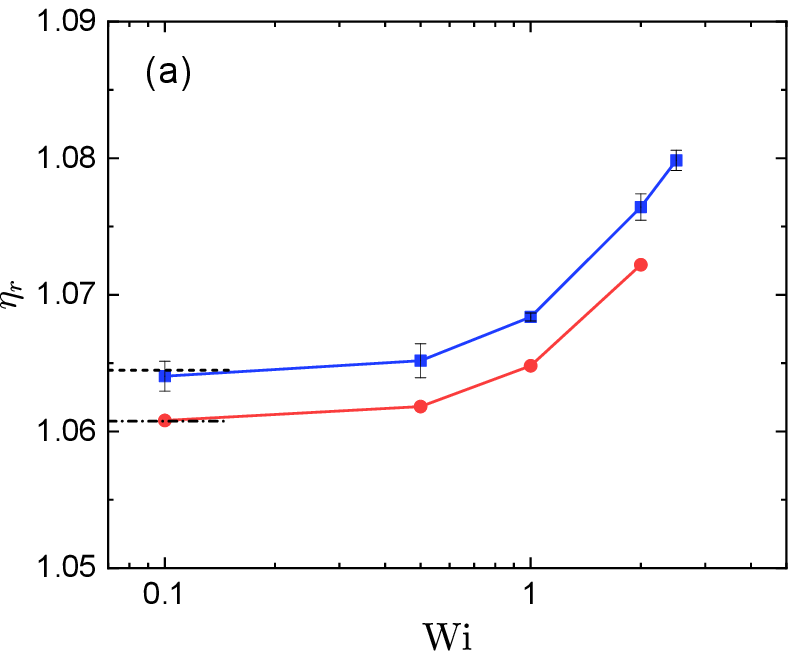}
  \end{center}
  \begin{center}
\includegraphics[scale=0.7]{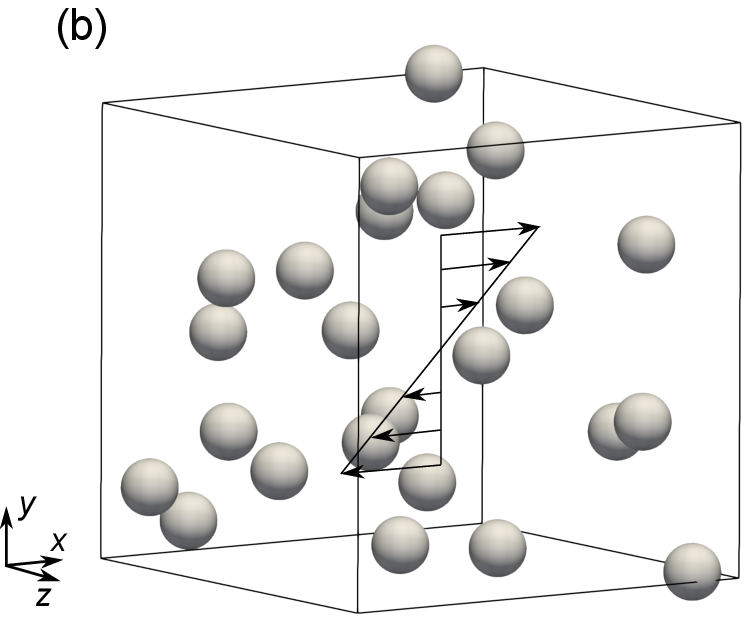}
  \end{center}
\caption{\label{fig:eta_r_phi0_025} {DNS results at 
$\phi_p=0.025$}: (a) suspension viscosity for single-particle (red circles) 
and many-particle (blue squares) systems; and (b) microstructure 
in a many-particle system at ${\rm Wi}=2.0$. In (a), the black 
lines are predictions for Newtonian suspensions according to 
the theories of Batchelor--Green~(dashed, \(\phi_{p}=0.02454\) 
 for many-particle systems) and Einstein~(dot-dashed, \(\phi_{p}=0.02430\)). 
Solid lines are guides to the eye.}
\end{figure}

To demonstrate the difference between many-particle and 
single-particle systems at dilute conditions, a single-particle 
simulation is conducted at $\phi_p\approx 0.025$ by setting the 
particle radius $a=23\Delta$ and system size $L=128\Delta$ in the 
single-particle system shown in Fig.~\ref{fig:system}(a); the 
Reynolds number is kept small~($\rm{Re}=0.076$).
Because of the periodic boundary conditions, this single-particle 
system corresponds to the sheared cubic array system shown in 
Fig.~\ref{fig:system}(b).
In Fig.~\ref{fig:eta_r_phi0_025}{(a)}, the suspension viscosity 
between single-particle~(cubic array structure) and 
many-particle~(random structure) systems is compared.
The single-particle result indicates lower viscosity, whereas the 
shear-thickening behavior is almost the same as that of the 
many-particle system. At $\rm{Wi}\rightarrow 0$, the viscosity 
from the single-particle system agrees with the Einstein 
prediction. This also agrees with the results of a cubic array 
system in a Newtonian medium~\citep{Nunan1984,Phan-Thien1991}. 
Correspondingly, $\langle S_{xy}\rangle$ for the single-particle 
system agrees with the Einstein stresslet~(the inset of 
Fig.~\ref{fig:scaling}(a)).
Fig.~\ref{fig:eta_r_phi0_025}(b) shows the microstructure of the 
many-particle system in a sheared steady state at $\phi_p=0.025$ 
and ${\rm Wi}=2$. 
In many-particle systems, particles are randomly dispersed and 
occasionally get very close to each other, which induces the 
large stresslet contribution. On the other hand, in 
the single-particle system, the inter-particle distance remains above 
a certain level as shown in Fig.~\ref{fig:system}(b).
Therefore, the viscosity shift between the two systems is 
attributed to the difference in the stresslet contribution by 
microstructures. 
Note that particle alignment, which is sometimes observed 
experimentally in suspensions with viscoelastic 
fluids~\citep{Michele1977,Scirocco2004}, is not be observed at 
all $\phi_p$ and ${\rm Wi}$ in our study. This suggests that our 
simulation conditions are out of range for an alignment critical 
condition predicted by DNS using Oldroyd-B and Giesekus matrices 
~\citep{Hwang2011,Jaensson2016}.
The result from this work, showing that the suspension 
microstructure affects the viscosity even at dilute conditions, 
is consistent with the results of a previous 
study~\citep{Vazquez-Quesada2019}. Furthermore, similar 
shear-thickening behavior independent of the microstructures 
suggests that the shear-thickening at dilute conditions is mainly 
originated from the polymer stress in the vicinity of a particle, 
which is consistent with a previous 
study~\citep{Yang2018,Yang2018a}.

\subsubsection{Relaxation mode decomposition of polymer stress}
\label{sc:many_mode}

\begin{figure}
  \begin{center}
\includegraphics[scale=0.75]{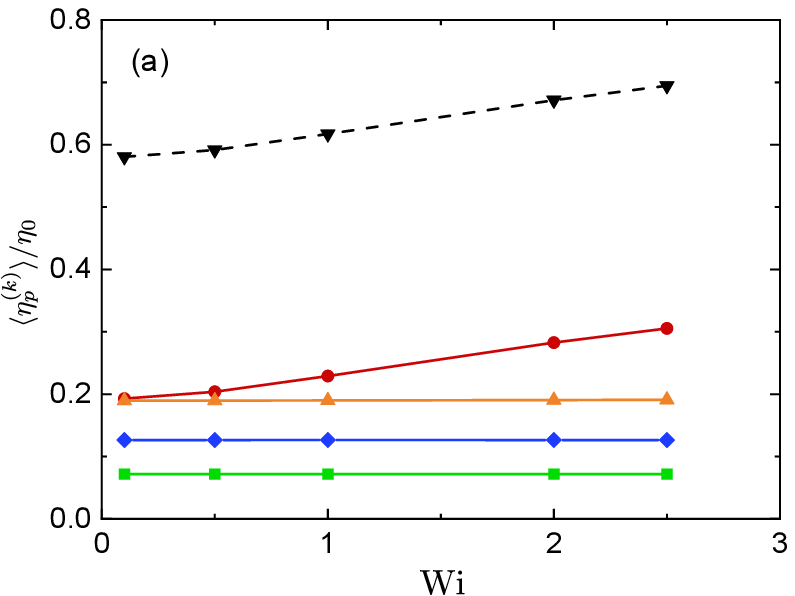}
  \end{center}
  \begin{center}
\includegraphics[scale=0.75]{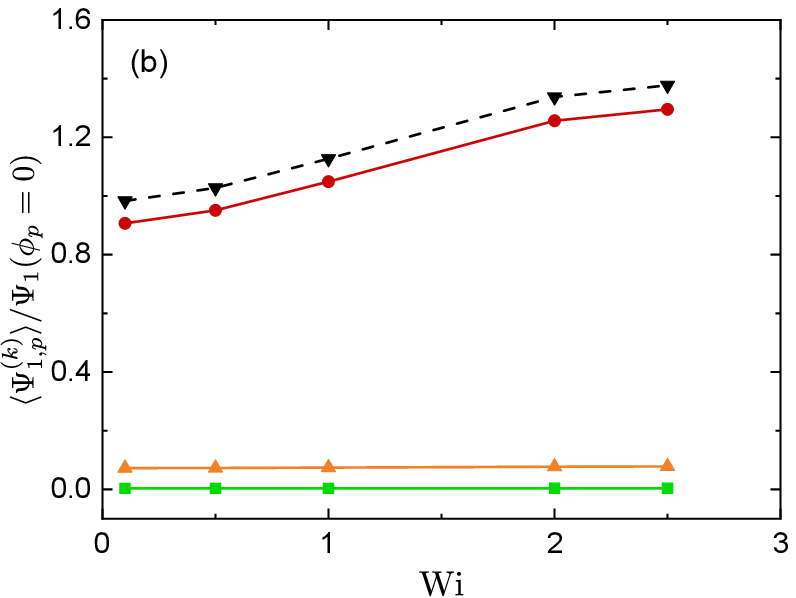}
  \end{center}
\caption{\label{fig:mode_decomp} Proportions of each relaxation 
mode in the polymer stress contribution (for $k$=1 (red circles), 2 
(orange triangles), 3 (green squares), and 4 (blue diamonds), and the 
sum of the mode contributions (black lower triangle)): (a) shear 
viscosity and (b) first NSD coefficient at $\phi_p=0.1$. 
The values of $\langle\eta_{p}^{(k)}\rangle$ and 
$\langle\Psi_{1,p}^{(k)}\rangle$ are normalized by $\eta_0$ and 
$\Psi_1(\phi_p=0)=2\sum_{k=1}^4\eta_p^{(k)}\lambda^{(k)}$, 
respectively. Note that the stresslet contributions are not 
included in the figure. Lines are guides to the eye. {By 
definition, the order of $\langle\eta_{p}^{(k)}\rangle$ and 
$\langle\Psi_{1,p}^{(k)}\rangle$ at $\rm{Wi}\rightarrow0$ 
corresponds to the order of $\eta_p^{(k)}$ and 
$\eta_p^{(k)}\lambda^{(k)}$ in Table.~\ref{tab:yang}, 
respectively. That is why 
$\langle\eta_{p}^{(4)}\rangle>\langle\eta_{p}^{(3)}\rangle$ in (a). 
In (b), $\langle\Psi_{1,p}^{(4)}\rangle$ is not shown because it 
is smaller than $\langle\Psi_{1,p}^{(3)}\rangle$.}}
\end{figure}

\noindent In the modelling of the suspensions in a Boger fluid, 
the four-mode Oldroyd-B model is used for the suspending medium. 
The separate contributions from each relaxation mode to the 
suspension shear-thickening is discussed.
The viscosity and the first NSD coefficient from the $k$-th mode 
are defined 
as $\langle\eta_{p}^{(k)}\rangle\equiv[\int_{D_V}(1-\lfloor\phi\rfloor)\sigma_{p,xy}^{(k)}d\bm{r}/V]/\dot{\gamma}$ and 
$\langle\Psi_{1,p}^{(k)}\rangle\equiv[\int_{D_V}(1-\lfloor\phi\rfloor)(\sigma_{p,xx}^{(k)}-\sigma_{p,yy}^{(k)})d\bm{r}/V]/\dot{\gamma}^2$ ($k=1,2,3,4$), respectively. 
Figure~\ref{fig:mode_decomp} shows the $k$-th viscosity 
normalized by $\eta_0$and the $k$-th first NSD coefficient 
normalized by $\Psi_1$ at $\phi_p=0$ as functions of $\rm{Wi}$ at 
$\phi_p=0.1$.
Both for the viscosity (Fig.~\ref{fig:mode_decomp}(a)) and for 
the first NSD coefficient (Fig.~\ref{fig:mode_decomp}(b)), 
only the first mode exhibits shear-thickening, whereas the other faster modes show a rate-independent contribution.
This is expected, because the $\rm{Wi}$ considered here is much 
smaller than $\lambda^{(1)}/\lambda^{(2)}=12.3$; the elastic 
stress from the second and subsequent modes fully relaxes to show 
a zero-shear response.

The results in Fig.~\ref{fig:mode_decomp} suggest that 
single-mode modelling for the suspending medium is likely to be 
sufficient to predict the rheological response at 
the $\rm{Wi}\leq 2.5$ considered in the current simulation.
If only the first mode is responsible for the polymer stress, the 
effective parameters for a single-mode Oldroyd-B fluid are 
determined from Table~\ref{tab:yang} to be $\lambda^{\rm 
eff}=\lambda^{(1)}=3.2$~s, $\eta_p^{\rm 
eff}=\eta_p^{(1)}=0.67$~Pa$\cdot$s, and $\eta_{s}^{\rm 
eff}=\eta_{s}+\sum_{k=2}^4\eta_{p}^{(k)}=2.81$~Pa$\cdot$s, 
resulting in $\beta^{\rm{eff}}=\eta_{s}^{\rm{eff}}/\eta_0=0.807$.
This effective $\beta$ value is smaller than the $\beta=0.9$ used 
in DNS~\citep{Yang2018a}, which underpredicted the experimental 
suspension rheology.
In the inset of Fig.~\ref{fig:compare_Yang}, the DNS result of 
the presented effective single-mode model~(black squares) is 
compared with that of the multi-mode model~(red circles), showing 
good agreement with the multi-mode results and thus experimental 
results~\citep{Yang2018a}.
This difference between the $\beta$ values originates from the 
difference in the estimation of the zero-shear NSD coefficient of 
the Boger fluid that was mentioned in 
Sec.~\ref{sc:suspensions_viscosity}.

In the system considered in this work, only $\lambda^{(1)}$ is relevant to the studied range of $\rm{Wi}$.
Whether single-mode modelling can be used for the quantitative 
prediction of suspension rheology for other types of suspending 
media depends on both the relaxation time distribution of the 
fluid and the distribution of the local shear-rate in the fluid, 
which is dependent on the fluid rheology as well as $\phi_{p}$. 
In Sec.~\ref{sc:many_flow}, we study how the local shear-rate 
distribution, flow pattern, and the elastic stress development 
change with $\phi_{p}$ and $\rm{Wi}$.

\subsubsection{Decomposition of the total suspension stress}\label{sc:many_decomp}

\begin{figure} 
  \begin{center}
\includegraphics[scale=0.67]{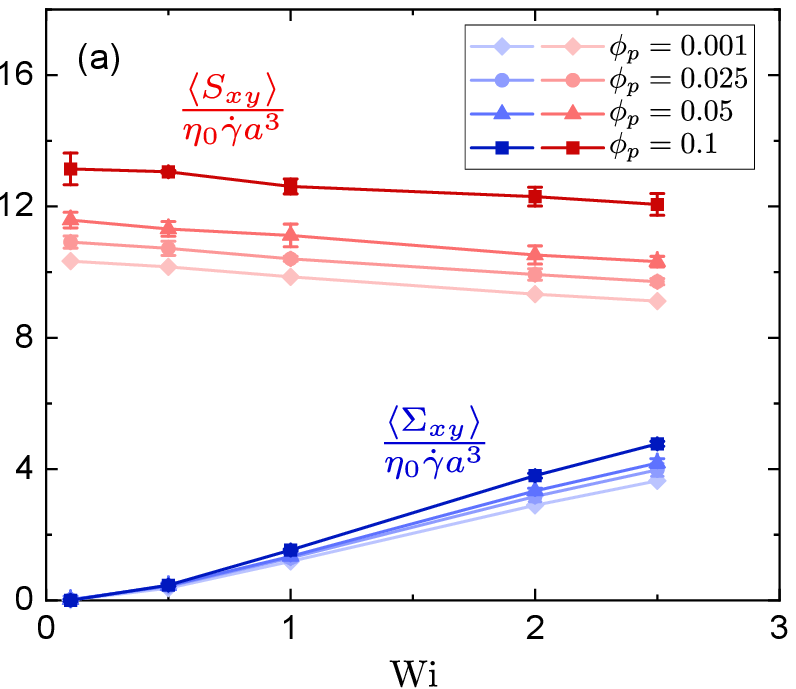}
  \end{center}
  \begin{center}
\includegraphics[scale=0.66]{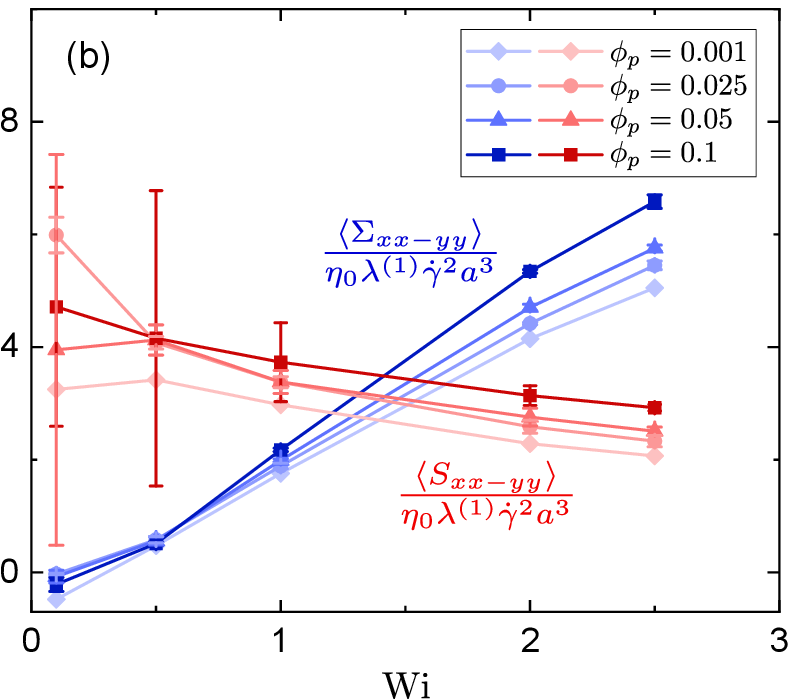}
  \end{center}
\caption{\label{fig:decomp_cont} Contributions to the total 
suspension stress from stresslet $\bm{S}$ (red) and 
particle-induced fluid stress $\bm{\Sigma}$ (blue) at 
$\phi_p=0.001$ to 0.1 (from light to dark color): 
(a)~Contributions to the total shear stress normalized by 
$\eta_0\dot{\gamma}a^3$; and (b)~contributions to the total first 
NSD normalized by $\eta_0\lambda^{(1)}\dot{\gamma}^2a^3$. Solid 
lines are guides to the eye.}
\end{figure}

\begin{figure}
  \begin{center}
\includegraphics[scale=0.75]{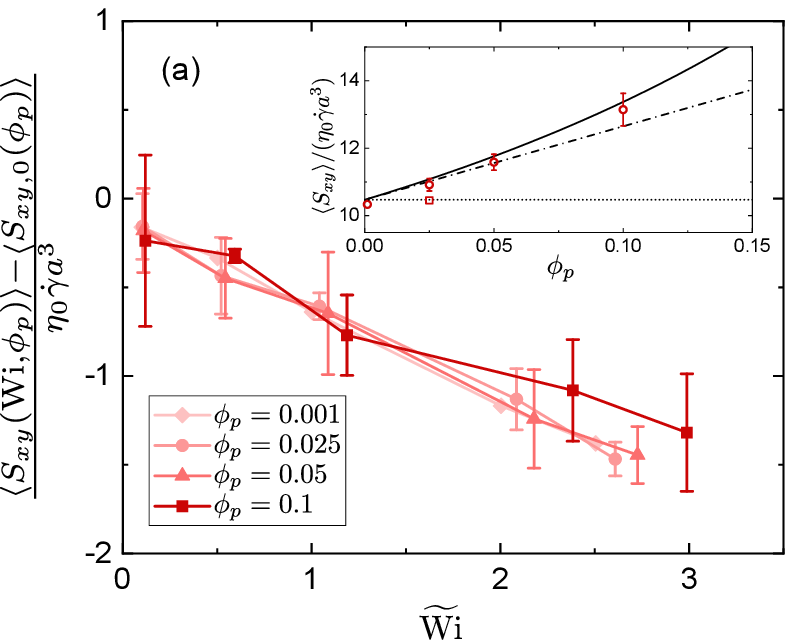}
  \end{center}
  \begin{center}
\includegraphics[scale=0.75]{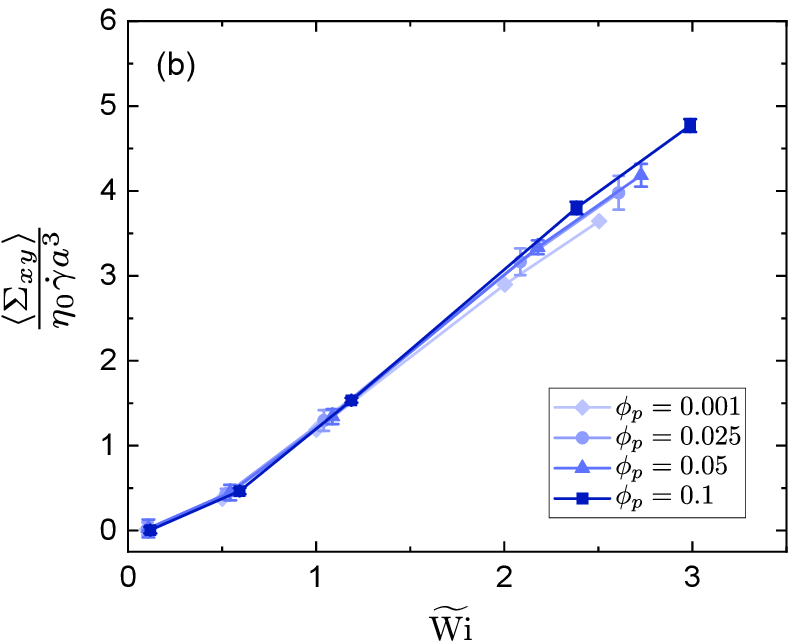}
  \end{center}
 \captionsetup{width=.9\linewidth}\caption{\label{fig:scaling} 
Viscoelastic contributions to the total suspension shear stress 
as a function of the effective Weissenberg number at 
$\phi_p=0.001$ to 0.1 (from light to dark color). (a)~Stresslet 
contribution to shear stress. The ordinate represents the 
polymeric part of $\langle S_{xy}\rangle$;~$\langle 
S_{xy}\rangle-\langle S_{xy,0}\rangle$ where $\langle 
S_{xy,0}\rangle$ is the Newtonian part of $\langle S_{xy}\rangle$ 
represented by Eq.~(\ref{S_xy_newton}). The inset shows the 
$\phi_p$ dependence of $\langle S_{xy}\rangle$ at $\rm{Wi}=0.1$ 
(red circles). The result from single-particle simulation at 
$\phi_p=0.025$ is also shown (red square). The black lines are 
predictions according to theories of Einstein~(dotted), 
Batchelor--Green~(dot-dashed), and Eilers fit by Haddadi \& 
Morris~(solid). The contributions for the shear stress are 
normalized by $\eta_0\dot{\gamma}a^3$. (b)~Particle-induced fluid 
stress contributions for shear stress. The effective Weissenberg 
number $\widetilde{\rm{Wi}}$ is defined with the average 
strain-rate in the fluid region at each $\phi_p$. In both panels, 
 the red and blue 
solid lines are guides to the eye.}
\end{figure}

\begin{figure}
\centering
\includegraphics[scale=.75]{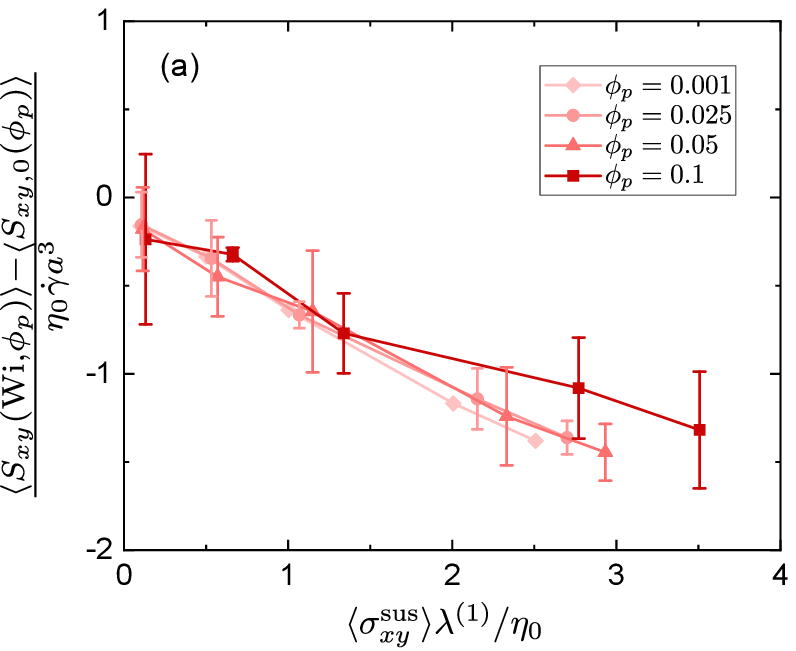}
\includegraphics[scale=.75]{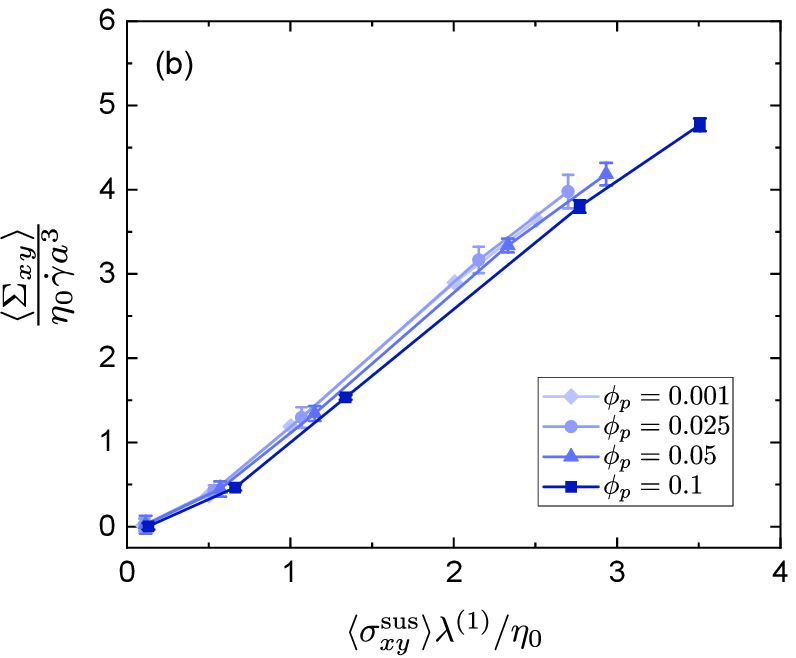}
\centering
\centering
\includegraphics[scale=.75]{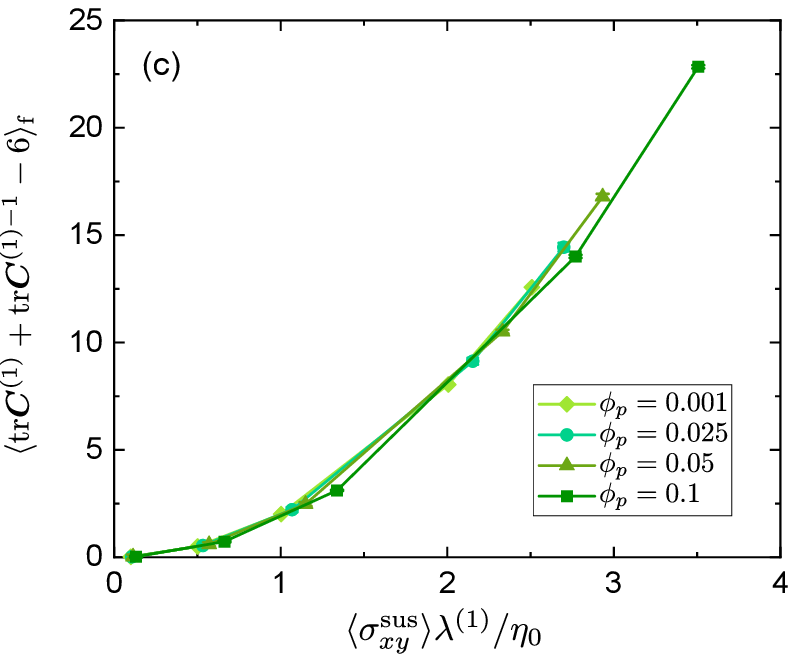}
\includegraphics[scale=.75]{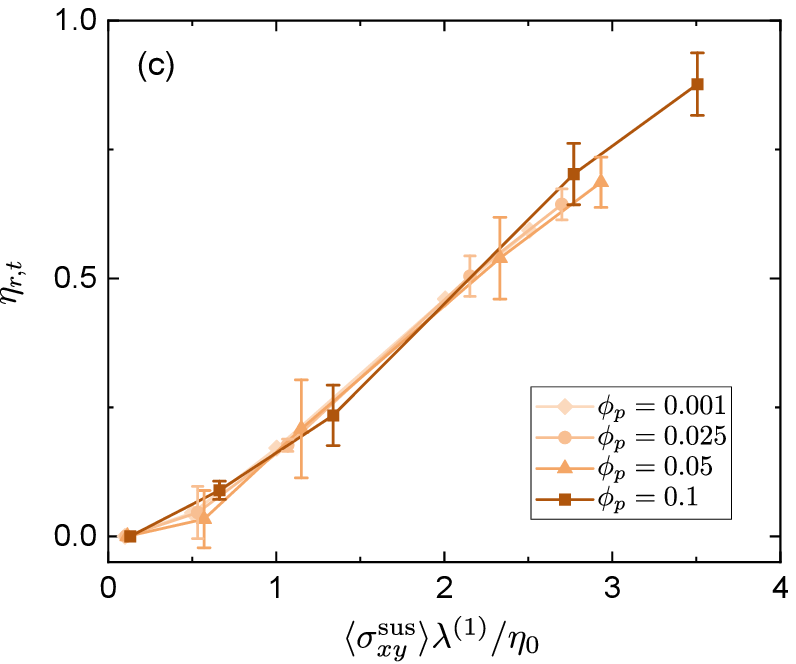}
\caption{\label{fig:scaling_Yang} Viscoelastic contributions to 
the total suspension shear stress as a function of the suspension 
shear stress at $\phi_p=0.001$ to $0.1$ (from light to dark color): 
(a)~stresslet, (b)~particle-induced fluid stress, (c)~normalized 
polymer dissipation function of the first relaxation mode, and 
(d)~the thickening portion of the relative viscosity. The 
suspension shear stress in the abscissa in each panel is 
non-dimensionalized as 
$\langle\sigma_{xy}^{\rm{sus}}\rangle\lambda^{(1)}/\eta_0$. Solid 
lines are guides to the eye.}
\end{figure}

\noindent The $\phi_p$ dependence of the shear-thickening of the 
suspension in the Oldroyd-B medium is discussed.
The contributions from the stresslet, $\bm{S}$, and the 
particle-induced fluid stress, $\bm{\Sigma}$, to the suspension 
rheology are shown in Fig.~\ref{fig:decomp_cont}, where the shear 
component is normalized by $\eta_0\dot{\gamma}a^{3}$ to 
correspond to a non-dimensional viscosity, and the first NSD component 
is normalized by $\eta_0\lambda^{(1)}\dot{\gamma}^{2}a^{3}$ to 
correspond to the non-dimensional NSD coefficient.
For the viscosity component in Fig.~\ref{fig:decomp_cont}(a), as $\rm{Wi}$ increases, the stresslet viscosity, $\langle S_{xy}\rangle/(\eta_0\dot{\gamma}a^{3})$, decreases, and the particle-induced fluid viscosity, $\langle\Sigma_{xy}\rangle/(\eta_0\dot{\gamma}a^{3})$, increases more than the change in the stresslet viscosity.
Specifically, at $\rm{Wi}=2.0$, the decrease in $\langle S_{xy}\rangle/(\eta_0\dot{\gamma}a^{3})$ is less than two, but the increase in $\langle\Sigma_{xy}\rangle/(\eta_0\dot{\gamma}a^{3})$ is more than three for all $\phi_{p}$ considered.
This result clearly demonstrates that the shear-thickening of the 
suspension viscosity originates from an increase in 
$\langle\Sigma_{xy}\rangle$, which is consistent with what has 
been reported in previous 
work~\citep{Yang2018,Yang2018a,Matsuoka2020}.
As $\phi_p$ increases, the increase in $\langle\Sigma_{xy}\rangle/(\eta_0\dot{\gamma}a^{3})$ with $\rm{Wi}$ is enhanced, whereas the decrease in $\langle S_{xy}\rangle/(\eta_0\dot{\gamma}a^{3})$ with $\rm{Wi}$ remains slow, which explains the enhancement of the shear-thickening with $\phi_{p}$ shown in Fig.~\ref{fig:compare_Yang}(a).
For the first NSD component (Fig.~\ref{fig:decomp_cont}(b)), the general trends with respect to $\rm{Wi}$ and $\phi_p$ are similar to that of the viscosity component.
These trends were also reported in a previous numerical study up to $\rm{Wi}\leq1.0$~\citep{Yang2018a}.
Because $N_{1}$ is very small and $N_{1}\propto\langle S_{xx-yy}\rangle$ at the $\rm{Wi}\to 0$ limit, the numerical fluctuation in calculating such a small value is large for $\langle S_{xx-yy}\rangle/(\eta_0\lambda^{(1)}\dot{\gamma}^2a^3)$ at $\rm{Wi}\leq 0.5$.

The reduction rate of the stresslet viscosity with $\rm{Wi}$ does not strongly depend on $\phi_{p}$.
Therefore, $\langle S_{xy}\rangle/(\eta_0\dot{\gamma}a^{3})$ is mainly determined by that at the $\rm{Wi}\to 0$ limit.
This reduction of $\langle\bm{S}\rangle/(\eta_0\dot{\gamma}a^{3})$ with $\rm{Wi}$ indicates the reduced viscous traction on the particles that originates from the increased fraction of the elastic energy dissipation with $\rm{Wi}$, which is also related to the slowdown of the particle rotation rate with $\rm{Wi}$ discussed in Sec.~\ref{sc:rotation}.
The change of $\langle S_{xy}\rangle$ to that at the $\rm{Wi}\to 0$ limit, $\langle S_{xy,0}(\phi_{p})\rangle$, is plotted in Fig.~\ref{fig:scaling}(a) versus an effective Weissenberg number explained later; in Fig.~\ref{fig:scaling_Yang}(a), it is plotted against the suspension shear stress $\langle\sigma_{xy}^{\rm{sus}}\rangle$ normalized by $\eta_0/\lambda^{(1)}$.
The numerical result for $\langle S_{xy}\rangle$ at $\rm{Wi}=0.1$ depicted in the inset of Fig.~\ref{fig:scaling}(a) almost agrees with the theoretical Batchelor--Green stresslet, $\langle S_{xy,0}\rangle/\left(\eta_0\dot{\gamma}a^3\right)=\left(4\pi/3\right) (2.5+\alpha\phi_p)$ for $\phi_p\leq 0.05$,
and with the empirical Eilers stresslet fitted for numerical results by \citet{Haddadi2014}:
\begin{align}
    \label{S_xy_newton}
    \frac{\langle S_{xy,0}\rangle}{\eta_0\dot{\gamma}a^3}=\frac{4\pi}{3\phi_p}\left[\left(1+\frac{\frac{1}{2}[\eta]\phi_p}{1-\phi_p/\phi_{p,m}}\right)^2-1\right],
\end{align} 
for $\phi_p \leq 0.1$. 
Based on this observation, $\langle S_{xy,0}\rangle$ in Fig.~\ref{fig:scaling}(a) 
is calculated with Eq.~(\ref{S_xy_newton}).
In the suspension, a local shear rate can be larger than the applied rate $\dot{\gamma}$. 
To take this into account, the effective Weissenberg number $\widetilde{{\rm Wi}}=\tilde{\dot{\gamma}}\lambda^{(1)}$ is defined by using the average shear rate 
$\tilde{\dot{\gamma}}(\phi_p,{\rm Wi})=\sqrt{\langle2\bm{D}:\bm{D}\rangle_{\rm{f}}}$, 
where $\langle A\rangle_{\rm{f}}=\int_{D_V}\left(1-\lceil\phi\rceil\right)Ad\bm{r}/\left[\left(1-\phi_{p}\right)V\right]$ represents the volume average of a local variable $A$ over the fluid region and $\lceil\cdot\rceil$ indicates the ceiling function. 
For dilute cases ($\phi_p\leq 0.05$), the changes of the 
stresslet viscosity as a function of $\widetilde{\rm{Wi}}$ in 
Fig.~\ref{fig:scaling}(a) nearly coincide.
For a semi-dilute case~($\phi_p=0.1$), the the stresslet viscosity change agrees with the dilute cases for $\widetilde{\rm{Wi}}\lesssim 1.5$. At higher $\widetilde{\rm{Wi}}\gtrsim 1.5$, the negative slope of the stresslet viscosity becomes smaller than that in the dilute cases, though this change is not large compared to that at $S_{xy,0}(\phi_{p})/(\eta_0\dot{\gamma}a^{3})$.
The change of $\langle S_{xy}\rangle/(\eta_0\dot{\gamma}a^{3})$ 
as a function of 
$\langle\sigma_{xy}^{\rm{sus}}\rangle\lambda^{(1)}/\eta_{0}$ in 
Fig.~\ref{fig:scaling_Yang}(a) shows a similar trend to that 
presented in Fig.~\ref{fig:scaling}(a). {
Although both ${\rm Wi}$ and $\phi_p$ increase 
$\widetilde{\rm{Wi}}$ and thus the elastic contribution in the 
fluid, the stresslet changes with $\phi_p$ and 
$\widetilde{\rm{Wi}}$ at $\phi_p=0.1$ are in opposite directions.
This suggests the stresslet change due to microstructure at 
$\phi_p=0.1$ in addition to the change induced by polymer stress 
around individual particles.}
For $\langle S_{xx-yy}\rangle$, the large error at $\rm{Wi}=0.1$ 
makes it difficult to evaluate the analysis as it is done for 
$\langle S_{xy}\rangle$.

Next, the particle-induced fluid viscosity 
$\langle\Sigma_{xy}\rangle/(\eta_0\dot{\gamma}a^{3})$ that 
directly accounts for the elastic stress is discussed.
At $\widetilde{\rm{Wi}}\lesssim 1$ in Fig.~\ref{fig:scaling}(b), the particle-induced fluid viscosity does not depend on $\phi_p$ because the elastic stress almost relaxes at $\widetilde{\rm{Wi}}\lesssim 1$.
This region of $\widetilde{\rm{Wi}}$ corresponds to the zero-shear plateau of the suspension viscosity.
At $\widetilde{\rm{Wi}}> 1$, the increase of $\langle\Sigma_{xy}\rangle/(\eta_0\dot{\gamma}a^{3})$  with $\widetilde{\rm{Wi}}$ is 
enhanced as $\phi_p$ increases, indicating increased 
elastic stress with $\phi_p$.
Since the elastic stress is dependent on flow-history and is not a 
simple function of the shear rate, the rate of increase of 
$\langle\Sigma_{xy}\rangle/(\eta_0\dot{\gamma}a^{3})$ with respect to 
$\widetilde{\rm{Wi}}$ changes with $\phi_{p}$.
The plot of $\langle\Sigma_{xy}\rangle/(\eta_0\dot{\gamma}a^{3})$ 
as a function of 
$\langle\sigma_{xy}^{\rm{sus}}\rangle\lambda^{(1)}/\eta_{0}$ in 
Fig.~\ref{fig:scaling_Yang}(b) does not depend on $\phi_p$ for 
dilute conditions~($\phi_p\leq0.05$), which is consistent with 
the previous work~\citep{Yang2018a}. At a semi-dilute 
condition~($\phi_p=0.1$), 
$\langle\Sigma_{xy}\rangle/(\eta_0\dot{\gamma}a^{3})$ is slightly 
lower than that in the dilute cases, but the rate of increase is 
almost the same as that in the dilute condition. 
In Fig.~\ref{fig:scaling_Yang}(b), after a slow increase at 
$\langle\sigma_{xy}^{\rm{sus}}\rangle\lambda^{(1)}/\eta_{0}\ll 1$, the particle-induced fluid 
viscosity increases linearly to $\langle\sigma_{xy}^{\rm{sus}}\rangle\lambda^{(1)}/\eta_{0}\gtrsim 0.5$.
{
The purely elastic contribution is directly evaluated by the polymer dissipation function, $\Phi_p^{(k)}=
\left(\eta_p^{(k)}/(2(\lambda^{(k)})^2)\right)\left\{{\rm tr}\bm{C}^{(k)}+{\rm tr}\bm{C}^{(k)-1}-6\right\}$.
By using $\Phi_p$, an extra elastic contribution compared to a pure Oldroyd-B fluid is discussed in ~\citet{Vazquez-Quesada2019}. 
By definition, the polymer dissipation function is a scalar of 
$\bm{C}$ and thus independent of the direction of $\bm{C}$; ${\rm 
tr}\bm{C}-3$ and ${\rm tr}\bm{C}^{-1}-3$ measure the stretch and 
compression of $\bm{C}$, respectively.
Fig.~\ref{fig:scaling_Yang}(c) shows the normalized polymer 
dissipation function of the first 
mode,~$2\langle\Phi_p^{(1)}\rangle_{\rm 
f}(\lambda^{(1)})^2/\eta_p^{(1)}$, as a function of 
$\langle\sigma_{xy}^{\rm{sus}}\rangle\lambda^{(1)}/\eta_{0}$.
Fig.~\ref{fig:scaling_Yang}(c) shows that the normalized polymer 
dissipation function at different $\phi_p$ collapses onto a 
single mastercurve, directly suggesting the similarity of the 
elastic contribution up to $\phi_p\leq 0.1$.
Fig.~\ref{fig:scaling_Yang}(d)} shows the shear-thickening part per particle defined 
as $\eta_{r,t}=[\eta_r(\phi_p,\rm{Wi})-\eta_r(\phi_p,\rm{Wi}\rightarrow0)]/\phi_p$ as a function of suspension shear stress, {where $\eta_r(\phi_p,{\rm Wi}\rightarrow0)$ is approximated by $\eta_r(\phi_p,{\rm Wi}=0.1)$ because $\eta_r$ almost reaches the zero-shear plateau even at ${\rm Wi}=0.1$. 
Up to semi-dilute cases ($\phi_p\leq 0.1$), the increases in 
$\eta_{r,t}$ with $\langle\sigma_{xy}^{\rm{sus}}\rangle$ nearly 
coincides.
Previous work~\citep{Yang2018a} reported that the variation of 
$\eta_{r,t}$ with $\langle\sigma_{xy}^{\rm{sus}}\rangle$ did not 
depend on $\phi_p$ for $\phi_{p}\leq 0.1$, which is also 
confirmed in this work.}

\subsubsection{Flow characterization of viscoelastic suspension}
\label{sc:many_flow}

\begin{figure}    
\centering
\vspace{1mm}
\includegraphics[scale=.75]{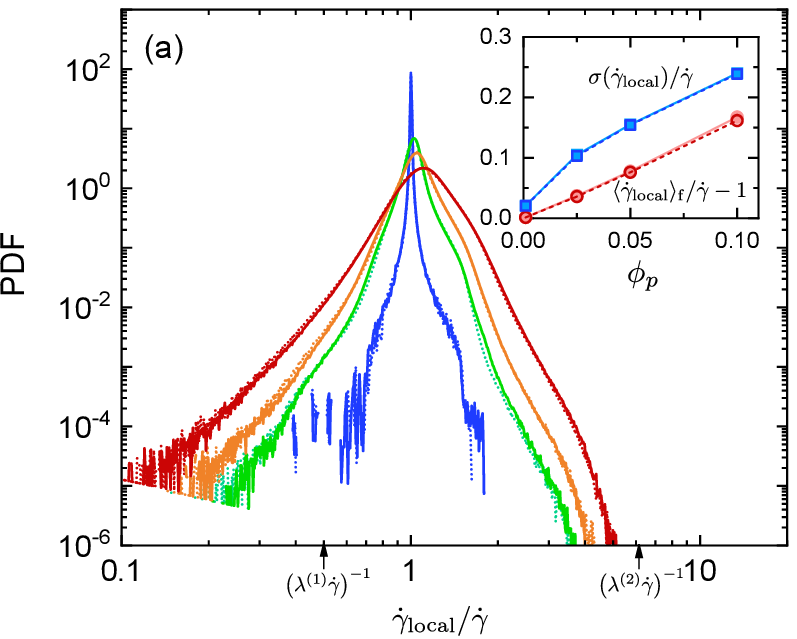}
\includegraphics[scale=.75]{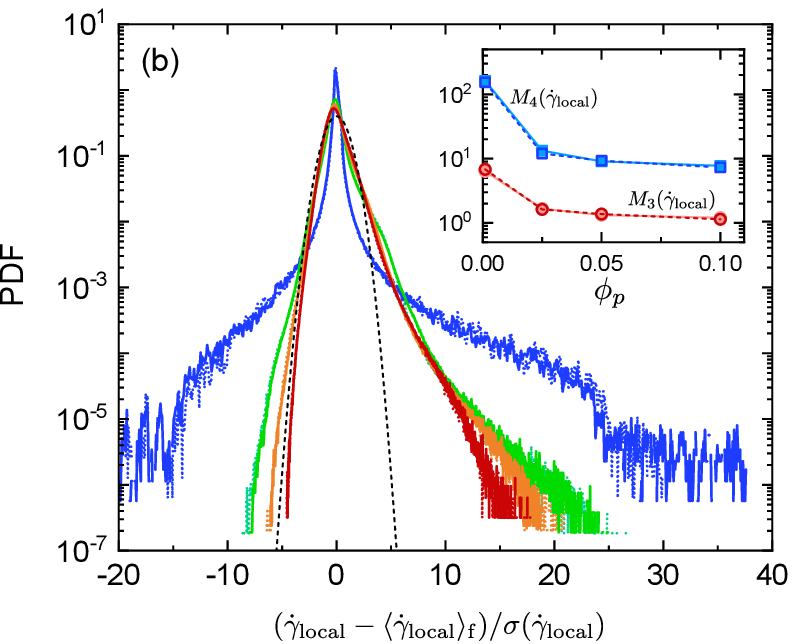}
\centering
\centering
\includegraphics[scale=.78]{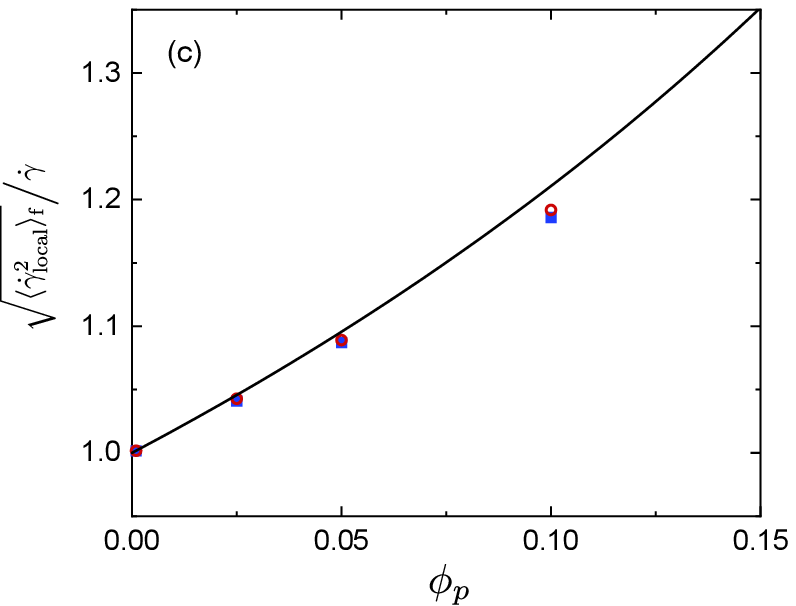}
\caption{\label{fig:pdf_sr} The $\phi_p$ and $\rm{Wi}$ dependence 
of the local strain-rate in the fluid region: (a) PDF of 
$\dot{\gamma}_{\rm{local}}$, where $\phi_p=0.1$ (red), 0.05 
(orange), 0.025 (green), and  0.001 (blue), and the dotted and 
solid lines represent PDFs at $\rm{Wi}=0.1$ and $2.0$, 
respectively. The strain-rate is normalized by the imposed shear 
rate $\dot{\gamma}$, and the arrows indicate the first and second 
relaxation rates at $\rm{Wi}=2.0$. The inset shows the $\phi_p$ 
dependence of the mean and the standard deviation for Wi$=0.1$ (open 
symbols) and 2.0 (closed symbols). (b) PDF of $\dot{\gamma}_{\rm 
local}$  centered at the mean and normalized by standard 
deviation. The inset shows the $\phi_p$ dependence of the 
skewness and kurtosis. The line types are the same as those in (a). 
The dashed line indicates the standard Gaussian distribution. (c) 
Average local strain-rate, where blue squares and red circles 
correspond to Wi$=$0.1 and 2.0, respectively, and the line is 
the result from homogenization theory.}
\end{figure}

\begin{figure}
  \begin{center}
\includegraphics[scale=0.75]{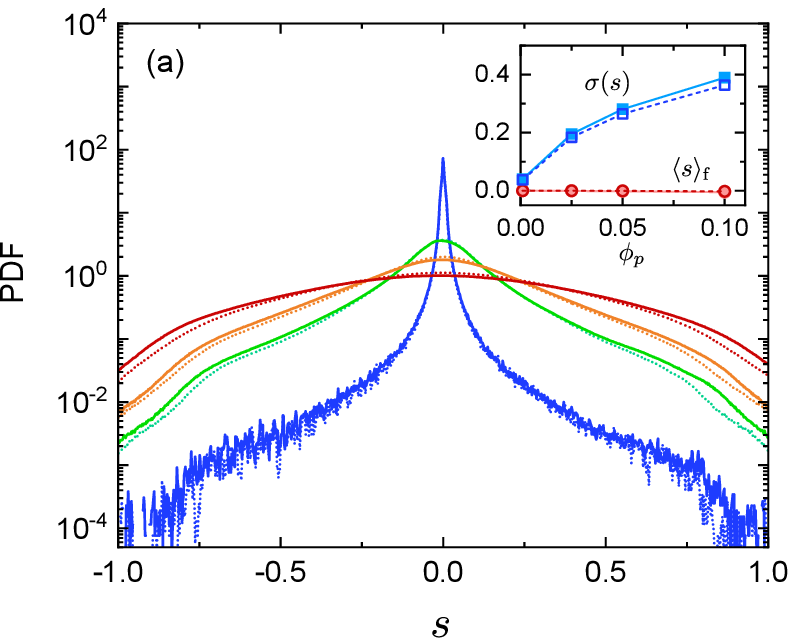}
  \end{center}
  \begin{center}
\includegraphics[scale=0.75]{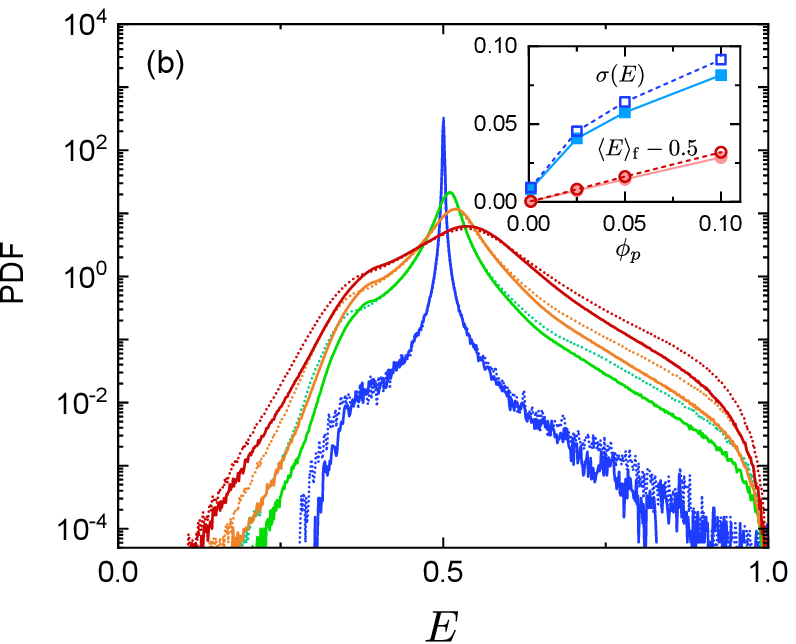}
  \end{center}
\caption{\label{fig:flow_pattern} PDF of (a) strain-rate state 
$s$ and (b) irrotationality $E$ in the fluid region at various 
values of $\phi_p$ (0.1 (red), 0.05 (orange), 0.025 (green), 
and 0.001 (blue)). For each $\phi_p$, \rm{Wi}=0.1 (dotted lines) and 
2.0 (solid lines). {The inset shows the $\phi_p$ dependence of 
the mean (red circles) and standard deviation (blue squares). The open and 
solid symbols in the inset indicate the results of ${\rm Wi}=0.1$ 
and $2.0$, respectively.}}
\end{figure}

\noindent The probability density functions (PDF) of the local 
shear-rate $\dot{\gamma}_{\rm{local}}=\sqrt{2\bm{D}:\bm{D}}$ in 
the fluid domain for different $\phi_p$ and $\rm{Wi}$ are 
presented in Fig.~\ref{fig:pdf_sr}(a). 
To sample the different particle configurations under flow for 
many-particle systems, the PDF is calculated from data over 25 
snapshots per sample~(in all, 75 snapshots) at the steady state 
by every $\dot{\gamma}\Delta t=0.215$ strain increment in three 
different initial particle configuration samples. 
Here $\dot{\gamma}_{\text{local}}\neq\dot{\gamma}$ is from the 
inhomogeneous flow near the particles, whereas 
$\dot{\gamma}_{\text{local}}=\dot{\gamma}$ is mainly from the 
region far from the particles where the flow is close to 
homogeneous shear flow.
For the same $\rm{Wi}$, {as $\phi_p$ increases, the shape of the 
PDF broadens and the peak position in the PDF gradually shifts 
towards large shear rate. This trend is clearly observed by the 
$\phi_p$ dependence of the mean $\langle\dot{\gamma}_{\rm 
local}\rangle_{\rm f}$ and standard deviation 
$\sigma(\dot{\gamma}_{\rm local})$~(the inset in 
Fig.~\ref{fig:pdf_sr}(a)).}
Specifically, $\dot{\gamma}_{\text{local}}/\dot{\gamma}\lesssim 
2$ for $\phi_{p}=0.001$ (single-particle result), while 
$\dot{\gamma}_{\text{local}}/\dot{\gamma}\lesssim 5$ for 
$\phi_{p}=0.1$.
In general, a large shear-rate is effective in exciting the fast 
relaxation mode.
At $\rm{Wi}=0.1$, the normalized first relaxation rate 
$\left(\lambda^{(1)}\dot{\gamma}\right)^{-1}=10$ is beyond the 
range of the local shear-rate for $\phi_{p}\leq 0.1$; therefore, 
the elastic response is irrelevant.
At $\rm{Wi}=2$, where the first mode is relevant, the normalized second relaxation rate is $\left(\lambda^{(2)}\dot{\gamma}\right)^{-1}=6.15$, thus indicating that the second mode is still irrelevant to the elastic response.
The PDF of the local shear rate which is centered at the mean and 
is normalized by the standard deviation, is shown in 
Fig.~\ref{fig:pdf_sr}(b). At $\phi_p=0.001$, the normalized PDF 
is highly skewed and has fat tails. This corresponds to large 
positive values of the skewness $M_3(\dot{\gamma}_{\rm local})$ 
and kurtosis $M_4(\dot{\gamma}_{\rm local})$~(the inset in 
Fig.~\ref{fig:pdf_sr}(b)), where $M_n(f)=\langle(f-\langle 
f\rangle_{\rm f})^n\rangle_{\rm f}/\sigma^n(f)$ is the normalized 
$n$-th-order statistics of $f$. As $\phi_p$ increases, the shape 
of PDF becomes closer to the Gaussian distribution~(the dashed 
line), which corresponds to the decrease of 
$M_3(\dot{\gamma}_{\rm local})$ and $M_4(\dot{\gamma}_{\rm 
local})$. 
However, even at $\phi_p=0.1$, the PDF remain positively skewed, 
suggesting the asymmetric nature of the local shear rate 
distribution. In addition, the shape of PDF in 
Fig.~\ref{fig:pdf_sr}~(a),(b) is not sensitive to the change in 
$\rm Wi$. 
Fig.~\ref{fig:pdf_sr}(c) shows the root-mean-square of the local 
shear-rate 
$\tilde{\dot{\gamma}}=\sqrt{\langle\dot{\gamma}_{\text{local}}^2\
rangle_{\rm f}}$ as a function of $\phi$ at $\rm{Wi}=0.1$ and 
$2.0$. 
This average shear-rate increases with $\phi_p$
because the deformable fluid volume decreases with $\phi_p$.
This phenomenon is expected to be common in solid suspensions.
For comparison, a prediction for the average shear-rate by a homogenization theory for viscous fluid~\citep{Chateau2008},
\begin{align}
    \tilde{\dot{\gamma}}=\dot{\gamma}\sqrt{\frac{\eta_r(\phi_p,\rm{Wi}\rightarrow0)}{1-\phi_p}},
    \label{eq:homo_theory}
\end{align} 
is drawn in Fig.~\ref{fig:pdf_sr}(b), where $\eta_r$ in Eq.~(\ref{eq:homo_theory}) is calculated with the Eilers fit by \citet{Haddadi2014}. Although the increasing trend of the average shear rate with $\phi_p$ is similar, the average shear-rate in the studied viscoelastic medium is slightly smaller than that predicted by Eq.~(\ref{eq:homo_theory}). 
This is partly because Eq.(\ref{eq:homo_theory}) does not consider suspension microstructures explicitly. 
In fact, even for a Newtonian medium, Eq.(\ref{eq:homo_theory}) was reported to overestimate the suspension viscosity obtained by DNS at high $\phi_p$~\citep{Alghalibi2018}.
From Fig.~\ref{fig:pdf_sr}, the level of shear rate is hardly 
affected by ${\rm Wi}$ for ${\rm Wi}\leq 2$, and the fluctuation 
of $\dot{\gamma}_{\rm{local}}$ is mainly dominated by the solid 
volume fraction.

Next, the local flow pattern is discussed for different 
$\phi_{p}$ and $\rm{Wi}$.
The topological aspect of the local flow pattern defined by 
$\nabla\bm{u}$ can be characterized by two scalars: 
multi-axiality of the strain-rate and irrotationality of 
$\nabla\bm{u}$~\citep{Nakayama2016}.
The multi-axiality of flow in the incompressible flow is 
conveniently identified by the strain-rate state, which is defined 
as
\begin{align}
    s=\frac{3\sqrt{6}\det\bm{D}}{(\bm{D}:\bm{D})^{(3/2)}},
\end{align} 
where $s\in[-1,1]$ by definition.
For uniaxial elongational flow, where stretching in one direction 
and compression in the other two directions occur, $s>0$, whereas 
for biaxial elongational flow, where compression in one direction 
and stretching in the other two directions occur, $s<0$. 
For planar flow, where stretching occurs in one direction, 
compression occurs in another direction and no strain is found in 
the other direction, $s=0$.
The magnitude of $s$ is determined by the relative magnitude of the three principal strain-rates of $\bm{D}$.
Fig.~\ref{fig:flow_pattern}(a) shows PDFs of $s$ for different $\phi_p$ and $\rm{Wi}$. 
Since homogeneous shear flow is planar flow, $s=0$ when $\phi_p\to 0$.
As $\phi_{p}$ and/or $\rm{Wi}$ increase, the fraction of the 
planar flow indicated by $s=0$ decreases, and the fraction of 
triaxial flow indicated by $s\neq 0$ increases. {This trend is 
also captured by the mean and standard deviation of $s$~(the 
inset in Fig.~\ref{fig:flow_pattern}(a)).}

The relative contribution of vorticity to the strain-rate is characterized by irrotationality, which is defined as
\begin{align} 
\label{eq:irrot}
E=\frac{\sqrt{\bm{D}:\bm{D}}}{\sqrt{\bm{D}:\bm{D}}+\sqrt{\bm{\Omega}:\bm{\Omega}^T}},
\end{align}
 where 
$\bm{\Omega}=(\nabla\bm{u}-\nabla\bm{u}^T)/2$ is the vorticity tensor. 
By definition, $E\in[0,1]$. 
For rigid-body rotation, $E=0$; and $E=1$ for irrotational flow. As the vorticity contribution decreases, $E$ increases.
Fig.~\ref{fig:flow_pattern}(b) shows the PDF of the irrotationality for different $\phi_p$ and $\rm{Wi}$.
In homogeneous simple shear flow at $\phi_{p}\to 0$, the flow is half rotational, {\itshape i.e.}, $E=1/2$.
As $\phi_p$ increases, the fraction of $E=1/2$ decreases, whereas the fraction of $E\neq 1/2$ increases.
In particular, the fraction of $E>1/2$ is larger than that of $E<1/2$, indicating that the region with more irrotational flow than homogeneous shear flow increases with $\phi_p$.
Since the vorticity contribution makes the fluid element avoid stretching, a large $E$ value suggests that the flow is strain-dominated to promote stretching of the conformation. 
As $\rm Wi$ increases, the width of the $E$ PDF gets narrower. 
The trend of the $E$ PDF with $\phi_p$ and ${\rm Wi}$ is 
summarized by the mean and standard deviation of $E$~(the inset 
in Fig.~\ref{fig:flow_pattern}(b)). The insets in 
Figs.~\ref{fig:flow_pattern}(a) and (b) indicate that the flow 
pattern as measured by $s$ and $E$ is mostly dominated by 
$\phi_p$. 
These changes in the PDFs of $s$ and $E$ reflect the 
modulation of the flow caused by the particle inclusion, which is 
further examined in the following section.

\begin{figure*}
\centering
\includegraphics[scale=1.]{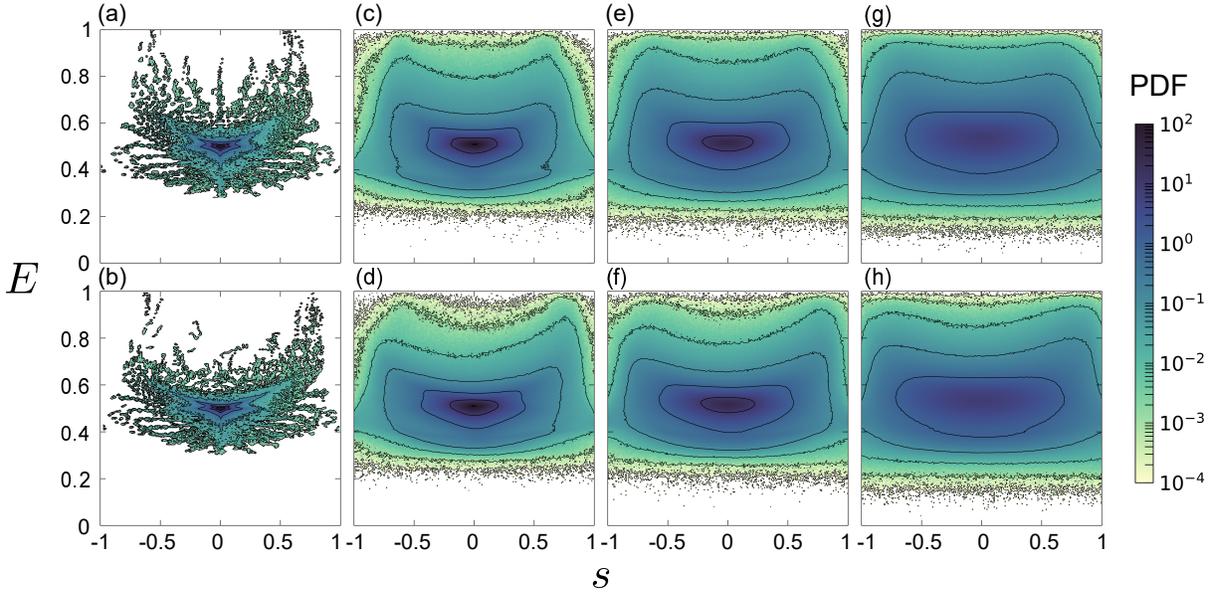}
\captionsetup{justification=centering}
\caption{\label{fig:scat_sE} The 2D PDFs of strain-rate state $s$ and 
irrotationality $E$ in the fluid region: (a),(b)~$\phi_p=0.001$; 
(c),(d)~$\phi_p=0.025$; (e),(f)~$\phi_p=0.05$; and 
(g),(h)~$\phi_p=0.1$. The top and bottom row indicate 
the results at $\rm{Wi}=0.1$ and $2.0$, respectively. The contour 
lines correspond to ${\rm PDF}=10^k: k=-4,-3,-2,-1,0,1,2$.}
\end{figure*}

To discuss the correlation between the strain-rate state and 
irrotationality and the spatial variation of the flow pattern, a 
joint PDF of $s$ and $E$ for different $\phi_p$ at $\rm{Wi}=0.1$ 
and $2.0$ is shown in Fig.~\ref{fig:scat_sE}; snapshots of $s$ 
and $E$ on a shear plane at different $\phi_p$ and $\rm{Wi}$ are 
presented in Figs.~\ref{fig:2d_dist_Wi0_5} 
and~\ref{fig:2d_dist_Wi2_0}, respectively.
The simple shear flow corresponds to $(s,E)=(0,1/2)$.
At $\rm{Wi}=0.1$ and $\phi_p=0.001$ and 
0.025~(Figs.~\ref{fig:scat_sE}(a) and (c), respectively), the 
$s-E$ distribution appears like the face of a fox; high-$E$ flow 
is actually non-planar high-$|s|$ flow, which forms the fox's 
ears.
At $\rm{Wi}=0.1$, the distribution of $s$ is almost symmetric for 
different $\phi_{p}$~(Fig.~\ref{fig:scat_sE}(a),(c),(e), and (g)), 
thus reflecting the fore-aft symmetry of the flow around a 
particle~($s$ and $E$ at $\phi_p=0.025$ in 
Fig.~\ref{fig:2d_dist_Wi0_5}).
For irrotational flow of $E>0.5$, the fraction of the planar flow 
of $s=0$ is relatively small, and hence, the triaxial flow of 
$s\neq 0$ is predominant.
This reflects the flow in the upstream and downstream regions of 
the particles~($s$ and $E$ at $\phi_p=0.025$ in 
Fig.~\ref{fig:2d_dist_Wi0_5}), where the flow is forced to avoid 
the particles to generate irrotational bifurcating (biaxial 
elongational) flow upstream and irrotational converging (uniaxial 
elongational) flow 
downstream~\citep{Einarsson2018,Yang2018,Vazquez-Quesada2019,Matsuoka2020}.
As $\rm{Wi}$ increases, the distribution of $s$ at $E>1/2$ becomes asymmetric~(Fig.~\ref{fig:scat_sE}(b),(d), and (f)); the fraction of $s>0$ is larger than that of $s<0$.
This corresponds to symmetry breaking in the upstream and downstream flows around the particles with an increase of $\rm{Wi}$. 
As shown in the $E$ distribution at $\phi_{p}=0.025$ and $\rm{Wi}=2.0$~(Fig.~\ref{fig:2d_dist_Wi2_0}), high-$E$ regions around a particle shift counter-clockwise with respect to the symmetric distribution at $\rm{Wi}=0.5$.
Because of this change, in the upstream region of the particle, 
the vorticity contribution increases with $\rm{Wi}$, leading to a 
decrease in $E$, whereas $E$ in the downstream region does not 
change significantly. 
This change of flow patterns with $\rm{Wi}$ is attributed to the 
local flow modulation by large polymer stress gradients around 
particles, which was examined in detail in our previous study for 
a single-particle system~\citep{Matsuoka2020}. 
Although ${\rm Wi}$ affects the local flow pattern around a 
particle, the microstructure does not change obviously with $\rm 
Wi$, as seen in Figs.~\ref{fig:2d_dist_Wi0_5} and 
\ref{fig:2d_dist_Wi2_0}.

In this study, our DNS of many-particle systems enables us to 
examine the effect of the particle volume fraction on the local 
flow patterns. As $\phi_p$ increases, the $s-E$ PDF spreads out 
widely~(from left to right panels in Figs.~\ref{fig:scat_sE}). In 
addition to the increase in the fraction of the characteristic 
flow field around single particles, this $s-E$ distribution also 
reflects the spatial overlap of the characteristic flow field between 
particles, which is shown in Figs.~\ref{fig:2d_dist_Wi0_5} 
and~\ref{fig:2d_dist_Wi2_0}.
Especially, the high-$E$ fox ears in the $s-E$ PDF are smeared 
out {with increased $\phi_p$} because the interaction between 
particles becomes predominant to modify the flow between 
particles.
Fig.~\ref{fig:3d_sE} shows the color contour of the strain-rate 
state at the highly irrotational region of $E\geq0.65$ at 
$\rm{Wi}=2.0$. 
At the dilute condition of $\phi_p=0.025$, the highly 
irrotational region adjacent to each particle is isolated over 
most of the time~(Fig.~\ref{fig:3d_sE}(b)). 
In contrast, as $\phi_p$ increases, additional bifurcating 
irrotational regions develop between particles when two particles 
get closer~(Fig.~\ref{fig:3d_sE}(c)(d)).

\begin{figure*}
\centering
\includegraphics[scale=0.8]{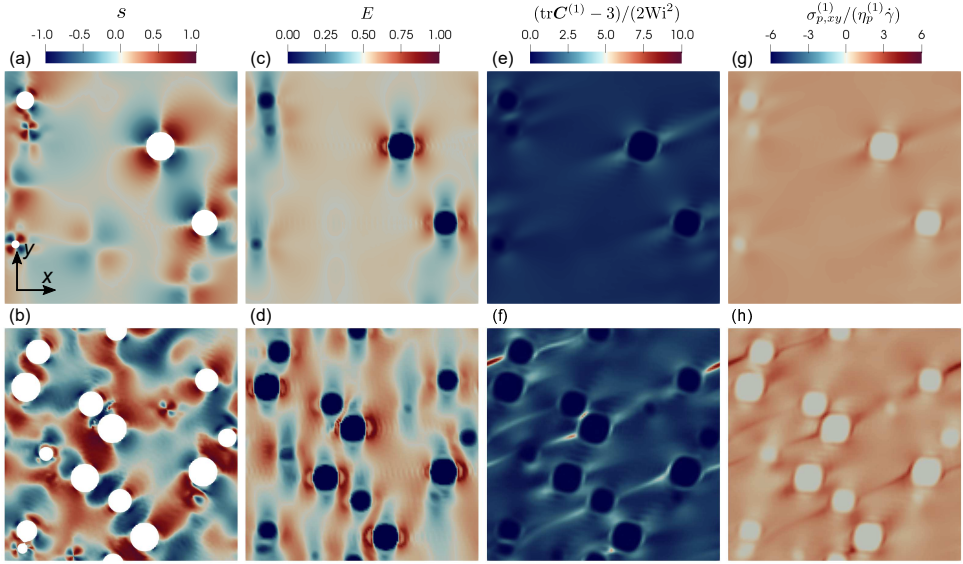}
\caption{\label{fig:2d_dist_Wi0_5} Spatial distribution of the 
flow pattern characterized by (a),(b) the strain-rate state $s$; 
(c),(d) irrotationality $E$; (e),(f) normalized polymer stretch 
of the first mode $(\text{tr}\bm{C}^{(1)}-3)/(2\rm{Wi}^2)$; and 
(g),(h) the normalized shear stress of the first mode 
$\sigma_{p,xy}^{(1)}/(\eta_{p}^{(1)}\dot{\gamma})$ on a shear 
plane~($x, y$ plane) at $\rm{Wi}=0.5$. The top~(a, c, e, g) 
and  bottom~(b, d, f, h) rows are the results at 
$\phi_p=0.025$ and $0.1$, respectively. }
\end{figure*}

\begin{figure*}
\centering
\includegraphics[scale=0.8]{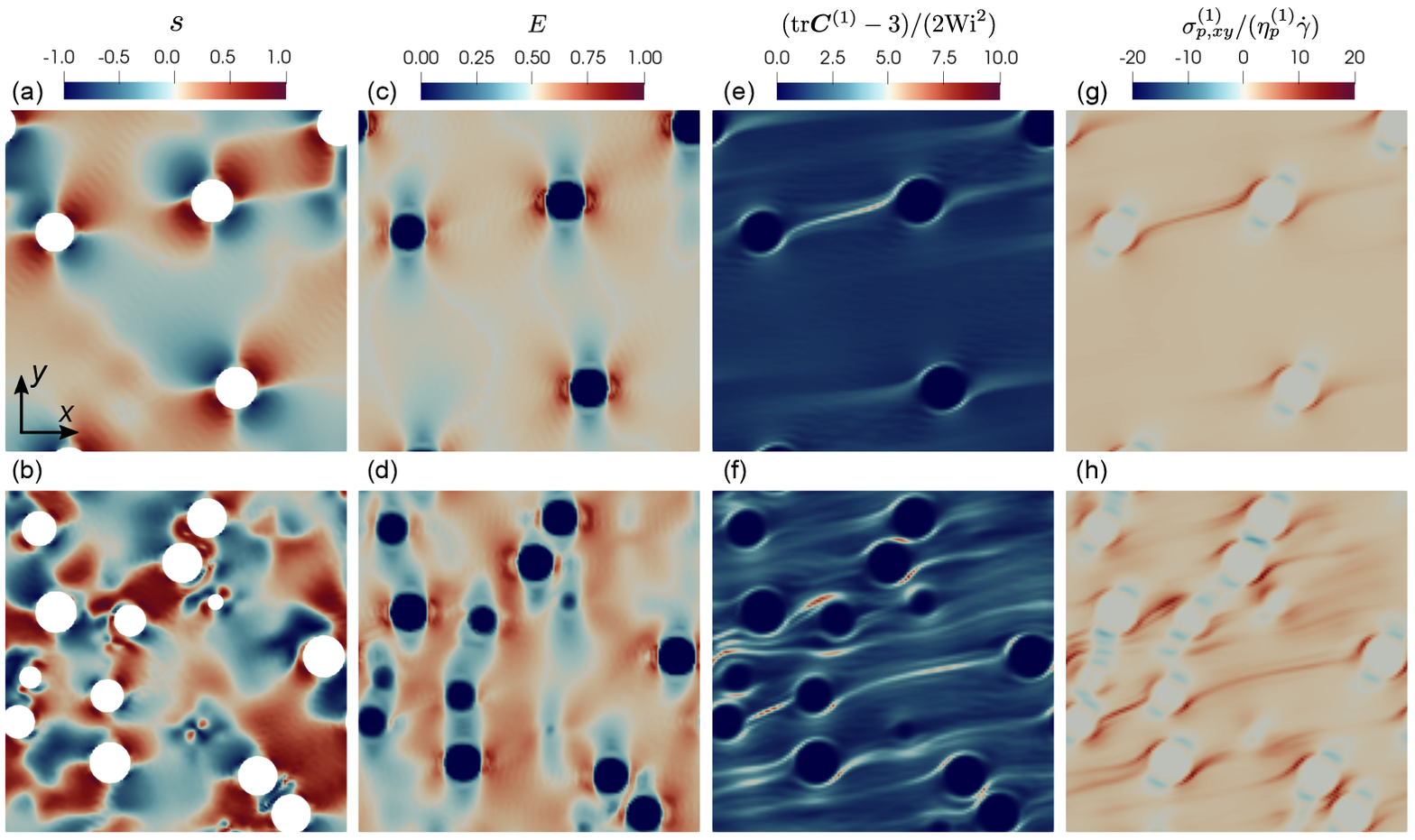}
\caption{\label{fig:2d_dist_Wi2_0} Spatial distribution of the 
flow pattern characterized by (a),(b) the strain-rate state $s$; 
(c),(d) irrotationality $E$; (e),(f) normalized polymer stretch 
of the first mode $(\text{tr}\bm{C}^{(1)}-3)/(2\rm{Wi}^2)$; and 
(g),(h) the normalized shear stress of the first mode 
$\sigma_{p,xy}^{(1)}/(\eta_{p}^{(1)}\dot{\gamma})$ on a shear 
plane~($x, y$ plane) at $\rm{Wi}=2.0$. The top~(a, c, e, g) 
 and  bottom~(b, d, f, h) rows are the results at 
$\phi_p=0.025$ and $0.1$, respectively.}
\end{figure*}

\begin{figure*}
\centerline{
\includegraphics[scale=1.]{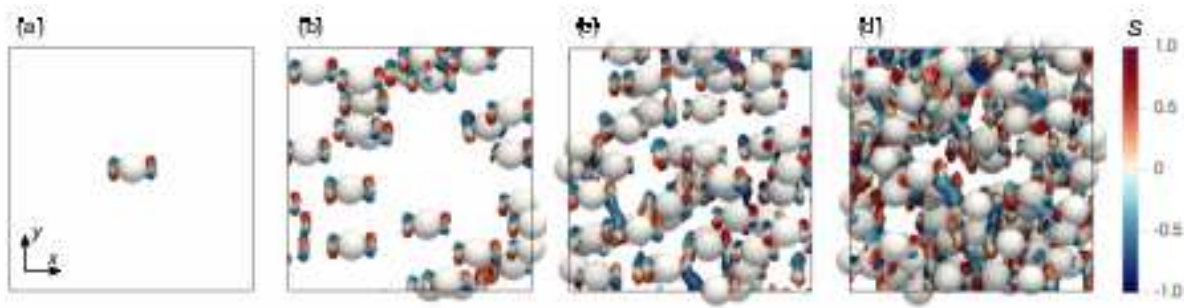}}
\caption{\label{fig:3d_sE} The isovolume visualization of a 
highly irrotational region at $\rm{Wi}=2.0$: (a)~$\phi_p=0.001$, 
(b)~$0.025$, (c)~$0.05$, and (d)~$0.1$. The isovolume represents 
a region where $E\geq 0.65$, and the color represents the 
strain-rate state $s$.}
\end{figure*}

\begin{figure*}
\centerline{
\includegraphics[scale=1.]{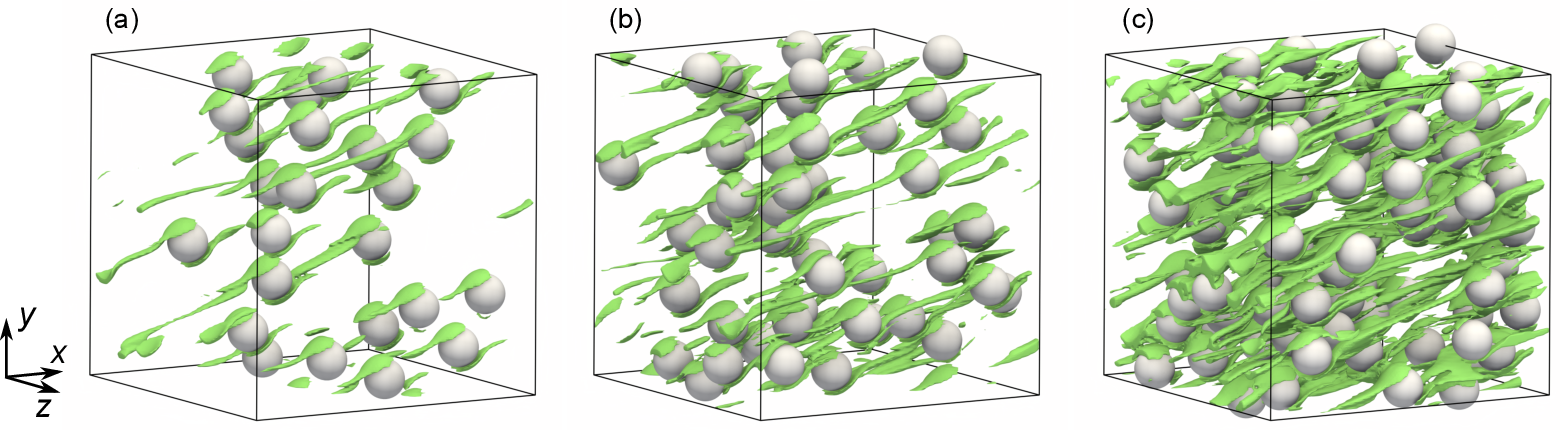}}
\caption{\label{fig:trC_3d_Wi2_0} Polymer high-stretch region at 
$\rm{Wi}=2.0$: (a)~$\phi_p=0.025$, (b)~$0.05$, and (c)~$0.1$. The 
isovolume of ${\rm tr}\bm{C}^{(1)}\geq 
2(2\widetilde{\rm{Wi}}^2+3)$ is visualized in green, where the 
threshold is twice the value of $\text{tr}\bm{C}$ in the 
Oldroyd-B fluid under homogeneous shear flow.}
\end{figure*}

Finally, the development of polymer stretch  and polymer shear stress at different $\phi_p$ and $\rm{Wi}$ is discussed with a focus on the longest relaxation mode ($k=1$) responsible for shear-thickening.
The snapshots of the normalized polymer stretch $({\rm tr}\bm{C}^{(1)}-3)/(2\rm{Wi}^2)$ and the normalized polymer shear stress $\sigma_{p,xy}^{(1)}/\eta_{p}^{(1)}\dot{\gamma}=C_{xy}^{(1)}/\rm{Wi}$ at $\phi_p=0.025$ and $0.1$ are shown in 
Fig.~\ref{fig:2d_dist_Wi0_5} for $\rm{Wi}=0.5$ and in 
Fig.~\ref{fig:2d_dist_Wi2_0} for $\rm{Wi}=2.0$.
The polymer shear stress distribution is similar to that of the normalized stretch.

At $\rm{Wi}=0.5$~(Fig.~\ref{fig:2d_dist_Wi0_5}), polymer stretch is promoted in the irrotational flow at the fore and aft of a particle. This results in two high-stretch regions; one is the recirculation region adjacent to the particle, and the other is downstream of the particle. 
In the recirculation flow around a particle, the polymer is subjected to repeated stretch and reorientation, thus resulting in a high-stretch region around the particle~\citep{Yang2018,Matsuoka2020}.
On the other hand, outside the recirculation flow, the polymer that has passed through the irrotational region 
around a particle is advected downstream to form another 
high-stretch region slightly diagonal to the flow direction. 

In a dilute condition of $\phi_p=0.025$, the high-stretch regions associated with different particles rarely interact with each other.
As $\phi_p$ increases, a downstream high-stretch region shared by two particles is observed that occurs after the two particles pass each other.
In the case of $\rm{Wi}=0.5$ in Fig.~\ref{fig:2d_dist_Wi0_5}, the downstream high-stretch region relaxes and does not reach far; hence the structure of the elastic stress at $\phi_p=0.1$ is similar to that in dilute cases.
This is consistent with what was observed in Fig.~\ref{fig:scaling}(b); the relationship between the particle-induced fluid stress $\langle\Sigma_{xy}\rangle$ and $\widetilde{\rm{Wi}}$ does not depend on $\phi_p$ when $\widetilde{\rm{Wi}}\lesssim 1.5$. 

On the other hand, at $\rm{Wi}=2.0$~(Fig.~\ref{fig:2d_dist_Wi2_0}), 
the downstream high-stretch region between particles does not 
relax immediately and extends over a long distance. 
Fig.~\ref{fig:trC_3d_Wi2_0} shows a 3D view of the high-stretch 
region at different $\phi_p$ and $\rm{Wi}=2.0$, where the 
isovolume of ${\rm tr}\bm{C}^{(1)}$ that is more than twice the 
stretch in an Oldroyd-B fluid under homogeneous shear flow is 
visualized.
As $\phi_p$ increases, the streak-shaped high-stretch regions 
bridging two separated particles become more evident. 
At $\phi_p=0.1$, most particles share high-stretch regions with 
other particles. 
This result suggests that the development of elastic stress at 
$\phi_p=0.1$ and $\rm{Wi}\gtrsim 2$ is qualitatively different 
from that in dilute cases. 
However, despite this distinctive microscopic picture observed in 
the polymer stretching, the effect of such polymer stretching 
structures on the averaged bulk polymer stress is still not 
significant in the scope of the present study, as seen in 
Figs.~\ref{fig:scaling} and~\ref{fig:scaling_Yang}. 
The polymer stretching structure between many particles 
identified in Fig.~\ref{fig:trC_3d_Wi2_0} would cause a 
qualitative change in the suspension rheology at higher $\phi_p$ 
and/or ${\rm Wi}$ where such structures would become more 
frequent and persistent.

\section{Conclusions}
\label{sc:summary}\noindent To elucidate the key factor for the 
quantitative prediction of the shear-thickening in suspensions in 
Boger fluids, DNS of many-particle suspensions in a multi-mode 
Oldroyd-B fluid is performed using SPM.
To evaluate the suspension rheology in bulk systems, rather than 
applying a wall-driven confined system, simple shear flow is 
imposed by Lees--Edwards periodic boundary conditions for the 
particle dynamics; a time-dependent moving frame that evolves 
with the mean shear flow is applied to create simple shear flow 
for the fluid dynamics.
Our DNS is validated by analyzing the viscoelastic flow in a 
single-particle suspension in an Oldroyd-B fluid under simple 
shear. Good agreement is obtained with analytical solutions as 
well as with numerical results for the shear-thickening in the 
suspension viscosity as well as in the viscosity from the 
particle-induced fluid stress, and the shear-thinning in the 
viscosity from the stresslet.

The shear rheology of many-particle systems is investigated from dilute to semi-dilute conditions up to $\phi_p\leq 0.1$ and $\rm{Wi}\leq 2.5$.
Based on previous experimental work on a suspension in a Boger fluid~\citep{Yang2018a}, a four-mode Oldroyd-B fluid is used as a matrix to mimic the linear modulus of the Boger fluid.
The presented many-particle, multi-mode results for the shear-thickening behavior of a suspension quantitatively agree with the experimental results.
Furthermore, for $\rm{Wi}\leq 2.5$, an effective set of parameters is derived for single-mode Oldroyd-B modelling for the matrix by considering a relevant mode in the four-mode modelling. 
The many-particle results with this effective single-mode model 
also reproduce the experimentally observed shear-thickening 
behavior in a suspension; this is in contrast to the 
underestimation obtained by another DNS study that used a 
different set of the fluid parameters~\citep{Yang2018a}.
The presented results elucidate that, with an accurate estimation 
of $N_1$ of the matrix in the shear-rate range where the 
shear-thickening starts to occur, shear-thickening in a 
suspension in a Boger fluid at around $\rm{Wi}=1$ can be 
predicted with a relevant mode Oldroyd-B model. 
This finding in our study prompts us to consider 
shear-thickening of suspensions in more complex viscoelastic 
media showing strong non-linearity in viscosity and $N_1$. In 
such cases, a proper estimation of nonlinear matrix $N_1$ as well 
as viscosity should be required to predict suspension 
rheology. 
Understanding the effects of matrix nonlinearity on suspension 
rheology is our future work.

At a dilute suspension, the single-particle and many-particle 
systems are compared, clarifying that the single-particle 
simulation underestimates the stresslet contribution due to the 
lack of relative motion between particles, {which is another 
factor affecting the quantitative prediction of the suspension 
rheology. 
The underestimation of the suspension viscosity in 
a single-particle calculation was pointed out in a previous work 
with a wall-driven system~\citep{Vazquez-Quesada2019}.
We revealed that the cause of the quantitative discrepancy comes 
from the stresslet contribution by the suspension microstructure.
The suspension stress decomposition into the stresslet and the  
particle-induced fluid stress demonstrated the scaling of the 
polymer contribution to the total shear-thickening as was 
reported in a previous DNS result up to $\phi_p\leq0.1$ and ${\rm 
Wi}\leq1.0$~\citep{Yang2018a}.
The underlying similarity of the elastic contribution at different 
$\phi_p\leq0.1$ was directly confirmed by the scaling relation of 
the normalized polymer dissipation function with respect to the 
suspension shear stress.
Lastly, the flow pattern and the elastic stress development are examined for different values of $\phi_p$ and $\rm{Wi}$. 
In dilute cases, shear-thickening is attributed to the elastic stress near each particle. 
As $\phi_p$ and/or $\rm{Wi}$ increase, the relative motion of the particles affects the local flow pattern and polymer stretch around the particles. 
At $\rm{Wi}\gtrsim 2$ in the semi-dilute case, the elastic stress between the passing particles does not fully relax to form an additional streak-shaped region of high elastic stress. {Although the impact of such polymer stretching structures on the bulk suspension rheology is likely to be small within the scope of this study, further study for the microstructures and corresponding polymer stretching structures at higher $\phi_p$ and ${\rm Wi}$ will be necessary.}

\vspace{\baselineskip}
\noindent{\bf Acknowledgements} 

\noindent The numerical calculations were mainly carried out using the computer facilities at the Research Institute for Information Technology at Kyushu University. This work was supported by Grants-in-Aid for Scientific Research (JSPS KAKENHI) under Grants No.~JP18K03563.

\vspace{\baselineskip}
\noindent{\bf Declaration of Interests}

\noindent The authors report no conflict of interest.

\appendix

\section{\label{sc:tensor_express}Tensorial representation of equations in an oblique coordinate system}

\subsection{oblique coordinate system}
\noindent To impose simple shear flow on the system, a time-dependent oblique coordinate $\bm{\hat{r}}$ evolving with mean shear 
velocity $\bm{U}=\dot{\gamma}r^2\bm{e}_1$ is introduced as
\begin{align}
    \hat{r}^1&=r^1-\dot{\gamma}tr^2,\label{eq:r1_trans}\\
    \hat{r}^2&=r^2,\\
    \hat{r}^3&=r^3,\\
    \hat{t}&=t\label{eq:time_trans},
\end{align} 
where the quantities with a caret $(\hat{\cdot})$ represent 
variables observed in the oblique coordinate system, and the 
upper indices 1,2, or 3 represent the shear-flow, velocity-gradient, and vorticity directions, respectively. 
By introducing an oblique coordinate system, advection by the 
mean flow, whose term explicitly depends on $r^2$, {{\itshape 
i.e.}, $(\bm{U}\cdot\nabla)=\dot{\gamma}r^2\partial/\partial r^1$,} 
is eliminated from the shear-enforced hydrodynamic equations. 
This enables the use of the periodic boundary 
conditions~\citep{Rogallo1981,Kobayashi2011,Molina2016}. From the 
coordinate transformation, the covariant and contravariant 
transformation matrices 
$[\bm{\Lambda}]_{\nu\mu}=\Lambda^{\nu}_{\;\;\;\mu}=\partial 
r^{\nu}/\partial \hat{r}^{\mu}$ and 
$[\bm{\Lambda}']_{\mu\nu}=\Lambda'^{\mu}_{\;\;\;\;\nu}=\partial 
\hat{r}^{\mu}/\partial r^{\nu}$ are derived as
\begin{align}
\bm{\Lambda}= \left(
    \begin{array}{ccc}
      1 & \gamma(t) & 0 \\
      0 & 1 & 0 \\
      0 & 0 & 1
    \end{array}
  \right), ~~
  \bm{\Lambda}'= \left(
    \begin{array}{ccc}
      1 & -\gamma(t) & 0 \\
      0 & 1 & 0 \\
      0 & 0 & 1
    \end{array}
  \right),
\end{align} 
respectively, where 
$\bm{\Lambda}\cdot\bm{\Lambda'}=\bm{\Lambda}'\cdot\bm{\Lambda}=\b
m{I}$ by definition.
Einstein's summation rule is applied hereafter. By using 
transformation matrices, the covariant and contravariant basis 
vectors $\hat{\bm{E}}_{\mu}$ and $\hat{\bm{E}}^{\mu}$, 
respectively, and the corresponding components of the position 
vectors 
$\bm{r}=r_{\mu}\bm{e}^{\mu}=r^{\mu}\bm{e}_{\mu}=\hat{r}^{\mu}\hat
{\bm{E}}_{\mu}=\hat{r}_{\mu}\hat{\bm{E}}^{\mu}$, are represented 
as
\begin{align}
\hat{\bm{E}}_{\mu}=\Lambda^{\nu}_{\;\;\;\mu}\bm{e}_{\nu},~\hat{\bm{E}}^{\mu}=\Lambda'^{\mu}_{\;\;\;\;\nu}\bm{e}^{\nu}\\
\hat{r}_{\mu}=\Lambda^{\nu}_{\;\;\;\mu}r_{\nu},~\hat{r}^{\mu}=\Lambda'^{\mu}_{\;\;\;\;\nu}r^{\nu}.
\end{align}
Since the oblique coordinate system is not an orthogonal system, covariant and contravariant bases are used,
where $\hat{\bm{E}}_{\mu}\cdot\hat{\bm{E}}^{\nu}=\delta_{\mu}^{\;\;\;\nu}$ holds. 
The lower and upper indices ($\mu,\,\nu=1, 2, 3$) of the tensor 
variables represent the covariant and contravariant components of 
the tensor, respectively.
Fig.~\ref{fig:obl} shows a schematic diagram of this transformation; a 2D diagram on the shear plane is used for the sake of explanation.  
At $t=0$, the basis vectors of the oblique coordinates 
$\hat{\bm{E}}_1$ and $\hat{\bm{E}}_2$ coincide with those of the 
static Cartesian coordinates $\bm{e}_1$ and $\bm{e}_2$.
At $t>0$, the second basis vector of the oblique coordinate changes with time. 
The contravariant metric tensor for the oblique coordinate is defined as
\begin{align}
\label{eq:metric}
    G^{\mu\nu}\equiv\hat{\bm{E}}^{\mu}\cdot\hat{\bm{E}}^{\nu}= \left(
    \begin{array}{ccc}
      1+\gamma^2(t) & -\gamma(t) & 0 \\
      -\gamma(t) & 1 & 0 \\
      0 & 0 & 1
    \end{array}
  \right).
\end{align}
Note that, in the static Cartesian coordinate system, there is no 
distinction between the covariant and contravariant 
expressions, {\itshape i.e.}~$r_{\mu}=r^{\mu}$ and $\bm{e}_{\mu}=\bm{e}^{\mu}$, and the 
metric tensor is identical to the unit tensor, {\itshape i.e.} ~$\bm{G}=\bm{I}$. 
In the oblique system in Fig.~\ref{fig:obl}, where the 
coordinates are non-orthogonal but linear and spatially 
homogeneous, the metric tensor is time-varying and spatially 
constant.
In this situation, the Christoffel term in the covariant 
differentiation is zero, and the covariant differentiation is 
represented by usual partial 
differentiation:~$\hat{\nabla}_{\mu}=\partial/\partial\hat{r}^{\mu}$.

\begin{figure}    
\centerline{\includegraphics[scale=1.0]{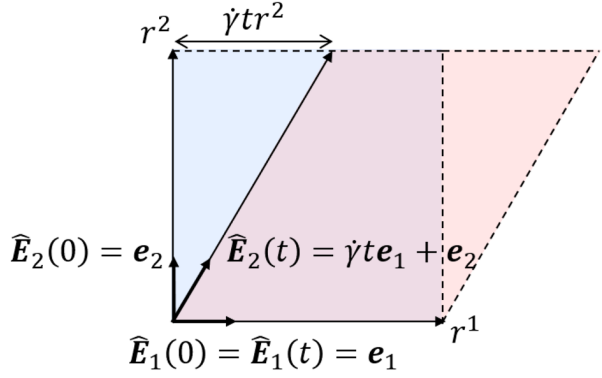}}
\caption{Schematic diagram of the oblique coordinate system. Here 
$\hat{\bm{E}}_1$ and $\hat{\bm{E}}_2$, and $\bm{e}_1$ and 
$\bm{e}_2$ are the basis vectors in the oblique system and those 
in the static Cartesian system, respectively; and $r^1$ and $r^2$ are 
the components of the position vector in the Cartesian system. 
The Cartesian system and the initial oblique system 
coincide~(blue square). At $t>0$, the oblique system is 
sheared~(pink parallelogram) by the strain of $\dot{\gamma}t$, 
where $\dot{\gamma}$ is the applied shear rate. The region of the 
oblique system outside the initial square~(right triangle) can be 
periodically transformed back into the square~(left 
triangle).}\label{fig:obl}
\end{figure}

The periodicity in the governing equations can also be achieved by 
only the coordinate 
transformation~(Eqs.~(\ref{eq:r1_trans})-(\ref{eq:time_trans})) 
with the orthogonal basis system~\citep{Rogallo1981,Onuki1997} 
without using the oblique dual-basis system. This single-basis 
formalism has the advantage that the tensorial representation for 
an equation is expressed uniquely but has the disadvantage that a 
spatial differential operator includes the cross oblique term 
explicitly. In contrast, the dual-basis formalism adopted in our 
method has the advantage that the forms of differential operators 
and governing equations in the oblique coordinate system are 
almost the same as that in the orthogonal system, as explained in 
Sec.~\ref{sec:obl_eqs}, although these forms have dual~(covariant 
and contravariant) expressions. This simple expression of 
the governing equations in the coordinate system is preferable for a 
convenient implementation of the practical simulation code.

\subsection{Governing equations in the oblique coordinate system}\label{sec:obl_eqs}
\noindent The tensorial component representation of the fluid 
momentum equation on the general coordinate 
system~\citep{Luo2004,Venturi2009,Molina2016} is
\begin{align}
\label{eq:ns_obl}
    \frac{\delta\hat{u}^{\mu}}{\delta \hat{t}}=\rho^{-1}\hat{\nabla}_\nu\hat{\sigma}^{\nu\mu}+\hat{\phi}\hat{f}_p^{\mu}.
\end{align}
The left-hand side of Eq.~(\ref{eq:ns_obl}) is the intrinsic time 
derivative in the general coordinate system,
\begin{align}
\label{eq:intrinsic_dt}
    \frac{\delta \hat{A}^{\mu}}{\delta \hat{t}}\equiv\frac{\partial \hat{A}^{\mu}}{\partial \hat{ t}}+(\hat{u}^{\nu}-\hat{U}^{\nu})\hat{\nabla}_{\nu}\hat{A}^{\mu}+\hat{A}^{\nu}\hat{\nabla}_{\nu}\hat{U}^{\mu},
\end{align}
where $\hat{U}^{\mu}\equiv-\partial\hat{r}^{\mu}/\partial t$ is 
the moving velocity of the coordinate and for simple shear flow 
$\bm{U}=\dot{\gamma}(t)r^2\bm{e}_1=\dot{\gamma}(t)\hat{r}^2\hat{\
bm{E}}_1$. Since $\hat{A}^{\mu}$ is defined in the moving system, 
the advection in the second term in Eq.~(\ref{eq:intrinsic_dt}) 
is by the relative velocity to the coordinate flow. The last term 
in Eq.~(\ref{eq:intrinsic_dt}) arises from the affine deformation 
caused by the coordinate flow. 
Introducing the relative fluid velocity to the coordinate flow 
$\bm{\xi}=\bm{u}-\bm{U}$,  Eq.~(\ref{eq:ns_obl}) becomes
\begin{align}
\label{eq:fluid_obl}
    (\hat{\partial}_{\hat{t}}+\hat{\xi}^{\nu}\hat{\nabla}_{\nu})\hat{\xi}^{\mu}=\rho^{-1}\hat{\nabla}_{\nu}\hat{\sigma}^{\nu\mu}+\hat{\phi}\hat{f}_p^{\mu}-2\dot{\gamma}(t)\hat{\xi}^2\delta^{\mu,1},
\end{align}
with the incompressibility condition $\hat{\nabla}_{\mu}\hat{\xi}^{\mu}=0$,
where $\hat{\partial}_{\hat{t}}\equiv\partial/\partial \hat{t}|_{\hat{r}^{\mu}}=\partial/\partial t|_{r^{\mu}}+\dot{\gamma}(t)r^2\partial/\partial r^1$. 
The last term in Eq.~(\ref{eq:fluid_obl}) arises from the spatial 
gradient of the coordinate flow. Since this equation does not 
explicitly depend on the coordinate components $\hat{r}^{\mu}$, 
periodic boundary conditions can be assigned to 
Eq.~(\ref{eq:fluid_obl}), and hence Eq.~(\ref{eq:fluid_obl}) can 
be solved by a spectral method~\citep{Rogallo1981,Canuto1988}. 
The stress tensor gradient in a Newtonian fluid is obtained as
\begin{align}
\label{eq:stress_grad}
    \hat{\nabla}_{\nu}\hat{\sigma}_n^{\nu\mu}=-G^{\nu\mu}\hat{\nabla}_{\nu}\hat{p}+\eta_sG^{\nu\gamma}\hat{\nabla}_{\nu}\hat{\nabla}_{\gamma}\hat{\xi}^{\mu}.
\end{align}
In a viscoelastic fluid, the polymer stress gradient term 
$\hat{\nabla}_{\nu}\hat{\sigma}_p^{\nu\mu}$ is considered in 
addition to Eq.~(\ref{eq:stress_grad}). In our method, the 
tensorial expression for the constitutive equation of the polymer 
stress is additionally introduced in a manner consistent with the 
previous Newtonian formulation~\citep{Molina2016}.

The intrinsic time derivative for conformation tensor $\bm{C}=\hat{C}^{\mu\nu}\hat{\bm{E}}_{\mu}\hat{\bm{E}}_{\nu}$, which is represented by its second-rank contravariant tensor, is expressed as~\citep{Venturi2009}
\begin{align}
\label{eq:dt_tensor}
\frac{\delta \hat{C}^{\mu\nu}}{\delta \hat{t}}\equiv\frac{\partial \hat{C}^{\mu\nu}}{\partial \hat{t}}+(\hat{u}^{\gamma}-\hat{U}^{\gamma})\hat{\nabla}_{\gamma}\hat{C}^{\mu\nu}+\hat{C}^{\mu\gamma }\hat{\nabla}_{\gamma}\hat{U}^{\nu}+\hat{C}^{\gamma\nu}\hat{\nabla}_{\gamma}\hat{U}^{\mu},
\end{align} and the upper-convected time derivative is expressed by
\begin{align}
\label{eq:old_dt}
    \frac{d_c\hat{C}^{\mu\nu}}{d\hat{t}}\equiv\frac{\delta \hat{C}^{\mu\nu}}{\delta \hat{t}}-\hat{C}^{\gamma\nu}\hat{\nabla}_{\gamma}\hat{u}^{\mu}-\hat{C}^{\mu\gamma}\hat{\nabla}_{\gamma}\hat{u}^{\nu}.
\end{align}
Substituting Eq.~(\ref{eq:dt_tensor}) into Eq.~(\ref{eq:old_dt}), 
one obtains
\begin{align}
\label{eq:old_dt_obl}
     \frac{d_c\hat{C}^{\mu\nu}}{d\hat{t}}=\frac{\partial \hat{C}^{\mu\nu}}{\partial \hat{t}}+\hat{\xi}^{\gamma}\hat{\nabla}_{\gamma}\hat{C}^{\mu\nu}-\hat{C}^{\mu\gamma}\hat{\nabla}_{\gamma}\hat{\xi}^{\nu}-\hat{C}^{\gamma\nu}\hat{\nabla}_{\gamma}\hat{\xi}^{\mu}.
\end{align} 
By using Eq.~(\ref{eq:old_dt_obl}), the single-mode Oldroyd-B constitutive equation in the general coordinate system is represented as
\begin{align}
\label{eq:old_obl}
    \frac{\partial \hat{C}^{\mu\nu}}{\partial \hat{t}}+\hat{\xi}^{\gamma}\hat{\nabla}_{\gamma}\hat{C}^{\mu\nu}=\hat{C}^{\mu\gamma}\hat{\nabla}_{\gamma}\hat{\xi}^{\nu}+\hat{C}^{\gamma\nu}\hat{\nabla}_{\gamma}\hat{\xi}^{\mu}-\frac{1}{\lambda}\left(\hat{C}^{\mu\nu}-G^{\mu\nu}\right),
\end{align}
\begin{align}
    \hat{\sigma}_p^{\mu\nu}=\frac{\eta_p}{\lambda}\left(\hat{C}^{\mu\nu}-G^{\mu\nu}\right).
\end{align}
Here, again, Eq.~(\ref{eq:old_obl}) is independent of the coordinate components and has the same form as that in the orthogonal coordinate system. Therefore, periodic boundary conditions can be assigned to Eq.~(\ref{eq:old_obl}).
\begin{figure*}
\centerline{
\includegraphics[scale=0.8]{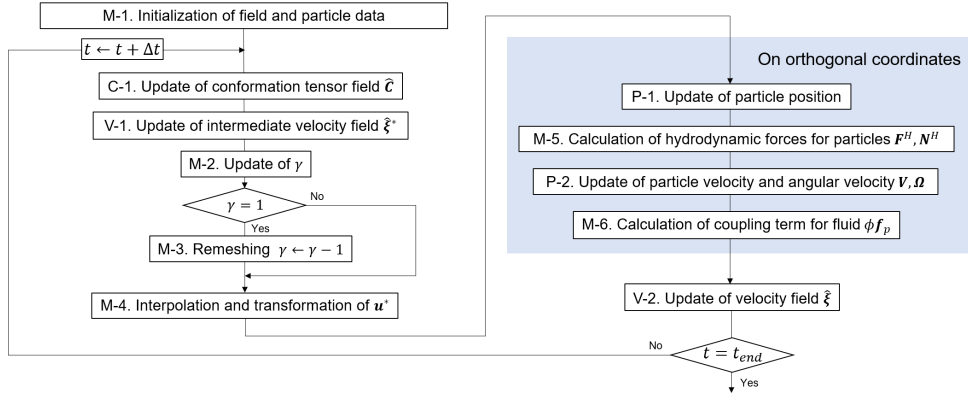}}
\caption{\label{fig:flow_dia} Flow chart of the main calculation procedure over one time step.}
\end{figure*}

\section{Numerical implementation}
\label{sc:numerical_implement}
\subsection{Time-stepping algorithm for the coupling between fluid and particles}

\noindent A flow chart showing the calculation procedure for one 
time step calculations is shown in Fig.~\ref{fig:flow_dia}. The 
couplings between the flow and conformation and between the fluid 
and particles are established in the following explicit 
fractional step approach. Throughout the evolution process, field 
variables are converted from real space to wavenumber space and 
{\itshape vice versa} as necessary. In this section, continuum variables in 
the Fourier space are denoted by the subscript $\bm{k}$, where 
$\bm{k}$ represents the wavenumber vector.
The discretized $n$-th time step is indicated by the superscript 
of a variable as $(\cdot)^{n}$. Here, the constitutive equation 
is a single-mode Oldroyd-B model to ease explanation, and the 
extension to the multi-mode constitutive equations is 
straightforward.
The calculation proceeds according to the following procedure.

\begin{enumerate}
    \item Initialization of variables (M-1).
Starting with the coordinate strain $\gamma=0$, the field 
variables are initialized as 
$\bm{u}=\bm{\xi}=\bm{0},\bm{C}=\bm{I}$ at $t=0$ over the entire 
domain. Correspondingly, the translational and angular velocities 
of the particles are set to zero. 
For a many-particle system, the positions of the particles are 
randomly generated to keep the distance between particle surfaces 
at least $2\Delta$, where $\Delta$ is the grid size. 
    
    \item Update of the conformation tensor field (C-1).
    The conformation tensor $\hat{\bm{C}}$ is updated to the next time step by integrating Eq.~(\ref{eq:old_obl}) or Eq.~(\ref{eq:psi_evo_obl}) over time to obtain the polymer stress field.
    As mentioned in Sec.~\ref{sc:stress}, the small error of $\hat{\bm{C}}$ accumulates in the inner particle region according to the time evolution. This error can be eliminated by resetting $\hat{\bm{C}}=\bm{G}$ over the $\hat{\phi}=1$ region if necessary. In this study, this reset operation is safely omitted because the error is sufficiently small. 
    
    \item Update of the intermediate velocity field (V-1). 
Equation~(\ref{eq:fluid_obl}) without the 
    $\hat{\phi}\hat{\bm{f}}_p$ term is time-integrated to obtain
    an intermediate velocity field $\hat{\bm{\xi}}^*$. In this step, the fluid stress, {\itshape i.e.}, the solvent and polymer stresses, are considered, and 
    the solid--fluid coupling is not considered. 
    
    \item Update of the shear strain (M-2). After the field calculations, the shear strain is updated: $\gamma^{n+1}=\gamma^n+\dot{\gamma}\Delta t$, where $\Delta t$ is the time increment. This corresponds to the deformation of the oblique coordinate system. In this step, if the apparent shear strain equals the threshold value $\gamma_{\rm{th}}$, the remeshing process is conducted. In this study, $\gamma_{\rm{th}}=1$. 
    
    \item Remeshing (M-3). Practically, as time evolves, the 
oblique mesh is gradually distorted, which can lead to a decrease 
in accuracy. 
To continue the simulation as the strain increases 
infinitely while maintaining accuracy, the strained oblique 
coordinates should be reset to a less strained state or to the 
static Cartesian coordinates at some finite shear 
strain~\citep{Rogallo1981}. In this study, the oblique coordinate 
system is reset to the orthogonal Cartesian coordinate system 
when $\gamma$ reaches $\gamma_{\rm{th}}=1$. First, the shear 
strain of the oblique system is reset as 
$\gamma\leftarrow\gamma-\gamma_{\rm{th}}$. Then, the field 
variables $\hat{\bm{\xi}}^*,\hat{\bm{C}}$ on the oblique grid 
outside of the initial orthogonal grid are remapped through the 
periodic boundary in the flow direction. Simultaneously, the 
components of the variables in the oblique coordinate system are 
transformed to those in the reset coordinate system by the 
transformation matrix. Correspondingly, the metric tensor 
(Eq.~(\ref{eq:metric})) and the norm of the wavenumber vector in 
the spectral scheme, $\hat{\bm{k}}\cdot\hat{\bm{k}}$, are updated. 
The norm of wavenumber vector in the wavenumber space 
corresponds to the Laplacian operator in real space, {\itshape i.e.} 
$G^{\mu\nu}\hat{\nabla}_{\mu}\hat{\nabla}_{\nu}\Leftrightarrow-G^
{\mu\nu}\hat{k}_{\mu}\hat{k}_{\nu}=-\hat{\bm{k}}\cdot\hat{\bm{k}}
$, and then
\begin{align}
    \hat{\bm{k}}\cdot\hat{\bm{k}}=\hat{k}_1^2+(\hat{k}_2-\gamma\hat{k}_1)^2+\hat{k}_3^2,
\end{align} where $\hat{k}_i$ is the $i$-th component of the covariant wavenumber vector in the oblique coordinate system.

    \item Interpolation and transformation of the intermediate velocity field (M-4). To simplify the reconstruction of the $\phi$ field based on particle positions, the coupling between the fluid and particle is treated on the usual static orthogonal coordinate system. The grid points in the oblique coordinate system do not always coincide with those in the static orthogonal coordinate system. Therefore, the intermediate velocity field $\hat{\bm{\xi}}^*$ on the oblique grids should be interpolated to the static orthogonal grids. This is done by using a periodic cubic spline interpolation~\citep{Molina2016}. After the interpolation, the oblique-basis components of $\hat{\bm{\xi}}^*$ are transformed to those in Cartesian basis. In this step, the absolute velocity field $\bm{u}^*$ is constructed using the transformed $\bm{\xi^*}$ and the base flow $\dot{\gamma}r^2\bm{e}_1$:
\begin{align}
    \bm{u}^*=(\dot{\gamma}r^2\delta^{\mu,1}+\Lambda^{\mu}_{\;\;\nu}\hat{\xi}^{\nu*})\bm{e}_{\mu}.
\end{align}

    \item Update of the particle position (P-1).
    Hereafter, the calculation is conducted in the orthogonal coordinate system (blue block in Fig.~\ref{fig:flow_dia}). Using the particle velocity at the previous time step $\bm{V}^n$, the position of the $i$-th particle is updated:
\begin{align}
\label{eq:r_particle}
\bm{R}_i^{n+1}=\bm{R}_i^n+\int_{t^n}^{t^{n+1}}\bm{V}_i^ndt.
\end{align} 
In this step, if the updated particle position crosses the top and bottom boundaries, the position and velocity of the particle are modified according to the Lees--Edwards boundary conditions~\citep{Lees1972,Kobayashi2011,Molina2016}. In this time, the $\phi$ field is also updated by using the new particle positions consistent with Lees--Edwards boundary conditions. Then, the intermediate particle velocity field $\bm{u}_p^*$ is calculated:
\begin{align}
\phi^{n+1}\bm{u}_p^*=\sum_i\phi_i^{n+1}\left[\bm{V}_i^n+\bm{\Omega}_i^n\times\bm{r}_i^{n+1}\right],
\end{align} where $\bm{r}_i=\bm{r}-\bm{R_i}$.
This corresponds to the mapping of the Lagrangian particle velocity on the Euler velocity field.

    \item Calculation of hydrodynamic forces acting on particles 
	  (M-5). The hydrodynamic force and torque exerted on the 
	  particles $\bm{F}^H$ and $\bm{N}^H$ are calculated by the change in momentum in the particle domain:
\begin{align}
\label{eq:F_hydro}
\int_{t^n}^{t^{n+1}}\bm{F}_i^Hdt&=\int\rho\phi_i^{n+1}(\bm{u}^*-\bm{u}_p^*)d\bm{r},\\
\label{eq:N_hydro}
\int_{t^n}^{t^{n+1}}\bm{N}_i^Hdt&=\int\bm{r}_i^{n+1}\times\rho\phi_i^{n+1}(\bm{u}^*-\bm{u}_p^*)d\bm{r}.
\end{align}

    \item Update of the particle velocity and angular velocity (P-2). Using Eqs.~(\ref{eq:F_hydro}) and (\ref{eq:N_hydro}), the particle velocities are updated as
\begin{align}
\label{eq:v_update}
\bm{V}_i^{n+1}&=\bm{V}_i^n+\frac{1}{M_i}\int_{t^n}^{t^{n+1}}\left[\bm{F}_i^H+\bm{F}_i^C\right]dt,\\
\label{eq:omega_update}
\bm{\Omega}_i^{n+1}&=\bm{\Omega}_i^n+\bm{I}_{p,i}^{-1}\cdot\int_{t^n}^{t^{n+1}}\bm{N}_i^Hdt.
\end{align}
    In this study, for the inter-particle force $\bm{F}^C$, the soft-core (truncated Lenard--Jones) potential, which produces the short-range repulsive force, is adopted:
\begin{align}
    \bm{F}^C_i(\bm{R})&=-\sum_{j\neq i}^NF_{\rm{soft}}(r_{ij})\frac{\bm{r}_{ij}}{|\bm{r}_{ij}|},\\
    F_{\rm{soft}}(r_{ij})&=-\left(\frac{\partial U_{\rm{soft}}}{\partial r}\right)_{r=r_{ij}},\\
    U_{\rm{soft}}(r)&=\begin{cases}
    4\epsilon\left[\left(\frac{2a}{r}\right)^{36}-\left(\frac{2a}{r}\right)^{18}\right]+\epsilon & (r<r_c) \\
    0 & (r\geq r_c),
    \end{cases}
\end{align}
where $\bm{r}_{ij}=\bm{R}_j-\bm{R}_i$ is the distance vector from the $i$-th particle to the $j$-th particle and $r_c=2^{1/18}(2a)$. 
Vector $\bm{r}_{ij}$ is modified according to periodic boundary conditions if necessary.
     This potential force is simply applied to avoid particle overlap. The force parameter $\epsilon$, which tunes the interaction strength, is set at $\epsilon/(\eta_0\dot{\gamma}a^3)=0.561$ in all many-particle calculations in this study. Under denser particle concentration conditions, where the particle collisions and/or friction and its contribution to the total stress can become significant, more realistic modelling of inter-particle force may be required.
 
    \item Calculation of the coupling term for fluid (M-6). Now that both the positions and velocities of particles have been updated, the final particle velocity field $\bm{u}_p$ is obtained as
\begin{align}
    \phi^{n+1}\bm{u}_p^{n+1}=\sum_i\phi_i^{n+1}\left[\bm{V}_i^{n+1}+\bm{\Omega}_i^{n+1}\times\bm{r}_i^{n+1}\right].
\end{align}
Then, the body force $\phi\bm{f}_p$ is calculated as
\begin{align}
\label{eq:body_force}
\int_{t^n}^{t^{n+1}}\phi\bm{f}_p(\bm{x},t)dt=\phi^{n+1}(\bm{u}_p^{n+1}-\bm{u}^*).
\end{align}
To calculate the stresslet (Eq.~(\ref{eq:S_SP})), Eq.~(\ref{eq:body_force}) is further transformed as
\begin{align}
\int_{t^n}^{t^{n+1}}\phi\bm{f}_pdt=\phi^{n+1}(\bm{u}_p^{n+1}-\bm{u}^*_p)-\phi^{n+1}(\bm{u}^*-\bm{u}_p^*).
\end{align}
The first term on the RHS is expressed by the changes in particle velocity $\Delta\bm{V}_i=\bm{V}_i^{n+1}-\bm{V}_i^n$ from Eq.~(\ref{eq:v_update}) and in angular velocity $\Delta\bm{\Omega}_i=\bm{\Omega}_i^{n+1}-\bm{\Omega}_i^n$ from Eq.~(\ref{eq:omega_update}), as
\begin{align}
    \phi^{n+1}(\bm{u}_p^{n+1}-\bm{u}_p^*)&=\sum_i\phi_i^{n+1}[\Delta\bm{V}_i^H+\Delta\bm{\Omega}_i^H\times\bm{r}_i^{n+1}]
\notag\\
 &~~~~+\sum_i\phi_i^{n+1}\Delta\bm{V}_i^C,
\end{align}
where $\Delta\bm{V}_i^H,\,\Delta\bm{\Omega}_i^H$, and $\Delta\bm{V}_i^C$ are the updates by the hydrodynamic forces $\bm{F}_i^H$ and $\bm{N}_i^H$ and the inter-particle force $\bm{F}_i^C$, respectively. Therefore, $\phi\bm{f}_p$ can be decomposed into the individual contributions from the hydrodynamic interactions $\phi\bm{f}_p^H$ and direct inter-particle interactions $\phi\bm{f}_p^C$, as $\phi\bm{f}_p=\phi\bm{f}_p^H+\phi\bm{f}_p^C$. The individual contributions of the body force are expressed to first order in time as
\begin{align}
    \phi\bm{f}_p^H\Delta t&=\sum_i\phi_i^{n+1}[\Delta\bm{V}_i^H+\Delta\bm{\Omega}_i^H\times\bm{r}_i^{n+1}]-\phi^{n+1}(\bm{u}^*-\bm{u}_p^*),\\
    \label{eq:body_force_c}
    \phi\bm{f}_p^C\Delta t&=\sum_i\phi_i^{n+1}\Delta\bm{V}_i^C.
\end{align}
In the calculation of the stresslet contribution from $\phi\bm{f}_p^C$, the direct virial expression was used instead of Eq.~(\ref{eq:body_force_c}) for computational efficiency~\citep{Molina2016}:
\begin{align}
    \bm{S}^C=-\frac{1}{N}\int_{D_V}\bm{r}\rho\phi\bm{f}_p^Cd\bm{r}=-\frac{1}{N}\sum_{i<j}\bm{r}_{ij}\bm{F}_{ij}^C,
\end{align} where $\bm{F}_{ij}^C$ is the inter-particle force on the $i$-th particle due to the $j$-th particle. In this study, for conditions up to $\phi_p=0.1$ and $\rm{Wi}=2.5$, the contribution of $S_{12}^C$ to the total stresslet $S_{12}$ is small compared to the hydrodynamic contributions~($\langle S_{12}^C\rangle/\langle S_{12}\rangle<0.05$ at $\phi_p=0.1$).

    \item Update of the velocity field (V-2). Finally, the integrated body force $\phi\bm{f}_p$ is remapped and transformed from the orthogonal coordinate system to the oblique coordinate system and added to the intermediate velocity field $\hat{\bm{\xi}}^*$:
\begin{align}
\label{eq:xi_update}
    \hat{\bm{\xi}}^{n+1}=\hat{\bm{\xi}}^*+\widehat{\left[\int_{t^n}^{t^{n+1}}\phi\bm{f}_pds\right]}.
\end{align} 
    At this stage, incompressibility is assigned in the Fourier space,
\begin{align}
    \hat{\bm{\xi}}_{\bm{k}}\leftarrow\hat{\bm{\xi}}_{\bm{k}}-\frac{(\hat{\xi}_{\bm{k}}\cdot\hat{\bm{k}})}{\hat{\bm{k}}\cdot\hat{\bm{k}}}\hat{\bm{k}}.
\end{align}
This solenoidal projection is also adopted after calculating $\hat{\bm{\xi}}^*$~(V-1). 
\end{enumerate}

The described fractional steps are repeated until the calculated 
time reaches the target final time. 
Further information about the time-stepping algorithm is detailed 
in previous work~\citep{Nakayama2008,Molina2016}. 

\subsection{Spatial discretization and time integral scheme}
\noindent Since the periodic boundary conditions are assigned in each direction, 
the continuum variables such as $\hat{\bm{\xi}}$, and $\hat{\bm{C}}$ are Fourier-transformed. 
In real space, the continuum variables are collocated on the 
equispaced mesh point with spacing $\Delta$. Spatial derivatives 
are calculated in Fourier space while the second-order terms like 
the advection term in Eq.~(\ref{eq:fluid_obl}) and 
(\ref{eq:old_obl}) are calculated by a transformation 
method~\citep{Orszag1969}.

For the integration of $\hat{\bm{\xi}}_{\bm{k}}$ over time, the 
exact linear part~(ELP) method, which is preferred for solving 
stiff equations~\citep{Beylkin1998}, is adopted, where the 
nonlinear part is discretized by the Euler method. 

For the polymer constitutive equation, the explicit Euler 
method~(first-order) is adopted. In this study, to evaluate a 
single-particle system that corresponds to very dilute 
suspensions~($\phi_p\sim0.001$), the discretized Eq.~(\ref{eq:old_obl}) 
is solved directly. 
This naive implementation has been stable and accurate in such 
dilute conditions. However, at high $\phi_p$ and ${\rm Wi}$ 
conditions, the large growth rate of the polymer stress around 
the particles violates the positive definiteness of the 
conformation tensor, thus resulting in an inaccurate solution or 
divergence. Therefore, to evaluate a many-particle system that 
corresponds to semi-dilute suspensions~($0.025\leq\phi_p\leq0.1$), 
the log-conformation formalism is used in which the time 
evolution equation of $\log\bm{C}$ rather than $\bm{C}$ is solved 
to guarantee the positive-definiteness of 
$\bm{C}$~\citep{Fattal2004,Hulsen2005}. A detailed description of 
the log-conformation formalism is provided in 
Appendix~\ref{sc:log_c}.

The particle position is updated by discretizing 
Eq.~(\ref{eq:r_particle}) by the Euler method~(first-order) at 
the first time step and the second-order Adams--Bashforth scheme 
later. In the update of the particle 
velocity~(Eqs.~(\ref{eq:v_update}) and (\ref{eq:omega_update})), 
the impulsive hydrodynamic force and torque are calculated by 
applying Eqs.~(\ref{eq:F_hydro}) and (\ref{eq:N_hydro}), 
respectively, and the potential force $\bm{F}^C$ in 
Eq.~(\ref{eq:v_update}) is discretized by the second-order Heun 
scheme because both particle positions at $t^n$ and $t^{n+1}$ are 
already obtained in that stage:
\begin{align}
   \frac{1}{M_i}\int_{t^n}^{t^{n+1}}\bm{F}_i^Cds=\frac{\Delta t}{2M_i}\left[\bm{F}_i^C(\bm{R}^{n+1})+\bm{F}_i^C(\bm{R}^n)\right].
\end{align}

The time increment is determined based on the stability given by the fluid momentum diffusion: $\Delta t=\rho/\eta_0K_{\rm max}^2$ ($K_{\rm max}$ is the largest wavenumber in the spectral scheme). As proven in the code validations in Sec. \ref{sc:single_rheo} and Appendix~\ref{sc:validation}, this choice is reasonable considering the conditions in this study.

\subsection{Log-conformation-based constitutive equation for Oldroyd-B model}
\label{sc:log_c}

\noindent Because the conformation tensor $\bm{C}$ is real-symmetric and positive-definite, $\bm{C}$ can be diagonalized as
\begin{align}
\label{eq:lambda}
    \bm{C}=\bm{R}\cdot\bm{\Lambda}\cdot\bm{R}^T,
\end{align} where $\bm{\Lambda}=\rm{diag}(\lambda_1,\lambda_2,\lambda_3)$ and $\lambda_i>0\,(i=1,2,3)$ are the eigenvalues of $\bm{C}$, and $\bm{R}$ is the rotation matrix composed of the eigenvectors of $\bm{C}$. Here, the new tensor variable $\bm{\Psi}$ is introduced~\citep{Fattal2004} as
\begin{align}
\label{eq:lambda_p}
    \bm{\Psi}=\bm{R}\cdot\bm{\Lambda}_{\Psi}\cdot\bm{R}^T,
\end{align} where $\bm{\Lambda}_{\Psi}=\rm{diag}(\ln\lambda_1,\ln\lambda_2,\ln\lambda_3)=\ln\bm{\Lambda}$. From Eqs.~(\ref{eq:lambda}) and (\ref{eq:lambda_p}), 
\begin{align}
\label{eq:exp_p}
    \bm{C}=\bm{R}\cdot\exp(\bm{\Lambda}_{\Psi})\cdot\bm{R}^T.
\end{align} Note that, when $\bm{C}$ is obtained through Eq.~(\ref{eq:exp_p}), $\bm{C}$ is strictly positive-definite by definition. Furthermore, utilizing the time evolution of $\bm{\Psi}$ instead of Eq.~(\ref{eq:conformation}), the exponential growth in $\bm{C}$ is translated to the linear growth of $\bm{\Psi}$, which enables numerical stability in the time evolution. Specifically, 
the stretching in the principal axes of $\bm{C}$ by the velocity gradient tensor $\bm{\nabla}\bm{u}$ is extracted as
\begin{align}
    \bm{\nabla u}&=\bm{R}\cdot\bm{M}\cdot\bm{R}^T,\\
    \bm{B}&=\bm{R}\cdot{\rm diag}(M_{11},M_{22},M_{33})\cdot\bm{R}^T,
\end{align} where $\bm{B}$ is symmetric and commutes with $\bm{C}$ by definition.

The residual component $\bm{\nabla}\bm{u}-\bm{B}$ can be decomposed as
\begin{align}
\label{eq:ban}
    \bm{\nabla u}-\bm{B}=\bm{A}+\bm{C}^{-1}\cdot\bm{N},
\end{align} 
with anti-symmetric tensors $\bm{A}$ and $\bm{N}$~\citep{Fattal2004}. 
Tensor $\bm{N}$ is proven to be irrelevant in the affine 
deformation of $\bm{C}$ by inserting Eq.~(\ref{eq:ban}) into the 
upper-convected time derivative of $\bm{C}$. 
On the other hand, $\bm{A}$ represents the rotation of the 
principal axes of $\bm{C}$. From the affine deformation of 
$\bm{C}$ in Eq.~(\ref{eq:conformation}), the explicit expression 
of $\bm{A}$ in the frame of the principal axes of $\bm{C}$ is 
derived by~\cite{Hulsen2005} as
\begin{align}
    A_{ij}=\frac{\lambda_iM_{ij}+\lambda_jM_{ji}}{\lambda_i-\lambda_j},~~i\neq j,~~\lambda_i\neq\lambda_j,
\end{align}
(the summation convention is not applied here). When 
$\lambda_i=\lambda_j$, $A_{ij}$ is not uniquely determined in the 
decomposition of $\bm{\nabla u}$ in Eq.~(\ref{eq:ban}), but the 
affine deformation of $\bm{C}$ and $\bm{\Psi}$ by $\bm{\nabla u}$ 
is still well defined, which case is explained next.

By using these tensors, the governing equation of $\bm{\Psi}$ for the single-mode Oldroyd-B model is expressed as
\begin{align}
\label{eq:psi_evo}
    \left(\frac{\partial }{\partial t}+\bm{u}\cdot\bm{\nabla}
\right)
\bm{\Psi}&=-\bm{A}\cdot\bm{\Psi}+\bm{\Psi}\cdot\bm{A}+2\bm{B}
\notag
\\&~~~~+\bm{R}\cdot\left[\frac{1}{\lambda}(\bm{\Lambda}^{-1}-\bm{I})\right]\cdot\bm{R}^T.
\end{align} 
When $\lambda_i=\lambda_j$, the corotational terms including $A_{ij}$ are reduced as (in the frame of the principal axes of \(\bm{C}\))
\begin{align*}
-A_{ij}\Psi_{jj}+\Psi_{ii}A_{ij} 
&=\left(\ln\lambda_i-\ln\lambda_j\right)
\frac{\lambda_iM_{ij}+\lambda_jM_{ji}}{\lambda_i-\lambda_j}
\\
&\to M_{ij}+M_{ji}=2D_{ij},
\end{align*}
(the summation convention is not applied here).
With this treatment, the evolution equation~(\ref{eq:psi_evo}) of 
\(\bm{\Psi}\) works safely even when \(\lambda_i=\lambda_j\) 
happens.

In the initial conditions, $\bm{u}=\bm{0}$ over the entire domain 
leads to $\bm{A}=\bm{0}$ and $\bm{B}=\bm{D}$, and 
$\bm{C}=\bm{I}$ results in $\bm{R}=\bm{I},\bm{\Lambda}=\bm{I}$, 
and $\bm{\Psi}=\bm{0}$.
The evolution of $\bm{\Psi}$ according to Eq.~(\ref{eq:psi_evo}) 
is solved by numerical simulation; and $\bm{C}$ is calculated 
from $\bm{\Psi}$ via Eqs.~(\ref{eq:lambda_p}) and (\ref{eq:exp_p}).
The contravariant tensor expression corresponding to 
Eq.~(\ref{eq:psi_evo}) in the oblique coordinates is the 
following:
\begin{widetext}
\begin{align}
\label{eq:psi_evo_obl}
\begin{split}
\left(
\frac{\partial}{\partial \hat{t}}+\hat{\xi}^{\gamma}\hat{\nabla}_{\gamma}
\right)
\hat{\Psi}^{\mu\nu}
=&-\dot{\gamma}\left(\hat{\Psi}^{\mu2}\delta^{1,\nu}+\hat{\Psi}^{2\nu}\delta^{1,\mu}\right)
-G_{\gamma\zeta}\hat{A}^{\mu\gamma}\hat{\Psi}^{\zeta\nu}
+G_{\gamma\zeta}\hat{\Psi}^{\mu\gamma}\hat{A}^{\zeta\nu}
+2\hat{B}^{\mu\nu}+\hat{R}^{\mu\gamma}\left[\frac{1}{\lambda}(
[\hat{\bm{\Lambda}}]^{-1}_{\gamma\zeta}
-G_{\gamma\zeta})\right]\hat{R}^{\nu\zeta},
\end{split}
\end{align}
\end{widetext} 
where 
$[\hat{\bm{\Lambda}}]^{-1}_{\gamma\zeta}$ represents the 
covariant matrix component of $\hat{\bm{\Lambda}}^{-1}$, which is 
simply the matrix inverse of the contravariant matrix 
$\hat{\Lambda}^{\gamma\zeta}$; and $G_{\gamma\zeta}$ is the covariant metric tensor, which is defined as
\begin{align}
    G_{\gamma\zeta}\equiv\hat{\bm{E}}_{\gamma}\cdot\hat{\bm{E}}_{\zeta}=\left(
    \begin{array}{ccc}
      1 & \gamma(t) & 0 \\
      \gamma(t) & 1+\gamma^2(t) & 0 \\
      0 & 0 & 1
    \end{array}
  \right).
\end{align}
Note that, in the oblique coordinates, there is an additional 
term originating from the moving coordinates~(the first term on 
RHS of Eq.~(\ref{eq:psi_evo_obl})).  
Since Eq.~(\ref{eq:psi_evo_obl}) does not explicitly depend on 
the coordinate variables, it can be discretized by a spectral 
method.

\begin{figure}   
  \begin{center}
\includegraphics[scale=0.8]{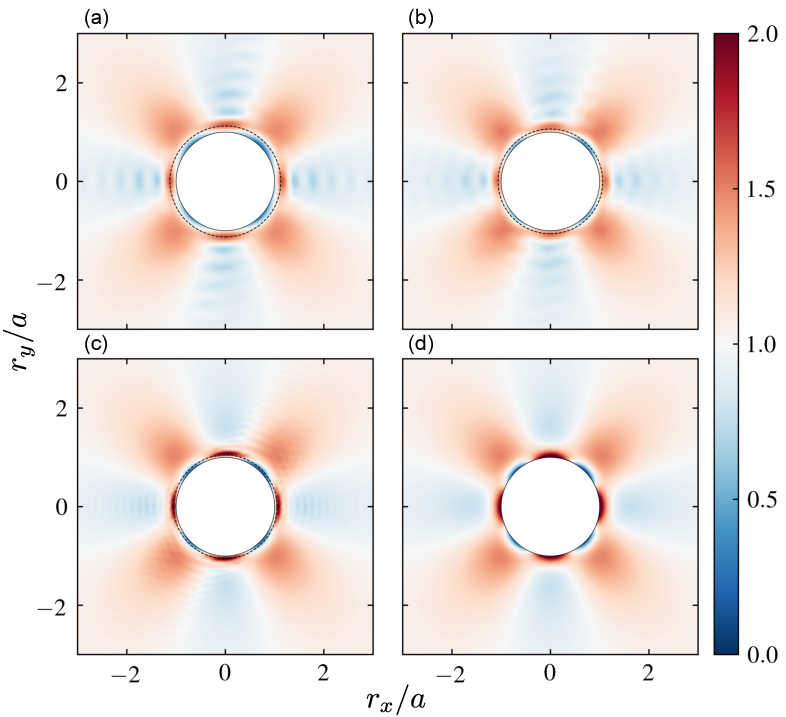}
  \end{center}
  \begin{center}
\includegraphics[scale=0.8]{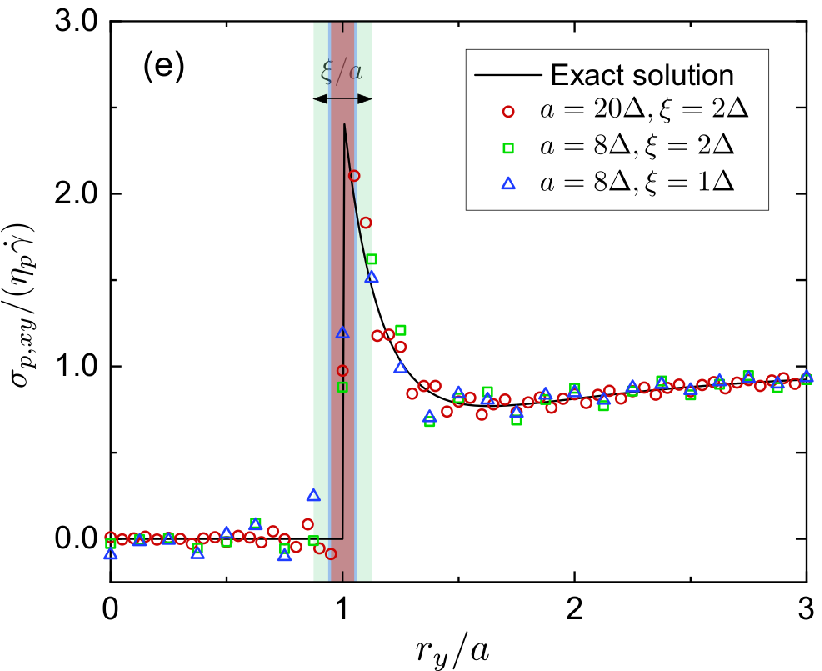}
  \end{center}
\caption{\label{fig:mesh_exam} Mesh resolution dependence of 
polymer shear stress distributions near the particle. 
In panels (a-d) the polymer shear stress normalized by 
$\eta_p\dot{\gamma}$ on the shear plane through the center of the 
particle is drawn by color contour in (a-d) for 
$(a/\Delta,\xi/\Delta)=(8,2),\,(8,1)$, and $(20,2)$ and for the 
analytical solution, respectively. 
The dotted lines around the particles for the DNS results in 
 panels (a-c) 
denote the radial location of $a+\xi/2$. 
In panel (e), the normalized polymer shear stress along the 
line from the particle center to the shear gradient direction $y$ 
is drawn: the exact solution~(analytical) of 
$\bm{\sigma}_p/(\eta_p\dot{\gamma})\approx2\tilde{\bm{D}}-2{\rm 
Wi}\tilde{\bm{D}}_{(2)}$ at ${\rm Wi}\to 0$ (solid line), the 
results for $(a/\Delta,\xi/\Delta)=(20,\,2)$ (red), $(8,2)$ 
(green), and $(8,1)$ (blue), respectively. 
In panel (e), the interface region indicated by $0<\phi<1$ at around 
$r_y/a=1$ with thickeness $\xi$ is colored in the same manner as 
that for the symbols.}
\end{figure}

\section{Validations of the developed method}\label{sc:validation}
\subsection{Polymer stress around a single particle}
\noindent To test the validity of the developed method, a 
single-particle system is set up where a neutrally buoyant 
spherical particle is suspended in a sheared Oldroyd-B 
fluid~(Fig.~\ref{fig:system}(a)). The cubic domain with a box 
length of $L$ is sufficiently large compared to the size of the 
particle used to represent the dilute particle system.
Hereafter, for simplicity, the directions of the Cartesian 
coordinate basis vectors are denoted by $x,y,$ and $z$ instead of 
the $1,2,$ and $3$ notation used in 
Appendices~\ref{sc:tensor_express} and 
\ref{sc:numerical_implement}, where $x,y$, and $z$ indicate the 
flow, velocity-gradient, and vorticity directions, respectively.
As shown in Fig.~\ref{fig:system}(a), because the particle is 
located at the center of simple shear flow, the net translational 
hydrodynamic force acting on the particle $\bm{F}^H$ vanishes 
while the hydrodynamic torque $\bm{N}^H$ rotates the particle.

A flow condition is considered in the $\beta\to 1$,  
small-$\rm{Wi}$ and small-$\rm{Re}$ limit where an analytical 
solution is available.
In this limit, the flow pattern is minimally affected by polymer 
stress, which is expressed 
analytically~\citep{CHEN-JUNGLIN1970,MIKULENCAK2004}. Furthermore, 
when $\rm{Wi}\ll 1$, the polymer stress distribution is 
approximated by the second-order fluid~(SOF) theory; 
$\bm{\sigma}_p=2\eta_p\bm{D}+4(\Psi_1+\Psi_2)\bm{D}\cdot\bm{D}-\P
si_1(\bm{D}_{(2)}+4\bm{D}\cdot\bm{D})$, where $\bm{D}_{(2)}$ is 
the upper-convected derivative of $\bm{D}$~\citep{Bird1987}. 
Considering an Oldroyd-B 
fluid~($\Psi_1=2\eta_p\lambda,\,\Psi_2=0$), the normalized 
polymer stress in the SOF limit is expressed as 
$\bm{\sigma}_p/(\eta_p\dot{\gamma})=2\tilde{\bm{D}}-2{\rm 
Wi}\tilde{\bm{D}}_{(2)}$. Here, $\beta=0.99,\rm{Wi}=0.001$, and 
$\rm{Re}=0.0142$. In this situation, the normalized polymer 
stress is approximated by 
$\bm{\sigma}_p/(\eta_p\dot{\gamma})\approx2\tilde{\bm{D}}$.
 
Different mesh resolutions of the particle interface are 
examined:~$(a/\Delta, \xi/\Delta)= (8, 2), (8, 1),$ and $(20, 2)$.
At first, the overall trend of the polymer shear 
stress~($\sigma_{p,xy}$) distribution is similar for different 
resolutions (Fig.~\ref{fig:mesh_exam}(a-c)) and the analytical 
solution ( Fig.~\ref{fig:mesh_exam}(d)). The only difference is 
that small $\sigma_{p,xy}$ oscillation is observed in the 
numerical solutions.
When comparing the results for $a=8\Delta$ and 
$\xi=2\Delta$~(Fig.~\ref{fig:mesh_exam}(a)) with that for 
$a=20\Delta$ and $\xi=2\Delta$~(Fig.~\ref{fig:mesh_exam}(c)), it 
is clear that as $a/\Delta$ increases, the wavenumber of the 
small ripple in $\sigma_{p,xy}$ increases, but its amplitude 
decreases. This is due to the slow convergence of the Fourier 
series caused by the discontinuous change in $\sigma_{p,xy}$ at 
the solid--liquid interface. {This artifact is a partly 
unavoidable intrinsic property of the spectral method.} 
Regarding the interface thickness, when comparing the results for 
$a=8\Delta$ and $\xi=2\Delta$~(Fig.~\ref{fig:mesh_exam}(a)) with that 
for $a=8\Delta$ and $\xi=\Delta$~(Fig.~\ref{fig:mesh_exam}(b)), no 
significant difference in the overall trend is observed. However, 
when $\xi=\Delta$, the distribution of ${\sigma}_{p,xy}$ near the 
interface is somewhat blurred, which is caused by the decrease in 
the number of mesh points that support the interface region.

For a detailed evaluation, a one-dimensional profile of the 
polymer shear stress in the velocity-gradient direction from the 
particle center is shown  in Fig.~\ref{fig:mesh_exam}(e). The 
steep increase in $\sigma_{p,xy}$ near the particle surface is 
reasonably reproduced as the mesh resolution increases, though 
the peak in $\sigma_{p,xy}$ is somewhat smeared due to the 
limited mesh points in the interface domain. 
Hereafter, considering a balance between the accuracy of the 
numerical solution and the required computational cost, the 
particle radius is set to $a=8\Delta$ and the interface thickness 
to $\xi=2\Delta$. As seen in Appendix~\ref{sc:rotation} and 
Sec.~\ref{sc:apps}, this resolution is sufficiently valid for the 
problems investigated in this study.

\begin{figure}   
  \begin{center}
 \vspace{-0.5mm}
\includegraphics[scale=0.755]{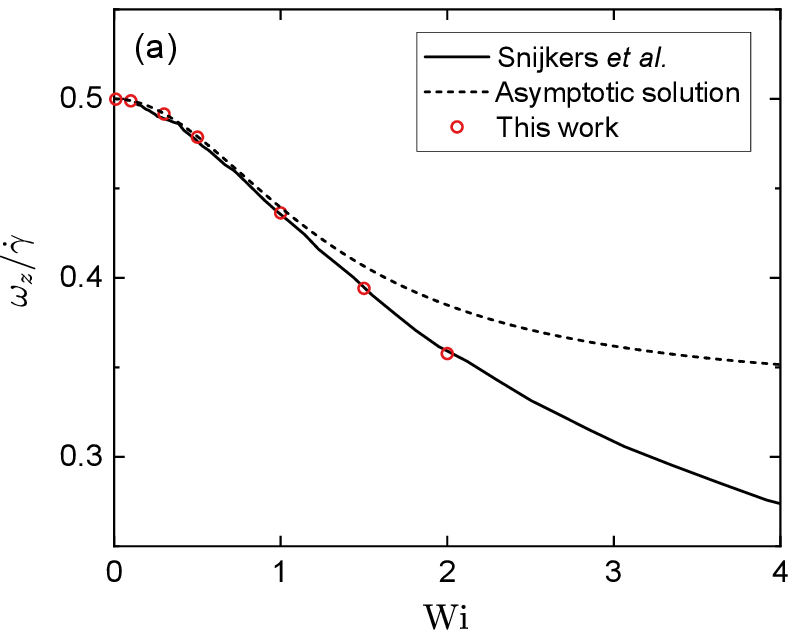}
\label{rot_a}
  \end{center}
  \begin{center}
\includegraphics[scale=0.75]{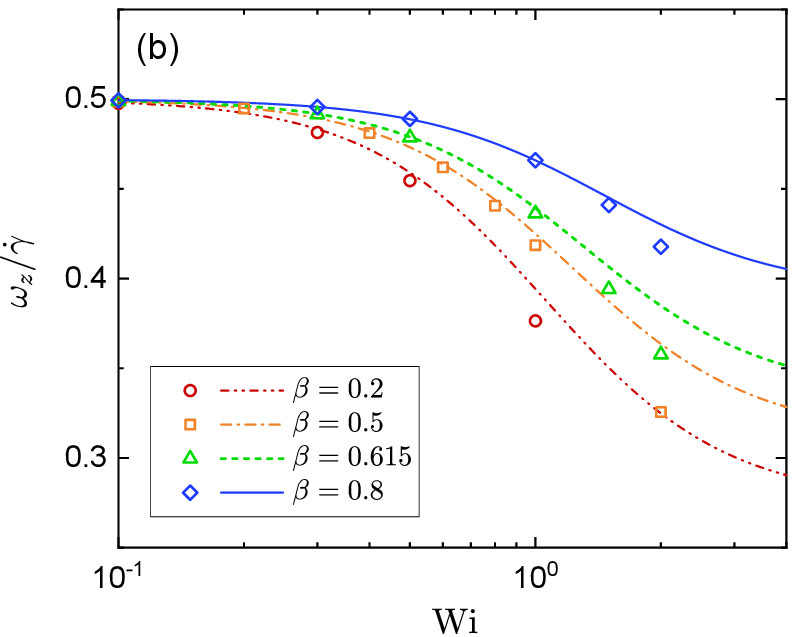}
\label{rot_b}
  \end{center}
\caption{\label{fig:rot} The $\rm{Wi}$ dependence of the normalized 
particle angular velocity $\omega_z/\dot{\gamma}$. (a) The 
result of this work~(red circles) compared with a previous 
numerical result (solid line)~\citep{Snijkers2009} and 
the asymptotic solution (dashed 
line)~\citep{Housiadas2011,Housiadas2011a,Housiadas2018}} at 
$\beta=0.615$. (b) The $\beta$ dependence of 
 $\omega_z/\dot{\gamma}$ for: 
$\beta=0.2$  (red circles), $0.5$ (orange squares), $0.615$ (green 
triangles), and $0.8$ (blue diamonds). The lines correspond to 
predictions of the asymptotic solution.
\end{figure}
 
\begin{figure}    
\centerline{\includegraphics[scale=0.9]{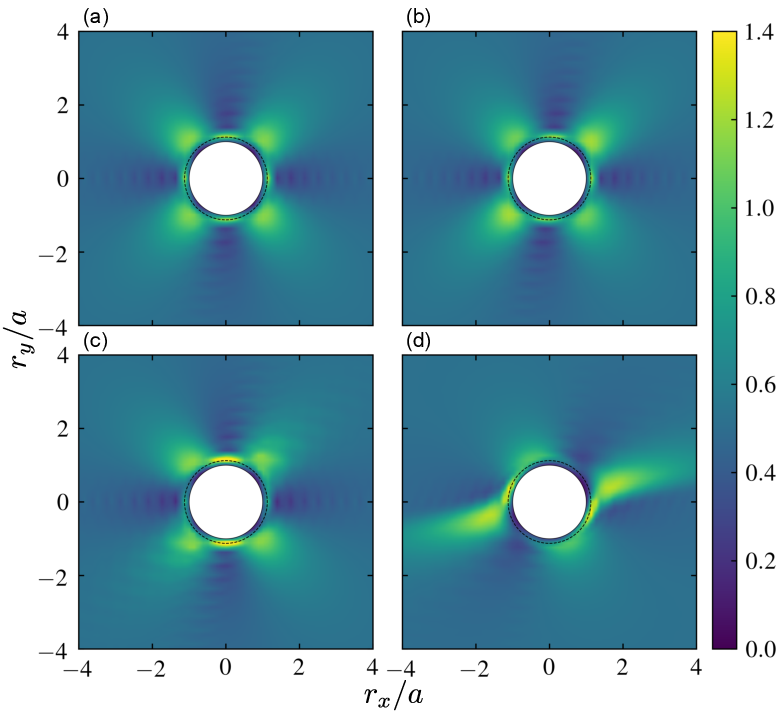}}
\caption{\label{fig:energy_disp} The $\rm{Wi}$ dependence of the 
normalized energy dissipation around a particle at $\beta=0.5$: 
(a)(b) $\rm{Wi}=0.1$ and (c)(d) $\rm{Wi}=1.0$. Panels (a)(c) and (b)(d) 
show the distributions of viscous dissipation $\Phi_{\rm 
s}/(\eta_0\dot{\gamma}^2)$ and polymer dissipation $\Phi_{\rm 
p}/(\eta_0\dot{\gamma}^2)$, respectively. The dotted lines around 
particles show the radial location of $a+\xi/2$.}
\end{figure}

\subsection{Rotation of a particle under simple shear flow}
\label{sc:rotation}\noindent Under simple shear in Stokes flow, 
as is well known, a suspended particle in a Newtonian medium 
rotates with an angular velocity that is half of the applied 
shear rate; $\omega_z/\dot{\gamma}=0.5$, where $\omega_z$ is the 
particle angular velocity in the vorticity direction. However, in 
viscoelastic fluids, this relative rotational speed decreases 
with increasing $\rm{Wi}$~\citep{Snijkers2009,Snijkers2011}. 
\citet{DAvino2008} and \citet{Snijkers2009,Snijkers2011} 
conducted numerical evaluation for this phenomenon using a finite 
element method~(FEM) and surface-conforming mesh, reproducing the 
experimental rotational slowdown data with DNS. 
They observed that the distribution of local torque and pressure 
on the particle surface becomes asymmetrical with $\rm{Wi}$. 
However, the physics of slowdown has not been elucidated.
This result is often referred to as the benchmark problem for a 
newly developed numerical scheme of viscoelastic 
suspensions~\citep{Ji2011,Yang2016,Vazquez-Quesada2017,Fernandes2019}. 
To validate the method developed in this work, the angular 
velocity of a particle in an Oldroyd-B fluid is evaluated at the 
same numerical conditions previously reported~\citep{Snijkers2009}; 
however, no walls are used in this study. 
The numerical setup is the same as that in Sec.~\ref{sc:single_rheo}.
 
Figures.~\ref{fig:rot}(a) and (b) show the $\beta$ and $\rm{Wi}$ dependence of the normalized particle angular velocity $\omega_z/\dot{\gamma}$. 
To compare with the previous numerical result, 
Fig.~\ref{fig:rot}(a) shows the result at $\beta=0.615$. As 
$\rm{Wi}$ increases, $\omega_z/\dot{\gamma}$ decreases. {At 
$\rm{Wi}\lesssim 1$, the result converges with the theoretical 
prediction up to $O(\rm{Wi}^4)$ made using asymptotic 
methods~\citep{Housiadas2011,Housiadas2011a,Housiadas2018}: 
$\omega_z/\dot{\gamma}=1/2-(1-\beta)\rm{Wi}^2/[4(1-4\rm{Wi}^2\tilde{\Omega}_4)]$, where $\tilde{\Omega}_4$ is the coefficient of the
$(1-\beta)\rm{Wi}^4$ term in the series solution.
At $\rm{Wi}\gtrsim 1$,} the asymptotic prediction starts to 
overestimate $\omega_z/\dot{\gamma}$.
In this region, the result agrees reasonably well with the 
previous FEM result~\citep{Snijkers2009}. Fig.~\ref{fig:rot}(b) 
shows the $\beta$ dependence of $\omega_z/\dot{\gamma}$. As 
$\beta$ decreases, which corresponds to an increase in the 
polymer stress contribution, the negative slope of 
$\omega_z/\dot{\gamma}$  increases. This trend is consistent with 
the asymptotic predictions shown in Fig.~\ref{fig:rot}(b).
 
The slowdown of rotation with increasing $\rm{Wi}$ and/or 
$1-\beta$ suggests that the energy partition from external work 
to elastic energy increases.
Fig.~\ref{fig:energy_disp} shows the normalized energy dissipation rate around the particle on the shear plane through the particle center at $\beta=0.5$ and $\rm{Wi}=0.1$ and $1.0$. The dissipation rate is decomposed into 
viscous~($\Phi_{\rm s}$) and elastic~($\Phi_{\rm p}$) contributions~\citep{Vazquez-Quesada2019} as $\Phi_{\rm{t}}=\Phi_{\rm{s}}+\Phi_{\rm{p}}$:
\begin{align}
     \Phi_{\rm s}&=2\eta_s\bm{D}:\bm{D},\\
     \Phi_{\rm p}&=\frac{\eta_p}{2\lambda^2}\left({\rm tr}\bm{C}+{\rm tr}\bm{C}^{-1}-6\right).
\end{align} 
At $\rm{Wi}=0.1$, $\Phi_{\rm s}$~(Fig.~\ref{fig:energy_disp}(a)) 
and $\Phi_{\rm p}$~(Fig.~\ref{fig:energy_disp}(b)) present 
similar distributions. Since 
$\bm{\sigma}_p=(\eta_p/\lambda)(\bm{C}-\bm{I})\to2\eta_p\bm{D}$ 
at $\rm{Wi}\to 0$, $\Phi_{\rm p}$ is reduced to 
$2\eta_p\bm{D}:\bm{D}$, which is proportional to $\Phi_{\rm s}$. 
In contrast, at $\rm{Wi}=1.0$, $\Phi_{\rm s}$ and $\Phi_{\rm p}$ 
develop differently; the high-$\Phi_{\rm p}$ region expands, 
whereas $\Phi_{\rm s}$ does not change much in comparison to the 
$\rm{Wi}=0.1$ case.
The distribution of $\Phi_{\rm p}$ expands towards the shear-flow 
direction and high-$\Phi_{\rm p}$ grows near the equator of the 
particle~(Fig.~\ref{fig:energy_disp}(d)), thus clearly showing an 
increase in the fraction of elastic energy dissipation at high 
$\rm{Wi}$.
Particle rotation is caused by viscous stress. An increase in 
elastic energy leads to a decrease in the relative fraction of 
viscous dissipation. As a result, the angular velocity of the 
particle in the viscoelastic medium decreases in comparison with 
that in viscous media. 
This slowdown of the particle rotation is enhanced with 
$\rm{Wi}$ and $1-\beta$.
 
In this section, the agreement between the numerical results of 
this work and the previously reported numerical and theoretical results verifies that the presented numerical scheme can successfully capture the dynamic coupling between particles and a viscoelastic fluid.
\end{document}